\documentclass[12pt]{article}

\usepackage{amssymb,amsmath,graphicx}
\usepackage{epsf}
\usepackage{graphicx,epsfig}
\usepackage{amsfonts}
\usepackage{amssymb}
\usepackage{cite}

\def\bk{{\bf k}}
\def\bp{{\bf p}}
\def\bq{{\bf q}}
\def\bx{{\bf x}}

\def\CC{{\cal C}}

\def\CH{{\cal H}}
\def\CL{{\cal L}}
\def\CO{{\cal O}}
\def\CP{{\cal P}}

\def\tx{{\tilde x}}

\def\ttau{{\tilde \tau}}

\def\high{\vphantom{\Biggl(}\displaystyle}

\def\half{\frac{1}{2}}

\makeatletter
\renewcommand\section{\@startsection {section}{1}{\z@}%
                                 {-3.5ex \@plus -1ex \@minus -.2ex}%
                                   {2.3ex \@plus.2ex}%
                                   {\normalfont\large\bfseries}}
\renewcommand\subsection{\@startsection{subsection}{2}{\z@}%
                                   {-3.25ex\@plus -1ex \@minus -.2ex}%
                                     {1.5ex \@plus .2ex}%
                                     {\normalfont\bfseries}}
\renewcommand\subsubsection{\@startsection{subsubsection}{3}{\z@}%
                                   {-3.25ex\@plus -1ex \@minus -.2ex}%
                                     {1.5ex \@plus .2ex}%
                                     {\normalfont\itshape}}
\makeatother

\newcommand{\Letter}{
\setlength{\textwidth}{16.5cm}
   \setlength{\textheight}{22.6cm}
    \hoffset=-0.5in
\voffset=-2.1cm }

\Letter

\begin{document}
\newcommand{\be}{\begin{equation}}
\newcommand{\ee}{\end{equation}}
\newcommand{\bea}{\begin{eqnarray}}
\newcommand{\eea}{\end{eqnarray}}
\newcommand{\barr}{\begin{array}}
\newcommand{\earr}{\end{array}}

\thispagestyle{empty}

\vspace*{0.3in}

\begin{center}
{\large \bf Quasi-Single Field Inflation and Non-Gaussianities}

\vspace*{0.5in} {Xingang Chen$^{1,2}$ and Yi Wang$^{3,4}$}
\\[.3in]
{\em
$^1$ Center for Theoretical Cosmology, \\
Department of Applied Mathematics and Theoretical Physics, \\
University of Cambridge, Cambridge CB30WA, UK
\\[.1in]
$^2$ Center for Theoretical Physics, \\
Massachusetts Institute of Technology, \\Cambridge, MA 02139, USA
\\[.1in]
$^3$ Physics Department, McGill University,\\ Montreal, H3A2T8, Canada
\\[.1in]
$^4$ Kavli Institute for Theoretical Physics China,\\
Key Laboratory of Frontiers in Theoretical Physics,\\
Institute of Theoretical Physics, Chinese Academy of Sciences,\\
Beijing 100190, P.~R.~China\\
[0.3in]}
\end{center}

\begin{center}
{\bf Abstract}
\end{center}
\noindent

In quasi-single field inflation models, massive isocurvature modes,
that are coupled to the inflaton and have mass of order the Hubble
parameter, can have nontrivial impacts on density perturbations,
especially non-Gaussianities. We study a simple example of quasi-single field inflation in terms of turning inflaton trajectory. Large bispectra with a
one-parameter family of novel shapes arise, lying between the
well-known local and equilateral shape. The trispectra can also be
very large and its magnitude $t_{NL}$ can be much larger than
$f_{NL}^2$.

\vfill

\newpage

\setcounter{page}{1}

\tableofcontents

\newpage

\section{Introduction and summary}
\setcounter{equation}{0}

Inflation \cite{Guth:1980zm} has become the leading paradigm of the early
universe. However, the detailed dynamics of inflation is still a
mystery. A major theme in cosmology is to build inflationary models and compare their predictions with experimental data. The theoretical predictions of general single field inflation models
have been well understood. Nonetheless, there exists another important possibility that the inflationary dynamics can involve multiple fields.
This leads to a
variety of new models and interesting phenomenologies.

When multiple fields are involved, the field
space can be decomposed into the inflationary direction and isocurvature
directions. The quanta in these field directions are called inflaton and isocurvatons, respectively. In this
paper, we investigate a class of models where there is one flat slow-roll
direction, and all the other isocurvature directions have mass at least of
order the Hubble parameter $H$. We call this class of models {\em quasi-single
field inflation}. If the inflaton decouples from the isocurvatons or
the isocurvaton mass are all much larger than $\CO(H)$,
quasi-single field inflation makes the same prediction as the single
field inflation.
However, once large couplings exist and the mass are of order $H$, we will show that these massive isocurvatons can have important effects on density perturbations.

\begin{figure}[htpb]
\begin{center}
\epsfig{file=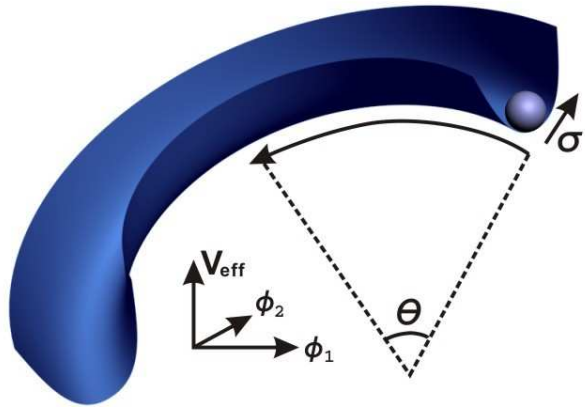, width=8cm}
\end{center}
\caption{\label{Fig:turn}This figure illustrates a model of quasi-single field
  inflation in terms of turning trajectory. The $\theta$ direction is the inflationary
  direction, with a slow-roll potential. The $\sigma$ direction denotes the isocurvature direction, which typically has mass of order $H$.}
\end{figure}

In this paper, we shall study a simple model of quasi-single field inflation \cite{Chen:2009we}. In this model, the coupling between the inflaton and the massive isocurvaton is introduced by a turning trajectory. The tangential direction of this turning trajectory is the usual slow-roll direction, while the orthogonal direction is lifted by a mass of order $H$. See Fig.~\ref{Fig:turn}.

The motivations for investigating quasi-single field inflation are as
follows.

$\bullet$ {\em UV completion and fine-tuning in inflation models.} To satisfy the conditions for inflation, fine-tunings or symmetries should generally be evoked. At least this is found to be the case for models that have reasonable UV completion in string theory and supergravity\cite{Copeland:1994vg,Chen:2008hz}.
For slow-roll inflation, this means that, in the inflationary background, the light fields will typically acquire mass of order the Hubble parameter $H$, which is too heavy to be the inflaton candidates. On the other hand, in a UV completed theory, multiple light fields arise
naturally. Taking these facts into
consideration, a natural picture of inflation emerges: There
is one inflation direction with mass $m\ll H$, and some other
directions in the field space with $m \sim H$. In contrast to the slow-roll potential, large higher order terms in
the potential
such as $V'''\sim H$ and $V''''\sim 1$ can arise naturally in these non-flat directions.
To have more than one flat direction needs extra fine-tuning. The above picture for inflation suggests the quasi-single field inflationary models.

$\bullet$ {\em Inflationary phenomenology.} When the inflaton trajectory
turns in the field space, isocurvature perturbation is converted to curvature perturbation.
Precisely understanding the effect of such a conversion on density perturbations is in general an important but difficult question. Quasi-single field inflation provides explicit
examples where such a conversion can be calculated from first principles and give non-trivial predictions, such
as new shapes of large non-Gaussianities and running of the density perturbations.

\begin{figure}[thpb]
\centering
\epsfig{file=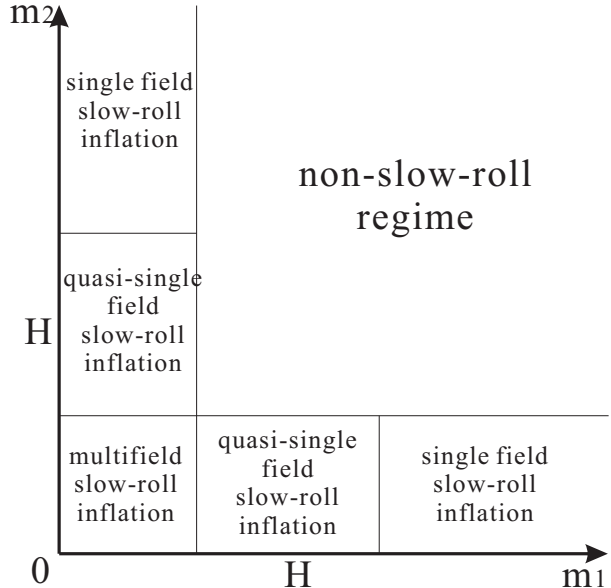, width=8cm}
\caption{\label{quasicompare} The relationship between single field
  inflation, quasi-single field inflation and multi-field
  inflation. Here we take inflation evolving two fields as an
  example. Quasi-single field inflation fills the gap between single
  field inflation and multiple field inflation.}
\end{figure}

$\bullet$ {\em Filling the gap between single and multiple field
inflation.} Quasi-single field inflation fills the gap between
single field inflation and multi-field inflation, with new
observational consequences. The relation between single field
inflation, quasi-single field inflation and multiple field inflation
is illustrated in Fig. \ref{quasicompare}.

$\bullet$ {\em Non-Gaussianity.} Non-Gaussianity is potentially one of the most promising probes of inflation. Generally speaking, observing shapes and running of non-Gaussianities can tell us about the details of inflatons and isocurvatons during inflation. In the most well-studied models,
scale-invariant non-Gaussianities usually take either local shape
or equilateral shape.
In quasi-single field inflation, large
non-Gaussianities with a family of new shapes lying between the local
and equilateral shape can arise naturally, and they are potentially observable in the near future. This provides new opportunities for interactions between model building and data analyses.

$\bullet$ {\em Methodology.}
We compute the correlation functions of the curvature perturbation using the in-in formalism \cite{inin,Weinberg:2005vy}. The methodology used here is inspiring in several
aspects.
Firstly, we use the transfer vertex to account for the transformation from the isocurvature mode to curvature mode, and use Feynman diagrams to compute the correlation functions perturbatively. As a result, the
leading order contribution comes from the fourth (for bispectrum) or
higher (for trispectrum) order perturbation theory.
Such methods can have broader applications in, for example, multi-field inflation models.
Secondly, we will use two representations of the in-in formalism, i.e.~the factorized form and commutator form. We find that each has its computational advantages and disadvantages and they are
complementary to each other.
Based on this, we develop a ``mixed form'' of the in-in formalism, which has the advantages of both forms.

This paper is organized as follows.

\begin{figure}[thpb]
\center
\epsfig{file=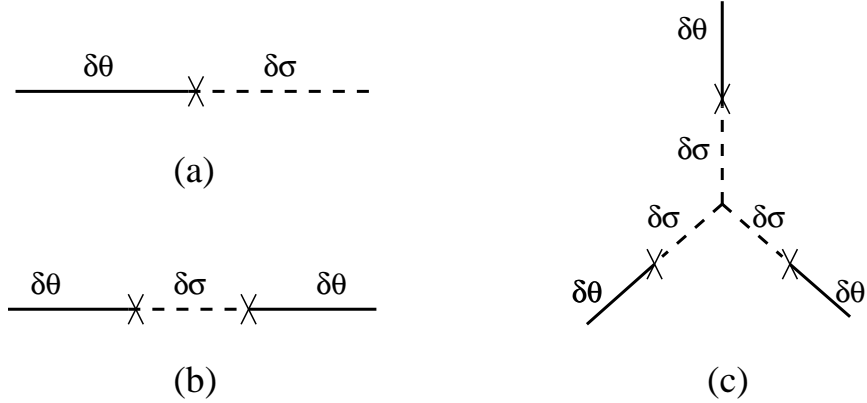, width=0.7\textwidth}
\caption{\label{Fig:fdiag} Feynman diagrams for the transfer vertex
  (a), corrections to the power spectrum from isocurvature modes (b), and the
  leading bispectrum (c).}
\end{figure}

In Sec.~\ref{Sec:model}, we study in detail an
explicit model of quasi-single field inflation with one isocurvature
direction, proposed in \cite{Chen:2009we}. We investigate the zero-mode solution, the perturbation theory, solve the mode
functions and discuss the transfer vertex (as illustrated in
Fig.~\ref{Fig:fdiag}(a)) between the inflaton
and isocurvaton.

In Sec.~\ref{Sec:power}, we study the power spectrum of this model. The power spectrum receives correction from two
transfer vertices, as illustrated in
Fig.~\ref{Fig:fdiag}(b). In the slow-turn case, we shall show that
the correction is of order
\begin{equation}
\delta P_\zeta \sim \left(\dot\theta/H\right)^2 P_\zeta~,
\end{equation}
where $P_\zeta$ is the power spectrum, $\dot\theta/H$ is the turning angular velocity in Hubble time.

\begin{figure}[thpb]
\begin{center}
\epsfig{file=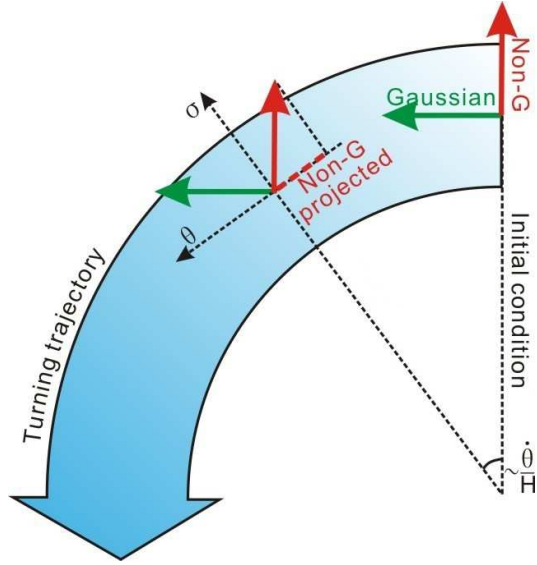, width=7cm}
\end{center}
\caption{
The physical interpretation of non-Gaussianity in quasi-single field
inflation. The non-Gaussianity in the isocurvature ($\sigma$) direction(s) is not suppressed
by slow roll parameters. This large non-Gaussianity is projected onto
the inflationary ($\theta$) direction if the inflaton trajectory
turns.}
\label{Fig:turnphys}
\end{figure}

In Sec.~\ref{Sec:bispectra}, we calculate the bispectra of
quasi-single field inflation. Since the slow-roll
condition is not required in the isocurvature direction, there is a sharp contrast between $V'''$ for the massive isocurvaton potential and $V_{\rm sr}'''$ for the slow-roll potential. The slow-roll condition requires
$V_{\rm sr}{'''}\sim {\cal O}(\epsilon^2) P_\zeta^{1/2} H$,
(where $\epsilon$ collectively denotes slow-roll parameters), which
contributes an ${\cal O}(\epsilon^2)$ term to $f_{NL}$ \cite{Maldacena:2002vr}. However $V'''$ is not subject to such conditions. For example, $V'''$ can naturally become of order $H$. To estimate the size of the bispectrum, we first note that $V'''$ contributes a factor of
$f_{NL}^{\rm iso}\sim P_{\zeta}^{-1/2}(V'''/H)$ to $f_{NL}$.
We also note that what we observe is not $f_{NL}^{\rm iso}$, but its projections onto the curvature perturbation (Fig.~\ref{Fig:turnphys}).
The efficiency of this
projection is determined by the turning angular velocity $\dot\theta/H$. For bispectra, we need three transfer vertices (Fig.~\ref{Fig:fdiag}(c)), so the bispectrum is of order
\begin{equation}\label{fnlestimate}
  f_{NL} \sim P_\zeta^{-1/2} \left(\dot\theta/H\right)^3
 \left(V'''/H\right) ~.
\end{equation}
In this section, we will derive
Eq. \eqref{fnlestimate} rigorously with explicit numerical coefficients. We will also compute the full shape functions using the mixed form of the in-in formalism.

In Sec.~\ref{Sec:squeezed}, we investigate the shapes of bispectra analytically by taking the squeezed limit, $p_3\ll p_1=p_2$, where $p_i$'s are the momenta in the three-point function. We find that the three-point function (up to
a momentum conservation delta function) scales as
\begin{equation}
  \langle\zeta^3\rangle \propto p_3^{-3/2-\nu}~,
\end{equation}
where $\nu=\sqrt{9/4-m^2/H^2}$, and $m$ is the effective mass of the
isocurvaton. Near $\nu=0$, the shape changes smoothly to
\begin{equation}
  \langle\zeta^3\rangle \propto \ln(p_3/p_1) p_3^{-3/2}~.
\end{equation}
This scaling behavior lies between that of
the local shape ($\langle\zeta^3\rangle \propto p_3^{-3}$)
and the equilateral shape ($\langle\zeta^3\rangle \propto p_3^{-1}$). We call these shapes the {\em intermediate shape}. We shall explain the underlying mechanisms that determine these different shapes.

\begin{figure}[thpb]
\center
\epsfig{file=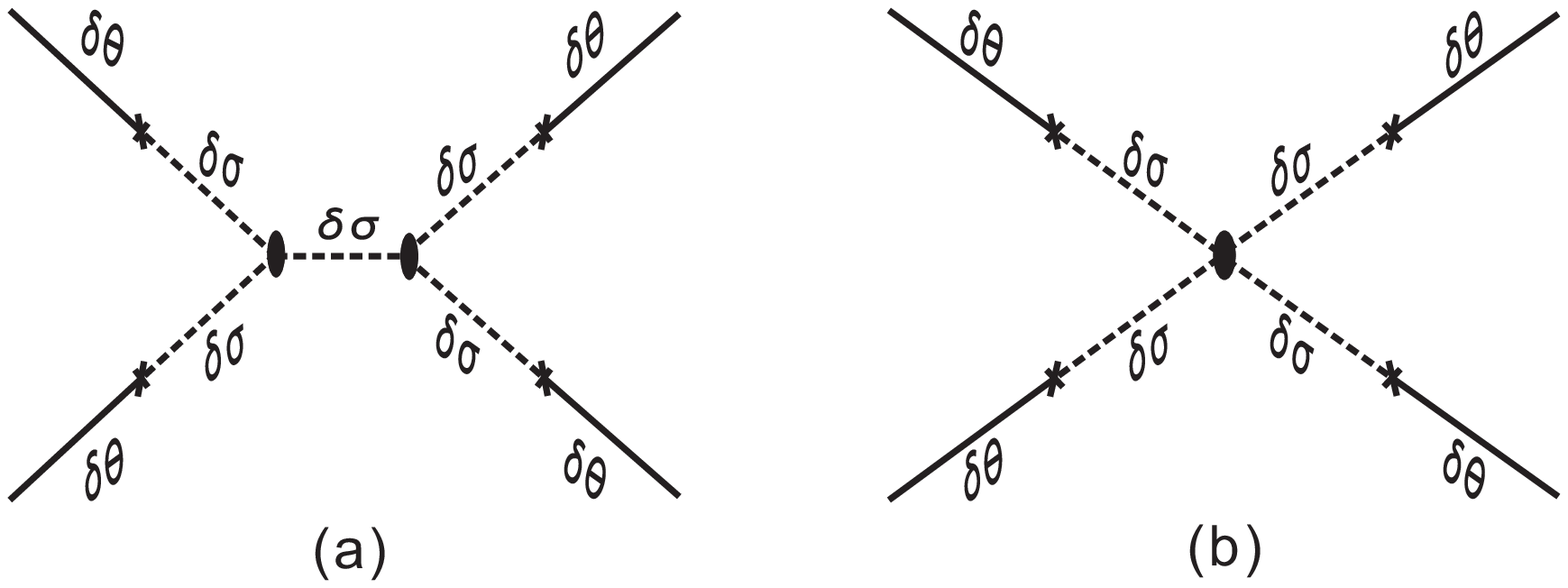, width=0.7\textwidth}
\caption{\label{Fig:fdiag4} Feynman diagrams for the scalar-exchange
  four-point correlation function (a) and contact-interaction four-point correlation function (b).}
\end{figure}

In Sec.~\ref{Sec:shapeAns}, we approximate the shape functions by simple analytical expressions, to facilitate future data analyses.

In Sec.~\ref{Sec:trispectra}, we discuss the trispectra. As
illustrated in Fig.~\ref{Fig:fdiag4}, we show that the trispectra
are of order
\begin{equation}
  t_{NL} \sim \max \left\{P_\zeta^{-1} \left(\dot\theta/H\right)^4
 \left(V'''/H\right)^2~,~
  P_\zeta^{-1} \left(\dot\theta/H\right)^4
  V''''\right\}~.
\end{equation}
Note that $t_{NL}\gg f_{NL}^2$ for $\dot\theta/H \ll 1$. So in future experiments where
angular multipoles $l$ can be measured as large as thousands,
trispectra may become a better probe for quasi-single field inflation.

We conclude in Sec.~\ref{Sec:conclusion} and discuss many future directions.

In Appendix \ref{App:thirdorderaction}, we present all terms in the Lagrangian up to the cubic order in two different gauges, and justify the calculation in the main text.
In Appendix \ref{App:allterms}, we present all terms used in our calculations in terms of the factorized, commutator, and mixed form, respectively. In Appendix \ref{App:wick}, we explain the Wick rotation and its several applications in this paper.
In Appendix \ref{App:inin}, we give general expressions for the series expansion in the in-in formalism.

\section{A model}
\label{Sec:model}
\setcounter{equation}{0}

In this section, we set up an explicit model of quasi-single field
inflation evolving two scalar fields.
The model configuration is illustrated in Fig.~\ref{Fig:turn}.
To consider non-trivial isocurvature
perturbation while keep things simple, we assume that the parameters characterizing the turning trajectory, such as the turning radius and various couplings, are constant.
We call this the ``constant turn'' case.

\subsection{Lagrangian, mode functions and transfer vertex}

When the inflaton trajectory moves along an arc, the action is
most conveniently written in the polar coordinates in terms of the tangential ($\theta$) and radial ($\sigma$)
directions of a circle with radius $\tilde R$,
\bea
S_m = \int d^4x \sqrt{-g} \left[
-\half (\tilde R+\sigma)^2 g^{\mu\nu} \partial_\mu \theta \partial_\nu \theta
- \half g^{\mu \nu} \partial_\mu \sigma \partial_\nu \sigma
- V_{\rm sr}(\theta) - V(\sigma) \right] ~,
\label{ModelAction}
\eea
where  $V_{\rm sr}(\theta)$ is a usual slow-roll potential, and $V(\sigma)$ is a potential that traps the isocurvaton at $\sigma_0$. The signature of the metric is $(-1,1,1,1)$.

The equations of motion for the spatial homogeneous background are
similar to those for single field inflation.
The Hubble equation and the continuity equation are
\begin{equation}
     3M_p^2 H^2=\frac{1}{2}R^2\dot\theta_0^2+V+V_{\rm sr}~,
\end{equation}
\begin{equation}
    -2M_p^2 \dot H= R^2\dot\theta_0^2 ~.
\end{equation}
The equations of motion for $\sigma_0(t)$ and $\theta_0(t)$ are
\begin{equation}
   \sigma_0={\rm const.} ~, \quad V'(\sigma_0)=R\dot\theta_0^2~,
\label{sigma_0}
\end{equation}
\begin{equation}
  R^2\ddot\theta_0+3R^2H\dot\theta_0+V_{\rm sr}'=0~.
\label{theta_0}
\end{equation}
We have absorbed the constant shift $\sigma_0$ into the net turning radius by defining $R\equiv\tilde R + \sigma_0$.
The inflaton $R\theta$ follows the usual equation of motion (\ref{theta_0}) determined by the slow-roll potential $V_{\rm sr}(\theta)$.

Equation (\ref{sigma_0}) indicates that, near $\sigma_0$, the potential has to contain a linear term to cancel the centrifugal force caused by the turning angular velocity. We therefore expand $V(\sigma)$ as
\bea
V(\sigma) = V'(\sigma_0)  (\sigma-\sigma_0) + \half V''(\sigma_0) (\sigma-\sigma_0)^2 + \frac{1}{6} V'''(\sigma_0) (\sigma-\sigma_0)^3 + \cdots
\eea
with $V''(\sigma_0) \gtrsim \CO(H^2)$.
Depending on the turning angular velocity, the minimum $\sigma_0$ of the effective potential,
\bea
V_{\rm eff} = - \frac{\dot\theta_0^2}{2} (R+\sigma-\sigma_0)^2 + V(\sigma) ~,
\eea
can be far away from the minimum of the original potential $V(\sigma)$. If their separation exceeds the typical field range over which the potential can be expanded perturbatively,
it is necessary to expand the potential around its effective minimum, as we did here. We emphasize that the mass $V''(\sigma_0)$ and coupling $V'''(\sigma_0)$ are evaluated at the new effective minimum $\sigma_0$, and are generally independent of their values at $\sigma=0$.
For simplicity, in the rest of the paper, we will just write $V'$, $V''$ and $V'''$ and omit their argument $\sigma_0$.

To study the leading order perturbations, we first present a simple intuitive approach \cite{Chen:2009we} and then justify it using a rigorous method.

We choose the {\em spatially flat gauge} in which the scale factor of the metric is homogeneous,
\bea
h_{ij}= a^2(t) \delta_{ij} ~,
\label{gauge1metric}
\eea
and the inflaton and isocurvaton are perturbed,
\bea
\theta(\bx,t) = \theta_0(t) + \delta\theta(\bx,t) ~, ~~~~~
\sigma(\bx,t) = \sigma_0 + \delta\sigma(\bx,t) ~.
\label{gauge1fields}
\eea
We expand only the matter part of the Lagrangian, namely (\ref{ModelAction}), while ignoring the perturbations in the gravity part. We neglect terms that are suppressed by the slow-roll parameters,
\begin{align}
\epsilon &\equiv - \frac{\dot H}{H^2} = \frac{R^2\dot\theta_0^2}{2H^2M_p^2} \approx \frac{M_p^2}{2} \left( \frac{V_{sr}'}{R V_{sr}}\right)^2 ~,
\cr
\eta &\equiv \frac{\dot\epsilon}{H\epsilon}
\approx -2 M_p^2 \frac{V_{sr}''}{R^2 V_{sr}}+ 2M_p^2\left( \frac{V_{sr}'}{R V_{sr}}\right)^2 ~.
\end{align}
For cubic terms in the perturbative expansion, we only consider the self-interaction in $V(\sigma)$.

This approach is intuitively correct because we expect that the main isocurvature contribution comes from the potential $V$ through the trajectory turning, and in particular, that the interactions involving the inflaton field have smaller contribution to the non-Gaussianities. Nonetheless, it is important to check this more rigorously, as we shall do in Sec.~\ref{Sec:twogauges}.

With the above prescription, we can easily get
\bea
\CL_2 = \frac{a^3}{2} R^2 \dot {\delta\theta}^2 - \frac{a R^2}{2}
  (\partial_i \delta\theta)^2
+ \frac{a^3}{2} \dot{\delta\sigma}^2 - \frac{a}{2} (\partial_i
\delta\sigma)^2 -\frac{a^3}{2} (V''-\dot\theta_0^2) \delta \sigma^2
~,
\label{CL2}
\eea
\bea
\delta \CL_2 &=& 2 a^3 R \dot \theta_0 ~\delta\sigma \dot{\delta\theta}
~,
\label{dCL2} \\
\CL_3 &=& -\frac{1}{6} a^3 V''' \delta\sigma^3 ~.
\label{CL3}
\eea
$\CL_2$ describes two free fields, one massless and another massive, in the inflationary background.
$\delta\CL_2$ is the coupling between them, and is the origin of the transfer vertex in Fig.~\ref{Fig:fdiag} (a). $\CL_3$ is the leading source for the three-point function for $\delta\theta$.

In order to treat $\delta\CL_2$ and $\CL_3$ as perturbations, we need
\begin{equation}
  \left(\frac{\dot\theta_0}{H}\right)^2 \ll 1 ~,
  \quad \frac{|V'''|}{H} \ll
  1~.
\end{equation}
This is because the correction to the leading power spectrum is suppressed by the factor $(\dot \theta_0/H)^2$, as we will show; and the requirements that the quadratic term dominates over the cubic term for $\delta\sigma\lesssim H$ gives the restriction $|V'''|/H \ll (V''/H)^2$.
However, it is worth to mention that these conditions are not the necessary conditions for the model-building.\footnote{In the absence of higher order derivative terms such as $V''''$, the $|V'''|/H$ cannot be much larger than $(V''/H)^2$ because otherwise the $\sigma$-field approaches a classical instability. But more generally even such kind of instability can be rescued by including large higher order derivative terms of $V$.} If they are not satisfied, it remains an interesting open question how this model can be solved non-perturbatively.

We define the conjugate momentum densities $\delta\pi_i = \partial \delta
\CL/\partial (\dot {\delta \phi_i})$, where $\delta \phi_i$ $(i=1,2)$
stand for
$\delta \theta$ and $\delta\sigma$ respectively, and $\delta \CL$
stands for the sum of (\ref{CL2}), (\ref{dCL2}) and (\ref{CL3}). We then work
out the Hamiltonian density in terms of $\delta \phi_i$ and $\delta
\pi_i$,
and separate them into the kinematic part $\CH_0$ and the interaction
part $\CH_I$. To
use the in-in formalism \cite{Weinberg:2005vy}, we replace $\delta\phi_i$'s and $\delta\pi_i$'s
in the Hamiltonian density
with the ones in the interaction pictures, $\delta\phi^I_i$'s and
$\delta\pi^I_i$'s. These latter variables satisfy the equations of
motion followed from the $\CH_0$.
Finally, we replaced $\delta \pi^I_i$ with $\dot{\delta\phi^I_i}$ using the relation $\dot {\delta\phi^I_i} = \partial
\CH_0/\partial (\delta \pi^I_i)$.
Following this procedure, we get
the following kinematic Hamiltonian density
\bea \label{H0}
\CH_0 = a^3 \left[ \half R^2 \dot {\delta\theta_I}^2 +
  \frac{R^2}{2a^2}
  (\partial_i \delta\theta_I)^2
+ \half \dot{\delta\sigma_I}^2 + \frac{1}{2a^2} (\partial_i
\delta\sigma_I)^2 + \half m^2 \delta \sigma_I^2
\right] ~,
\eea
and interaction Hamiltonian density
\bea
\CH^I_2 &=& -c_2 a^3 \delta\sigma_I \dot{\delta\theta_I} ~,
\label{CH2}
\\
\CH^I_3 &=&  c_3 a^3 \delta\sigma_I^3 ~,
\label{CH3}
\eea
where
\bea
c_2 = 2 R \dot\theta_0 ~, \quad c_3= \frac{1}{6} V''' ~, \quad m^2= V'' + 7\dot \theta_0^2 ~.
\eea
For the constant case, $c_2$, $c_3$ and $m^2$ are all constants. $m$ is the effective mass of the isocurvaton.\footnote{Although we mostly consider the case $m^2 \sim V'' \sim \CO(H^2)$ and $(\dot\theta_0/H)^2 \ll 1$, we comment that the non-perturbative case $(\dot\theta_0/H)^2\gtrsim 1$ with $m^2 \sim \CO(H^2)$ is also possible because $V''$ can be negative. As long as $m^2>0$ this does not introduce a tachyon mode at least at the quadratic level. Higher order non-perturbative corrections remain an open question.}

In the interaction picture, we quantize the Fourier components $\delta
\theta_\bk^I$ and $\delta\sigma_\bk^I$ of the free fields
$\delta\theta^I$ and $\delta\sigma^I$,
\bea
\delta\theta_\bk^I &=& u_\bk a_\bk + u_{-\bk}^* a_{-\bk}^\dagger ~,
\\
\delta\sigma_\bk^I &=& v_\bk b_\bk + v_{-\bk}^* b_{-\bk}^\dagger ~,
\eea
where $a_\bk$ and $b_\bk$ are independent of each other, and each
satisfies the usual commutation relation,
\bea
[a_\bk,a_{-\bk'}^\dagger] = (2\pi)^3 \delta^3 (\bk+\bk') ~,
\quad
[b_\bk,b_{-\bk'}^\dagger] = (2\pi)^3 \delta^3 (\bk+\bk') ~.
\eea
The mode functions $u_\bk$ and $v_\bk$ satisfy the linear equations of
motion followed from the kinematic Hamiltonian,
\bea
u_\bk'' - \frac{2}{\tau} u_\bk' + k^2 u_\bk &=&0 ~,
\label{modefun_u} \\
v_\bk'' - \frac{2}{\tau} v_\bk' + k^2 v_\bk +
\frac{m^2}{H^2 \tau^2} v_\bk &=& 0 ~,
\label{modefun_v}
\eea
where $\tau$ is the conformal time, $dt\equiv a d\tau$, and the prime
denotes the derivative with respect to $\tau$.

The solution for the mode functions are
\bea
u_\bk = \frac{H}{R\sqrt{2k^3}} ( 1+i k\tau)e^{-i k\tau} ~,
\label{mode_u}
\eea
and
\bea
v_\bk = -i e^{i(\nu+\half)\frac{\pi}{2}} \frac{\sqrt{\pi}}{2} H
  (-\tau)^{3/2} H^{(1)}_\nu (-k\tau) ~,
\quad {\rm for}~~ m^2/H^2 \le 9/4 ~,
\label{mode_v1}
\eea
where $\nu = \sqrt{9/4-m^2/H^2}$;
\bea
v_\bk = -i e^{ -\frac{\pi}{2}\tilde\nu + i\frac{\pi}{4}}
\frac{\sqrt{\pi}}{2} H
  (-\tau)^{3/2} H^{(1)}_{i\tilde\nu} (-k\tau) ~,
\quad {\rm for}~~ m^2/H^2 > 9/4 ~,
\label{mode_v2}
\eea
where $\tilde\nu = \sqrt{m^2/H^2-9/4}$.
The normalization of both mode functions have been chosen so that, when
the momentum $k/a$ is much larger than the Hubble parameter $H$ and the
mass $m$, we recover the Bunch-Davies vacuum
\bea
R u_\bk ~,~ v_\bk \to i \frac{H}{\sqrt{2k}} \tau e^{-i k\tau} ~.
\eea
The behavior of the mode functions after the horizon exit, $k\tau \to
0$, is also useful. For $m^2/H^2 \le 9/4$,
\bea
v_\bk \to \left\{
\begin{array}{lr}
\high{
-e^{i(\nu+\half)\frac{\pi}{2}} \frac{2^{\nu-1}}{\sqrt{\pi}}
\Gamma(\nu) \frac{H}{k^\nu} (-\tau)^{-\nu+\frac{3}{2}}
} ~,
& 0<\nu\le 3/2 ~,\\
\high{
e^{i\frac{\pi}{4}} \frac{1}{\sqrt{\pi}} H (-\tau)^{3/2} \ln (-k\tau)
} ~,
& \nu=0 ~;
\end{array}
\right.
\label{vk_small}
\eea
for $m^2/H^2 > 9/4$,
\bea
v_\bk \to -i e^{-\frac{\pi}{2}\tilde\nu+i\frac{\pi}{4}}
\frac{\sqrt{\pi}}{2} H
(-\tau)^{3/2}
\left[ \frac{1}{\Gamma(i\tilde\nu+1)}
\left(\frac{-k\tau}{2}\right)^{i\tilde\nu} -
i\frac{\Gamma({i\tilde\nu})}{\pi}
\left(\frac{-k\tau}{2}\right)^{-i\tilde\nu} \right]
~.
\label{vk_larger}
\eea
In the $k\tau\rightarrow 0$ limit, Eq.~\eqref{vk_small} contains
a decay factor $(-\tau)^{-\nu+3/2}$, while Eq.~\eqref{vk_larger} has both the decay factor $(-\tau)^{3/2}$ and an oscillation factor $\tau^{\pm i\tilde \nu}$. The decay factors
indicate that the perturbations for massive fields eventually roll back to zero. As the mass increases,
the behavior of the perturbations changes from that of an over-damped oscillator (corresponding to $m/H< 3/2$) to an under-damped oscillator (corresponding to $m/H>3/2$). In the under-damped case, the contribution of the massive isocurvaton to the curvature correlation functions is suppressed by factors of $e^{-m/H}$, analogous to the Boltzmann suppression, due to the oscillation factor in the mode function. The derivation of this Boltzmann suppression is given at the end of Appendix \ref{App:wick}.
[Note that this suppression factor is not the first factor $e^{-\pi \tilde\nu/2}$ in (\ref{vk_larger}), which is cancelled by $\Gamma(i\tilde\nu +1)$.]
Because of this suppression, in the remainder of this paper, we shall concentrate on the critical-damped and over-damped case, $0\le\nu<3/2$ (corresponding to $3H/2 \ge m >0$).

\subsection{Two gauges}
\label{Sec:twogauges}

In this subsection and Appendix \ref{App:thirdorderaction}, we use a rigorous approach to expand the perturbations. This will also justify the simple approach used above.

We first still work in the spatially flat gauge (\ref{gauge1metric}) and (\ref{gauge1fields}). Now we keep all terms, including those from the gravity sector, and get
\begin{align}
\CL_2=& \frac{a^3}{2}R^2 \dot{\delta\theta}^2 - \frac{a}{2} R^2(\partial_i \delta\theta)^2
-a^3 \left( \frac{V_{sr}''}{2R^2} - (3\epsilon-\epsilon^2+\epsilon\eta)H^2 \right)
R^2 \delta\theta^2
\cr
& +\frac{a^3}{2} \dot{\delta\sigma}^2 - \frac{a}{2} (\partial_i\delta\sigma)^2
-\frac{a^3}{2} (V''- \dot\theta_0^2) \delta\sigma^2
\label{L2_gauge1} \\
\delta\CL_2=& 2 a^3 R\dot\theta_0 \dot{\delta\theta} \delta\sigma
- 2\epsilon a^3 R\dot\theta_0 H \delta\theta \delta\sigma ~,
\label{dL2_gauge1} \\
\delta \CL_3 =& -\frac{a^3}{6} V''' \delta\sigma^3 + \cdots ~.
\label{L3_gauge1}
\end{align}
The derivation and the full cubic action can be found in Appendix \ref{Appgauge1}. There we also show that the $V'''$ term in Eq. \eqref{L3_gauge1} indeed gives the leading order contribution to bispectrum. Compared with (\ref{CL2}) and (\ref{dCL2}), extra terms that are suppressed by slow-roll parameters arise in the following two places. Firstly, the $\delta\theta$ field has a small mass, so it is no longer constant after the horizon exit. Secondly, there is an extra coupling term between $\delta\theta$ and $\delta\sigma$, i.e.~the second term in (\ref{dL2_gauge1}). This term is relatively smaller than (\ref{dCL2}) before and near the horizon exit due to the slow-roll suppression, but can become important afterwards because $\dot{\delta\theta}$ decays exponentially in real time. The detailed form of the coupling is also very important to the infrared behavior in the computation of the correlation functions. So we would like to clear up these issues by using a different gauge.

In single field inflation, we know that the quantity $\delta\theta$ is not the conserved quantity after the horizon exit, and there are correction terms suppressed by slow-roll parameters when we relate it to the conserved curvature perturbation $\zeta$. This conserved quantity is most explicit in the {\em uniform inflaton gauge}. Therefore we would also like to use a similar gauge in this model and see in particular how the corrections affect the second term in (\ref{dL2_gauge1}). We choose
\bea
\theta(\bx,t)=\theta_0(t)~, ~~~~~
\sigma(\bx,t)=\sigma_0 + \delta\sigma(\bx,t) ~,
\eea
and the spatial metric
\bea
h_{ij}(\bx,t) = a^2(t) e^{2\zeta(\bx,t)} \delta_{ij} ~.
\eea
Here we have done a position dependent time-shift with respect to the spatially flat gauge, so that the inflaton is homogenous but the scale factor has fluctuations.
In this gauge, the full quadratic Lagrangian is
\begin{align}
\CL_2 = & \epsilon a^3 \dot\zeta^2 - \epsilon a (\partial\zeta)^2
+ \frac{a^3}{2} \dot{\delta\sigma}^2 - \frac{a}{2} \left(\partial_i \delta\sigma \right)^2 - \frac{a^3}{2}\left(V''-\dot\theta_0^2\right) \delta\sigma^2 ~,
\label{L2_gauge2}
\end{align}
\begin{align}
\delta\CL_2=& - 2a^3 \frac{R\dot\theta_0^2}{H} \delta\sigma \dot\zeta ~.
\label{dL2_gauge2}
\end{align}
The cubic action is
\bea
\CL_3 = - \frac{a^3}{6} V''' \delta\sigma^3 + \cdots ~.
\label{L3_gauge2}
\eea
The derivation can be found in Appendix \ref{Appgauge2}.
There we also show that the single term listed in (\ref{L3_gauge2}) is the most important one.

In the uniform inflaton gauge, the physics is conceptually clearer. The linear equation of motion for the curvature mode $\zeta$ describes an exactly massless scalar field in inflationary background and has the following leading order solution,
\bea
\zeta_k = -\frac{H}{\sqrt{4\epsilon k^3}} (1+ik\tau) e^{-ik\tau} ~.
\label{zeta_k_mode}
\eea
Before considering the coupling to the isocurvature modes, $\zeta_k$ is constant after horizon exit with $H$ and $\epsilon$ evaluated at $k=aH$.
The linear equation of motion for the isocurvature mode $\delta\sigma$ describes a massive scalar field, and has the solution (\ref{mode_v1}) and (\ref{mode_v2}). $\delta\sigma$ eventually decays away. The transfer vertex is described by the single term in (\ref{dL2_gauge2}). In our setup, the curvature mode $\zeta$ seeds the large scale structures, as in the single field inflation, while the isocurvature perturbation $\delta\sigma$ is negligible at the reheating.

This also gives a justification of the simple method used previously. Comparing (\ref{L2_gauge2}), (\ref{dL2_gauge2}) and (\ref{zeta_k_mode}) with (\ref{CL2}), (\ref{dCL2}) and (\ref{mode_u}), we see that, as long as the parameters $H$ and $\epsilon$ are approximately constant, we can use the usual time-delay relation
\bea
\zeta \approx -\frac{R\delta\theta}{\sqrt{2\epsilon}} \approx -\frac{H}{\dot\theta_0}\delta\theta
\label{zetarelation}
\eea
to convert the correlation functions of $\delta\theta$ to those of $\zeta$. If $H$ and $\epsilon$ are not always constants, we should start with (\ref{L2_gauge2})-(\ref{L3_gauge2}).

We would also like to emphasize that, in our case, although $\zeta$ is eventually a constant, it may not be so right after the horizon exit, in contrast to the single field inflation. As we will demonstrate, depending on the mass of the isocurvaton, the isocurvature mode can decay slowly and be transferred into the curvature mode long after the horizon exit.

\section{Power spectrum}
\label{Sec:power}
\setcounter{equation}{0}

The turning trajectory leads to a correction to the power
spectrum. The term responsible for the transition from the isocurvature
mode to the curvature mode is (\ref{CH2}),
\bea
H^I_2 = \int d^3 \bx ~\CH^I_2 = -c_2 a^3 \int \frac{d^3\bk}{(2\pi)^3}
\delta\sigma^I_\bk \dot{\delta\theta}^I_{-\bk} ~.
\eea
This term introduces the transfer vertex (Fig.~\ref{Fig:fdiag}(a)).

The two-point function for $\delta\theta$ is given by the expectation value of $\delta\theta^2$ at the end of inflation. In the in-in formalism this is given by,
\bea
\langle \delta\theta^2 \rangle &\equiv&
\langle 0| \left[ \bar T \exp\left( i\int_{t_0}^t dt' H_I(t')\right)
  \right] \delta\theta_I^2(t)
  \left[T \exp\left( -i\int_{t_0}^t dt' H_I(t')\right)
  \right] |0\rangle
\\
&=&\langle 0| \delta \theta_I^2 |0\rangle
\cr
&+& \int_{t_0}^t d\tilde t_1 \int_{t_0}^t dt_1
\langle 0| H_I(\tilde t_1) ~\delta\theta_I^2~ H_I(t_1) |0\rangle
\label{2pt1}\\
&-& 2 ~{\rm Re} \left[
\int_{t_0}^t dt_1 \int_{t_0}^{t_1} dt_2
\langle 0| \delta\theta_I^2~ H_I(t_1) H_I(t_2) |0\rangle \right]
\label{2pt2} \\
&+& \cdots ~.
\nonumber
\eea
The terms (\ref{2pt1}) and (\ref{2pt2}) are the leading order corrections to the two-point function. According to the Feynman diagram Fig.~\ref{Fig:fdiag}(b), each $H_I$ should be replaced by $H_2^I$.

We evaluate these terms
using the technique of normal ordering. After normal ordering, only terms with all fields contracted survive. For example, a contraction between $\delta\sigma(p,t)$ on the left with $\delta\sigma(p',t')$ on the right gives $v_{p}(t)v_{p'}^*(t') (2\pi)^3\delta(\bp+\bp')$. We sum over all terms that represent the Feynman diagram Fig.~\ref{Fig:fdiag}(b).

The term
(\ref{2pt1}) gives
\bea
&& 2c_2^2 u_{p_1}^*(t) u_{p_1}(t)
~\Big| \int_{t_0}^{t} d\tilde t_1~ a^3(\tilde t_1) v_{p_1}(\tilde t_1)
\dot u_{p_1}(\tilde t_1) \Big|^2 ~ (2\pi)^3 \delta^3 (\bp_1+\bp_2)
\nonumber \\
&=& \frac{\pi}{8} \frac{c_2^2}{R^4} \frac{1}{p_1^3}
~\Big| \int_0^\infty dx~ x^{-1/2} H^{(1)}_\nu(x) e^{ix} \Big|^2 ~
(2\pi)^3 \delta^3 (\bp_1+\bp_2) ~. \label{powerint1}
\eea
The term (\ref{2pt2}) gives
\bea
&-& 4c_2^2 u^2_{p_1}(t)
{\rm Re} \left[ \int_{t_0}^t dt_1 a^3(t_1)v_{p_1}(t_1)
\dot u^*_{p_1}(t_1)
\int_{t_0}^{t_1} dt_2 a^3(t_2) v^*_{p_2}(t_2) \dot u^*_{p_2}(t_2)
\right]
\cr
&\times& (2\pi)^3 \delta^3 (\bp_1+\bp_2)
\nonumber \\
&=& -\frac{\pi}{4} \frac{c_2^2}{R^4} \frac{1}{p_1^3}
{\rm Re} \left[ \int_0^\infty dx_1 ~x_1^{-1/2} H^{(1)}_\nu(x_1)
  e^{-ix_1} \int_{x_1}^\infty dx_2 ~x_2^{-1/2} H^{(2)}_\nu(x_2)
  e^{-ix_2} \right]
\cr
&\times& (2\pi)^3 \delta^3 (\bp_1+\bp_2) ~. \label{powerint2}
\eea

\begin{figure}[t]
\center
\epsfig{file=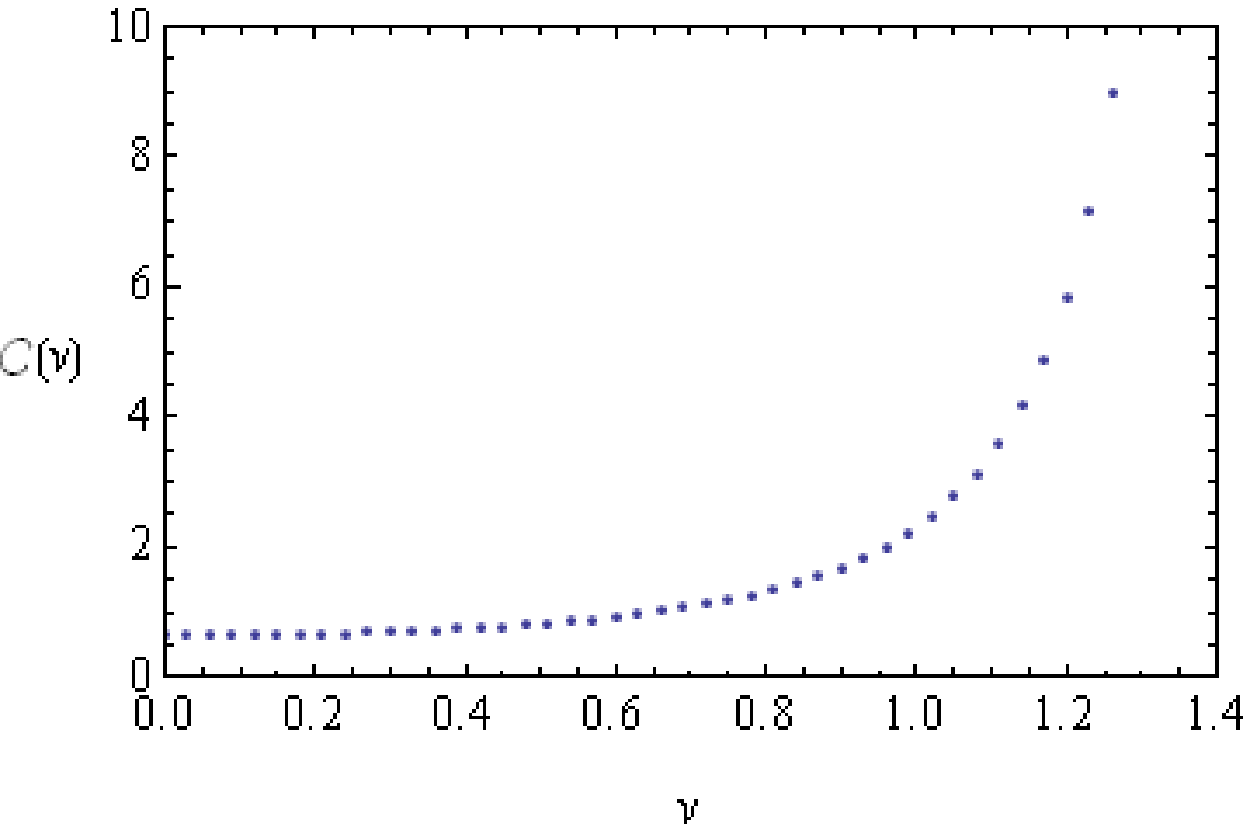, width=10cm}
\caption{\label{Fig:Cnu} The coefficient $\cal C (\nu)$, defined in
  Eq. \eqref{cdefinition}.}
\end{figure}

For $1/2 \le \nu<3/2$, there are divergences at $\tau\rightarrow 0$
in Eqs.~\eqref{powerint1} and \eqref{powerint2}. However these
divergences are spurious and are canceled by summing up
\eqref{powerint1} and \eqref{powerint2}. The final result can be
written as
\begin{align}
 (2\pi)^3 \delta^3 (\bp_1+\bp_2) \frac{c_2^2}{R^4}
 \frac{\cal C(\nu)}{p_1^3}~,
\end{align}
where the numerical factor $\cal C$ is defined as
\begin{align}
{\cal C}(\nu) \equiv \frac{\pi}{4}
{\rm Re} \left[ \int_0^\infty dx_1 \int_{x_1}^\infty dx_2
\left(
x_1^{-1/2} H^{(1)}_\nu(x_1) e^{ix_1}
x_2^{-1/2} H^{(2)}_\nu(x_2) e^{-ix_2}
\right. \right.
\cr
\left. \left.
-x_1^{-1/2} H^{(1)}_\nu(x_1) e^{-ix_1}
x_2^{-1/2} H^{(2)}_\nu(x_2) e^{-ix_2}
\right)
\right] ~.
\label{cdefinition}
\end{align}
It is not difficult to see that, as $x_2 \to x_1\to 0$, the leading
divergence in the integrand is imaginary and $\CC(\nu)$ is finite for $\nu<3/2$. We plotted ${\cal C}$ as a function of $\nu$ in Fig. \ref{Fig:Cnu}.

We use (\ref{zetarelation}) to relate $\langle \zeta^2 \rangle$ to $\langle \delta\theta^2 \rangle$, and get
\begin{equation}
  \langle\zeta^2\rangle = (2\pi)^5
  \delta^3({\bf p}_1+{\bf p}_2) \frac{1}{2p_1^3} P_\zeta ~,
\end{equation}
where the power spectrum is
\begin{align}
P_\zeta=&
\frac{H^4}{4\pi^2 R^2 \dot\theta_0^2}\left[
1+8 {\cal C} \left( \frac{\dot\theta_0}{H} \right)^2 \right]~.
\label{power}
\end{align}
The spectral index is
\begin{align}
  n_s-1\equiv\frac{d \ln P_\zeta}{d \ln k} =
  -2\epsilon - \eta + 8\CC \eta \left( \frac{\dot\theta_0}{H} \right)^2 ~.
\label{ns}
\end{align}
The last terms proportional to $( \dot\theta_0 / H )^2$ in (\ref{power}) and (\ref{ns}) are the corrections from turning trajectory. So the perturbation theory requires $(\dot\theta_0/H)^2\ll 1$.

If the turning of trajectory only happens for a period of time, transient large running of the power spectrum and spectral index becomes possible. We leave this non-constant turn case to future work.

Although the observational consequence on the power spectrum for the constant turn case is not very distinctive, this case serves as a warmup exercise to the bispectra and trispectra cases in the next sections. As we will see, the simple versions of the problems and solutions that we have encountered here will be significantly magnified.

\section{Bispectra}
\label{Sec:bispectra}
\setcounter{equation}{0}

The origin of the large non-Gaussianities is the self-interaction terms for the isocurvaton field $\sigma$. The leading cubic term in this model is
(\ref{CH3}). This gives the following term in the interaction
Hamiltonian,
\bea
H^I_3 =\int d^3\bx \CH^I_3
= c_3 a^3 \int \frac{d^3\bp}{(2\pi)^3} \frac{d^3\bq}{(2\pi)^3}
\delta\sigma^I_\bp(t) \delta\sigma^I_\bq(t)
\delta\sigma^I_{-\bp-\bq}(t) ~.
\eea
Through the transfer vertex (Fig. \ref{Fig:fdiag}(a)), this interaction is converted to
that of the curvature perturbation. The Feynman diagram is given in Fig. \ref{Fig:fdiag}(c).

The three-point function for $\delta\theta$ is given by,
\bea
\langle \delta\theta^3 \rangle &\equiv&
\langle 0| \left[ \bar T \exp\left( i\int_{t_0}^t dt' H_I(t')\right)
  \right] \delta\theta_I^3(t)
  \left[ T \exp\left( -i\int_{t_0}^t dt' H_I(t')\right)
  \right] |0\rangle ~.
\label{3ptinin}
\eea
Again using (\ref{zetarelation}), $\langle \zeta^3 \rangle$ is related to it by a factor of $(-H/\dot\theta_0)^3$.

One can expand (\ref{3ptinin}) and write it in two equivalent
forms. In the first form,
we simply expand the exponentials,
\bea
\langle \delta\theta^3 \rangle
&=& \int_{t_0}^t d\tilde t_1 \int_{t_0}^{\tilde t_1}
d\tilde t_2
\int_{t_0}^t dt_1 \int_{t_0}^{t_1} dt_2
~\langle H_I(\tilde t_2) H_I(\tilde t_1) ~\delta\theta_I^3~
H_I(t_1)H_I(t_2) \rangle
\label{3pt1} \\
&-& 2 ~{\rm Re} \left[
\int_{t_0}^{t}d\tilde t_1
\int_{t_0}^t dt_1 \int_{t_0}^{t_1} dt_2 \int_{t_0}^{t_2} dt_3
~\langle H_I(\tilde t_1) ~\delta\theta_I^3~
H_I(t_1) H_I(t_2) H_I(t_3) \rangle
\right]
\label{3pt2}\\
&+& 2 ~{\rm Re} \left[
\int_{t_0}^t dt_1 \int_{t_0}^{t_1} dt_2 \int_{t_0}^{t_2} dt_3
\int_{t_0}^{t_3} dt_4
~\langle \delta\theta_I^3~
H_I(t_1) H_I(t_2) H_I(t_3) H_I(t_4) \rangle
\right] ~.
\label{3pt3}
\eea
In each term, the interaction vertex $H^I_3$ can appear in one of the
four $H_I$,
and the rest three $H_I$'s should be replaced by the transfer
vertex $H^I_2$.
We refer to this form as the {\em factorized form} because there is no
cross time-ordering between the
integral from the left interaction vacuum and that from the right.
After contractions, we get
\bea
&-& 12 c_2^3 c_3 u_{p_1}^*(0) u_{p_2}(0) u_{p_3}(0)
\cr &\times&
{\rm Re} \left[ \int_{-\infty}^0 d\ttau_1~ a^3(\ttau_1)
  v_{p_1}^*(\ttau_1) u'_{p_1}(\ttau_1)
\int_{-\infty}^{\ttau_1} d\ttau_2~ a^4(\ttau_2) v_{p_1}(\ttau_2)
v_{p_2}(\ttau_2) v_{p_3}(\ttau_2) \right.
\cr &\times& \left.
\int_{-\infty}^0 d\tau_1~ a^3(\tau_1) v_{p_2}^*(\tau_1)
u_{p_2}^{\prime *} (\tau_1)
\int_{-\infty}^{\tau_1} d\tau_2~ a^3(\tau_2) v_{p_3}^*(\tau_2)
u_{p_3}^{\prime *} (\tau_2) \right]
\cr &\times&
(2\pi)^3 \delta^3(\sum_i \bp_i) + {\rm 9~other~similar~terms}
\cr
&+& {\rm 5~permutations~of~} \bp_i ~.
\label{FacForm_term}
\eea
We leave the details of the 9 other terms to Appendix
\ref{10terms_F}.

In the second form, (\ref{3ptinin}) can be
expressed in terms of the nested commutators \cite{Weinberg:2005vy},
\bea
\langle \delta \theta^3 \rangle =
\int_{t_0}^t dt_1 \int_{t_0}^{t_1} dt_2 \int_{t_0}^{t_2} dt_3
\int_{t_0}^{t_3} dt_4
\langle [ H_I(t_4), [ H_I(t_3), [ H_I(t_2), [ H_I(t_1),
    \delta \theta_I(t)^3]]]] \rangle ~,
\label{ComForm}
\eea
in which all the integrands are written under a single time-ordered
integral. We refer to this form as the {\em commutator form}.
Again replacing one of the $H_I$ with $H^I_3$ and the rest
with $H^I_2$, we get
\bea
\langle\delta\theta^3\rangle&=&12 c_2^3c_3
u_{p_1}(0) u_{p_2}(0) u_{p_3}(0)
\cr
&\times&
{\rm Re} \left[ \int_{-\infty}^0 d\tau_1 \int_{-\infty}^{\tau_1}
  d\tau_2 \int_{-\infty}^{\tau_2} d\tau_3 \int_{-\infty}^{\tau_3}
  d\tau_4~
\prod_{i=1}^4 a^3(\tau_i) \right.
\cr
&\times&
\left. \left( a(\tau_2)  A
+ a(\tau_3) B
+ a(\tau_4) C \right) \right]~
(2\pi)^3 \delta^3(\sum \bp_i)
\cr
&+& {\rm 5~perm.} ~.
\label{ComForm3}
\eea
We have used the fact that $u_{p_i}(0)$ are real.
$A$, $B$ and $C$ are three contributions corresponding to
replacing $H_I(t_2)$, $H_I(t_3)$ and $H_I(t_4)$, respectively, with $H^I_3$:
\bea
A &=& \left(u_{p_1}^{\prime}(\tau_1) - c.c. \right)
\left( v_{p_1}(\tau_1) v_{p_1}^*(\tau_2) - c.c. \right)
\left( v_{p_3}(\tau_2) v_{p_3}^*(\tau_4) u_{p_3}^{\prime *}(\tau_4) -
c.c. \right)
\cr
&& v_{p_2}(\tau_2) v_{p_2}^*(\tau_3) u_{p_2}^{\prime *}(\tau_3) ~,
\label{termA} \\
B &=& \left(u_{p_1}^{\prime}(\tau_1) - c.c. \right)
\left( u_{p_2}^\prime (\tau_2) - c.c. \right)
\left( v_{p_1}^*(\tau_1) v_{p_2}^*(\tau_2) v_{p_1}(\tau_3)
v_{p_2}(\tau_3) - c.c. \right)
\cr
&& v_{p_3}(\tau_3) v_{p_3}^*(\tau_4) u_{p_3}^{\prime *}(\tau_4) ~,
\label{termB} \\
C &=& -\left( u_{p_1}^\prime(\tau_1) - c.c. \right)
\left( u_{p_2}^\prime(\tau_2) - c.c. \right)
\left( u_{p_3}^\prime(\tau_3) - c.c. \right)
\cr
&& v_{p_1}^*(\tau_1) v_{p_2}^*(\tau_2) v_{p_3}^*(\tau_3)
v_{p_1}(\tau_4) v_{p_2}(\tau_4) v_{p_3}(\tau_4) ~.
\label{termC}
\eea

The factorized and commutator form each has its
computational advantages and disadvantages when evaluating the
integrals. To see this, let us
investigate the properties of these integrals in the IR ($\tau \to 0$)
and UV ($\tau\to -\infty$).

$\bullet$ {\em IR convergence.}
To see the IR behavior, we use the asymptotic forms
of the mode functions in
the $\tau \to 0$ limit,
\bea
u'_{p_i}(\tau) &\propto& (-\tau) (1-i p_i \tau + \cdots) ~,
\cr
v_{p_i}(\tau) &\propto& (-\tau)^{\frac{3}{2}-\nu} \left( 1+ \alpha_1 (-\tau)^2
+ \alpha_2 (-\tau)^{2\nu} + \cdots \right) ~,
\label{modeIR}
\eea
where we have ignored an overall phase in $v_{p_i}$ that will always be
cancelled. Also note that $\alpha_1$ is real, $\alpha_2$ is complex.

We first look at the factorized form.
It is easy to see that each term in
(\ref{FacForm_term}) has IR divergence for
$3/2>\nu>1/2$ ($0<m<\sqrt{2} H$). For smaller $m$, the
isocurvature mode is decaying slower, hence the conversion to the
curvature mode
lasts longer after the horizon exit.
But since the isocurvature mode eventually decays,
we do not expect any singular behavior starting from
$m=\sqrt{2}H$, as we have seen in the case of the
power spectrum. So these IR divergences should be cancelled when we sum
over all ten terms. It is possible to check this analytically,
but it takes
very long time even with the aid of Mathematica. This is because
we not only have the leading order spurious divergence, $\sim
\tau_1^{3-6\nu}$, but also have eight different orders of subleading
spurious divergence, such as $\sim
\tau_1^{4-6\nu},~\tau_1^{4-4\nu},\cdots$. Numerically, errors occurred
in the cancellation of these spurious divergence quickly dominate in
the final results as
$\nu$ approaches $1/2$ from below.

However this cancellation is made much more transparent in the
commutator form. In this form, all integrands are under the same
integral and their mutual cancellation in the IR is explicit.
Without loss of generality, let
us examine the term $A$ (\ref{termA}). Using
(\ref{modeIR}), we see that if we take $\tau_i\to 0$, the leading
powers of
$\tau_i$ in the whole integral go as follows: $(-\tau_1)^{3/2-\nu}$,
$(-\tau_2)^{3/2-3\nu}$, $(-\tau_3)^{1/2-\nu}$,
$(-\tau_4)^{1/2-\nu}$. For the multi-layer integral (\ref{ComForm}),
if an inner integral diverges in the IR, its
contribution to the outer integral is dominated by this IR behavior;
if it converges in the IR, its
contribution to the outer integral is $\CO(1)$ and complex. One can
then list all possibilities and examine them case by case. But a
quicker way to see the conclusion goes as follows. The
largest power for $\tau_i$ that we listed above is $3/2-\nu$. So by
considering the case $\nu=3/2$, we are considering the largest
possible IR contributions for all $0<\nu<3/2$. For this case, we can
just take the $\tau_i\to 0$ limit for all terms in the integral and
take
the limit $\tau_i \to \tau_1$. The terms in the first bracket in
(\ref{termA}) goes as $(-\tau_1)^2$ as the leading term in $u'_{p_1}$
is cancelled by the subtraction of its complex conjugate. The terms in
the second and third brackets are similar, they go as $(-\tau_1)^3$
and $(-\tau_1)^{5-2\nu}$, respectively. Since the first line of
(\ref{termA}) is pure imaginary, the second line,
$v_{p_2} v^*_{p_2} u^{\prime *}_{p_2}$, has to be imaginary to make
the
overall integrand real. Hence it goes as
$(-\tau_1)^{5-2\nu}$. Including
all the scale factors, $a(\tau_i) \propto 1/\tau_i$, the whole
integral goes as $(-\tau_1)^{6-4\nu}=1$ and is logarithmically divergent for $\nu=3/2$. So we conclude that the IR
divergence is explicitly absent for $\nu<3/2$ in the commutator form.

$\bullet$ {\em UV convergence.}
It is a common feature that the integrands in these correlation
functions are oscillatory in the UV, when modes are well within
the horizon. Their contribution is averaged out. For the Bunch-Davies vacuum, this regulation can be achieved by slightly tilting the integration contour into the imaginary plane, $\tau\to \tau(1\pm i\epsilon)$.

For the factorized form, we tilt the
contour clockwise, $\tau\to \tau(1-i\epsilon)$, for the
anti-time-ordered integral from the left, and
counter-clockwise, $\tau\to \tau(1+i\epsilon)$, for the right. This procedure works for all $\nu$.

For the commutator form, if we do not have the problem of the spurious IR divergence, a procedure similar to the above
still works. Now the left and right factors are mixed and the original tilts cannot be kept intact, but the effect of these tilt can be easily implemented. The effect is to suppress the oscillating contributions in UV. So we can re-choose a proper tilt for each term in the new integral
according to its convergent direction. The only exception
occurs
for some special momentum configurations where the oscillating
factor form the left and right happen to cancel each other. At these
places, one encounters the
slower convergence, or spurious divergence that are completely absent
in the point of view of the factorized form discussed above. These divergences
eventually will be
cancelled.\footnote{To illustrate this using a simple example, we look
at the integral $\int^0_{-\infty} dx_1 e^{-iax_1} \cdot
  \int^0_{-\infty} dx_2
  e^{i b x_2} = \int^0_{-\infty} dx_1 \int^{x_1}_{-\infty} dx_2 (e^{-i
    a x_1}
    e^{i b x_2} + e^{-i a x_2} e^{i b x_1} ) $. Tilting the integration contours at $x\to-\infty$ into the imaginary plane according to the signs of $a$, $b$ and $a-b$, the
      factorized form on the LHS simply gives $1/ab$, and the
      commutator form on the RHS gives $1/(a-b)b -
      1/(a-b)a$. Each of the two terms in the commutator form has a
      divergence at the special point $a=b$, but cancelled by
      each other. Cases of such UV spurious divergences are
      encountered, for example, in the study of trispectra
      \cite{Adshead:2009cb,Chen:2009bc,Seery:2008ax,Gao:2009gd} if one uses the commutator form. So the factorized form is more
      convenient in such cases.}

However with the problem of the spurious IR divergence, as for the
case
$1/2<\nu<3/2$, the problem in UV gets much
worse. We cannot even avoid the spurious UV divergence for {\em general} momentum
configurations, while preserving the explicit IR convergence. This is
because, for example after expanding the term $A$ (\ref{termA}), different
terms have
different convergent tilting directions as we mentioned.
But now we cannot re-choose the tilts for them
individually if they have to be grouped to achieve the explicit IR
convergence.

$\bullet$ {\em Mixed form.}
In summary, the factorized form is much more convenient to achieve the
explicit UV
convergence, while the commutator form is much more convenient to achieve
the explicit IR convergence.
For $0<\nu<1/2$, we do not have the IR problem, so we can use the
factorized form, but it starts to fail as $\nu \to 1/2$. For $1/2<\nu<3/2$, both types of spurious divergence exist. The best way to proceed is to combine the two forms. We introduce a cutoff $\tau_c$ (e.g.~$\tau_c= -2/p_1$), and write the IR part ($\tau_c<\tau\le 0$) of the integrals in terms of the commutator form, and the UV part ($\tau<\tau_c$) in terms of the factorized form,
\bea
\sum_i \int_{\tau_c}^0 d\tau_1 \cdots \int_{\tau_c}^{\tau_{i-1}} d\tau_i
~{\rm \{commutator~form\}} ~
\int_{-\infty}^{\tau_c} d\tau_{i+1} \cdots
\int_{-\infty}^{\tau_{n-1}} d\tau_n
~{\rm \{factorized~form\} } ~.
\eea
The detailed expressions of this mixed form are presented in Appendix \ref{App:mixedform}.

$\bullet$ {\em Wick rotation.} An additional subtlety is that the contour tilting is easy to do analytically, but difficult to implement numerically. In this paper,
we will use a much more efficient approach of Wick rotation to achieve
the fast convergence in UV.
As shown in Appendix \ref{App:wick}, one can
rotate the integration in the complex plane, $\tau_i \to \pm i x_i$, so that the oscillation factors for
$\tau_i \rightarrow -\infty$ become the exponential suppression
factors in the $x_i$-coordinates.

\begin{figure}[tpb]
\begin{center}
\epsfig{file=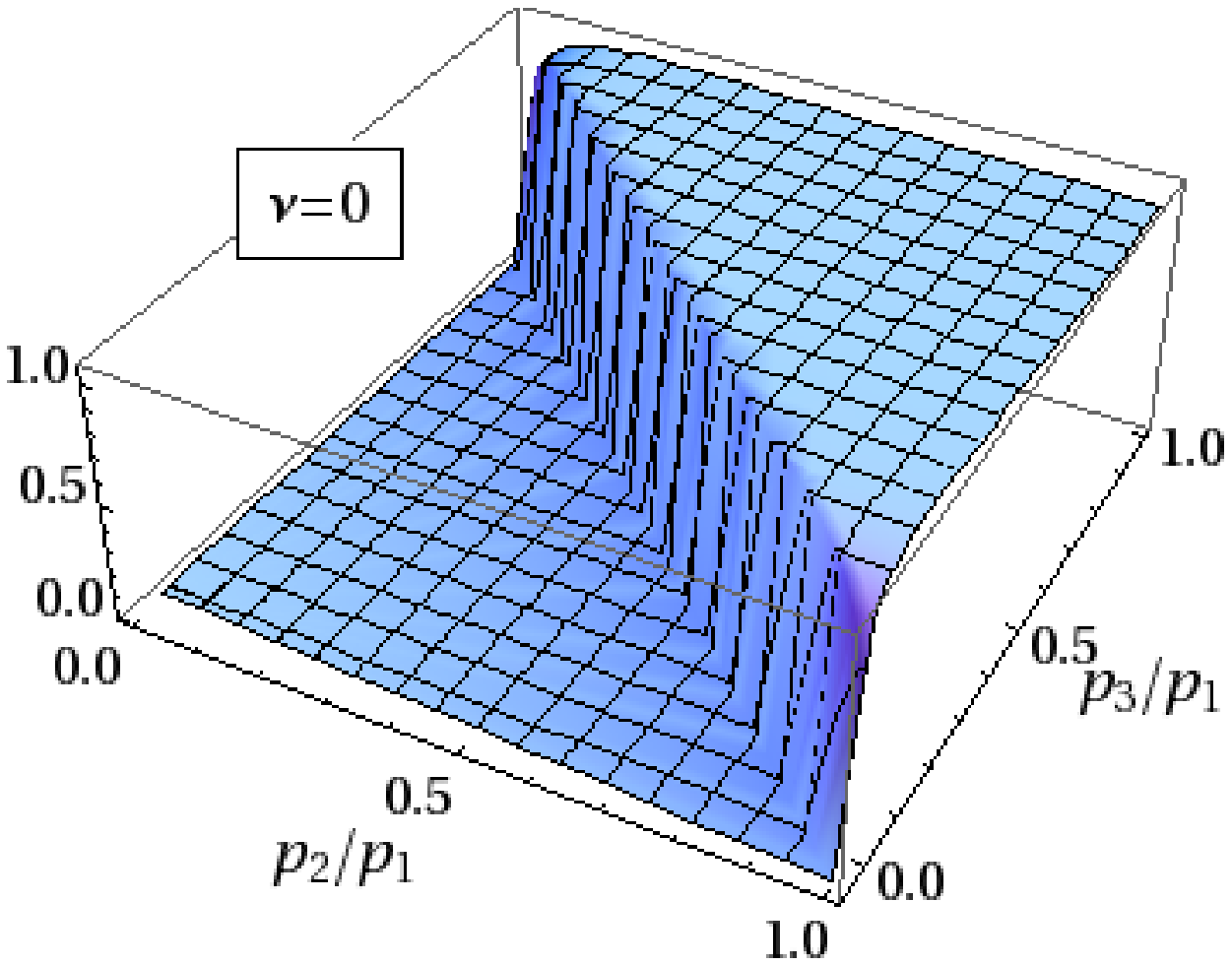, width=0.45\textwidth}
\epsfig{file=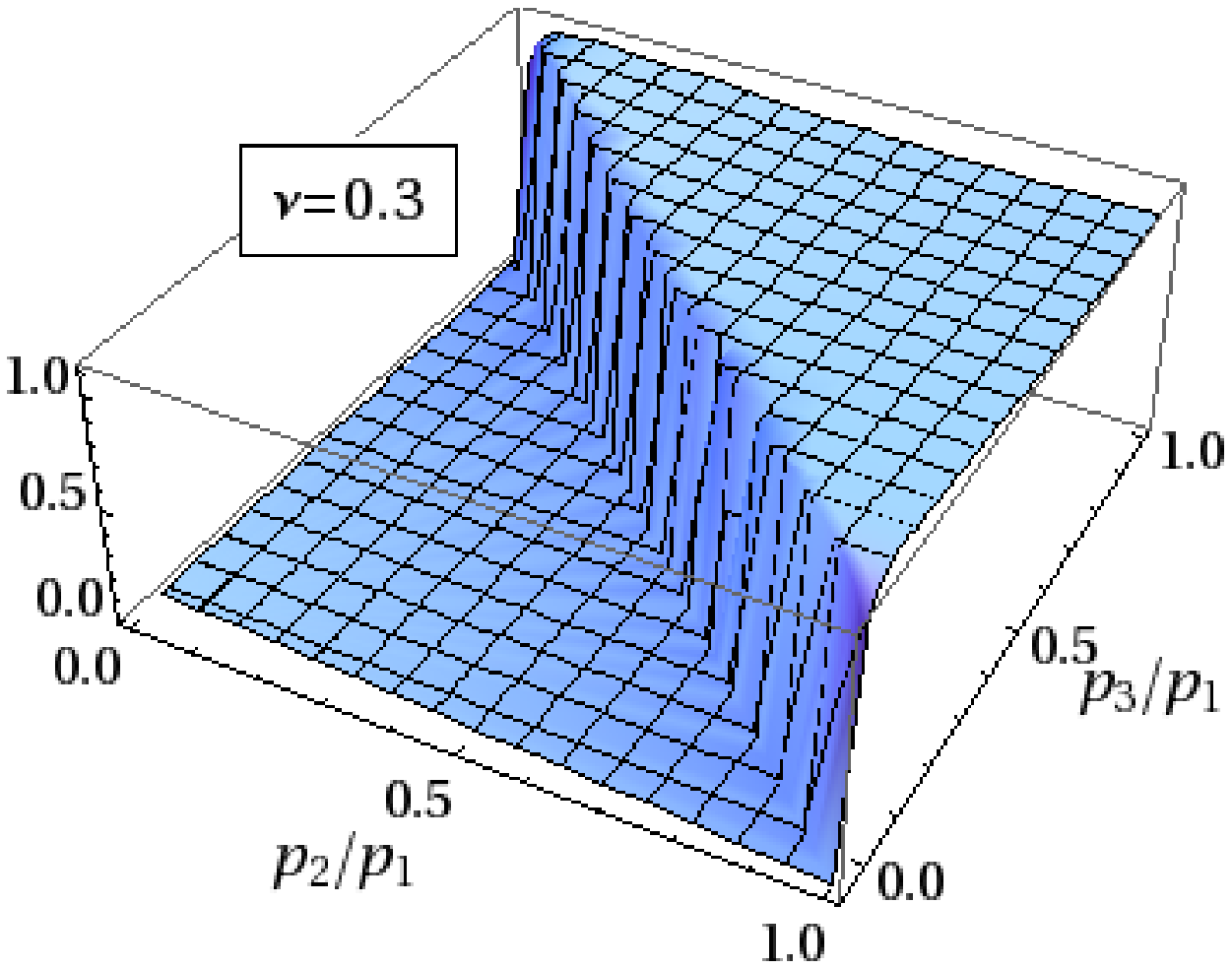, width=0.45\textwidth}
\epsfig{file=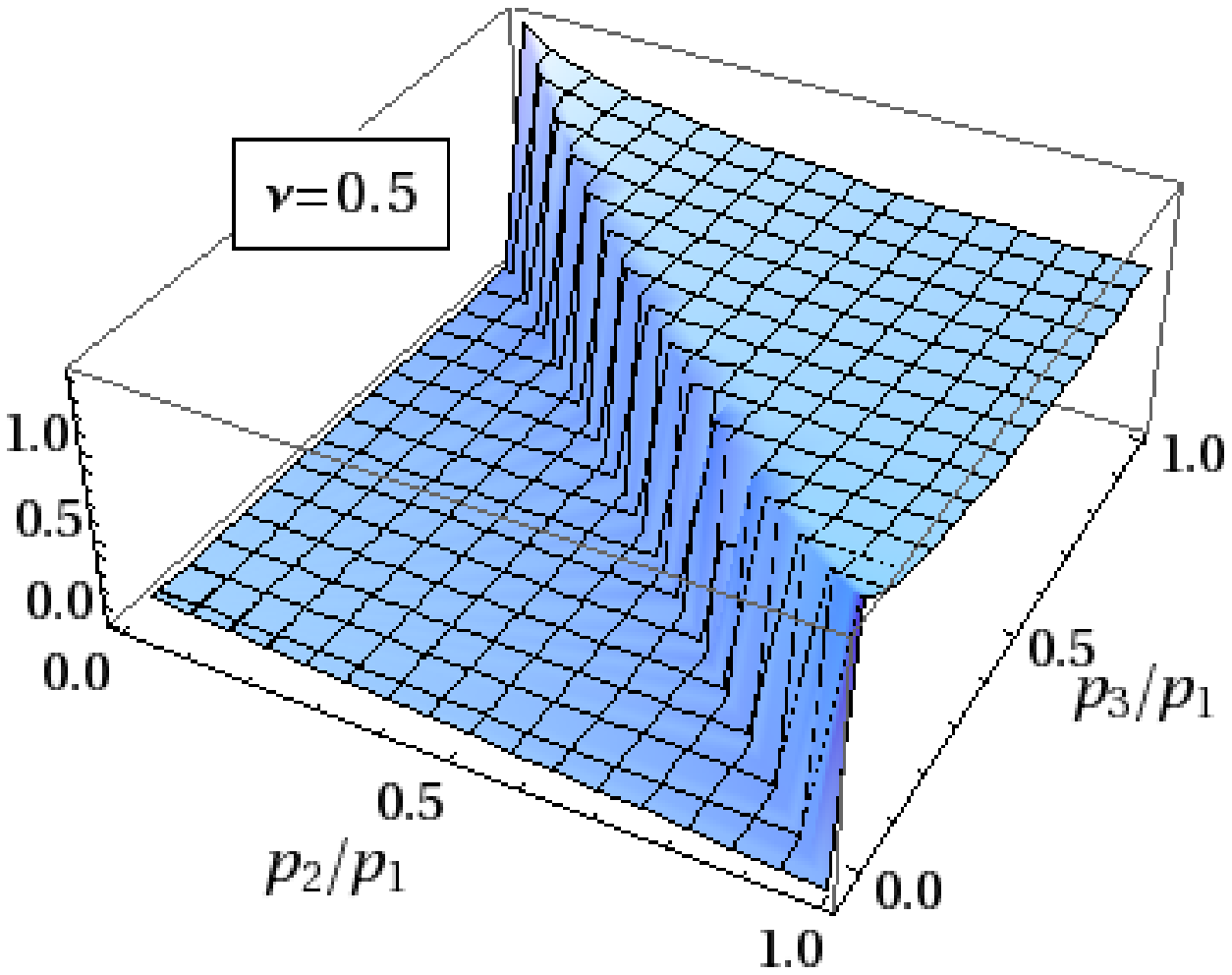, width=0.45\textwidth}
\epsfig{file=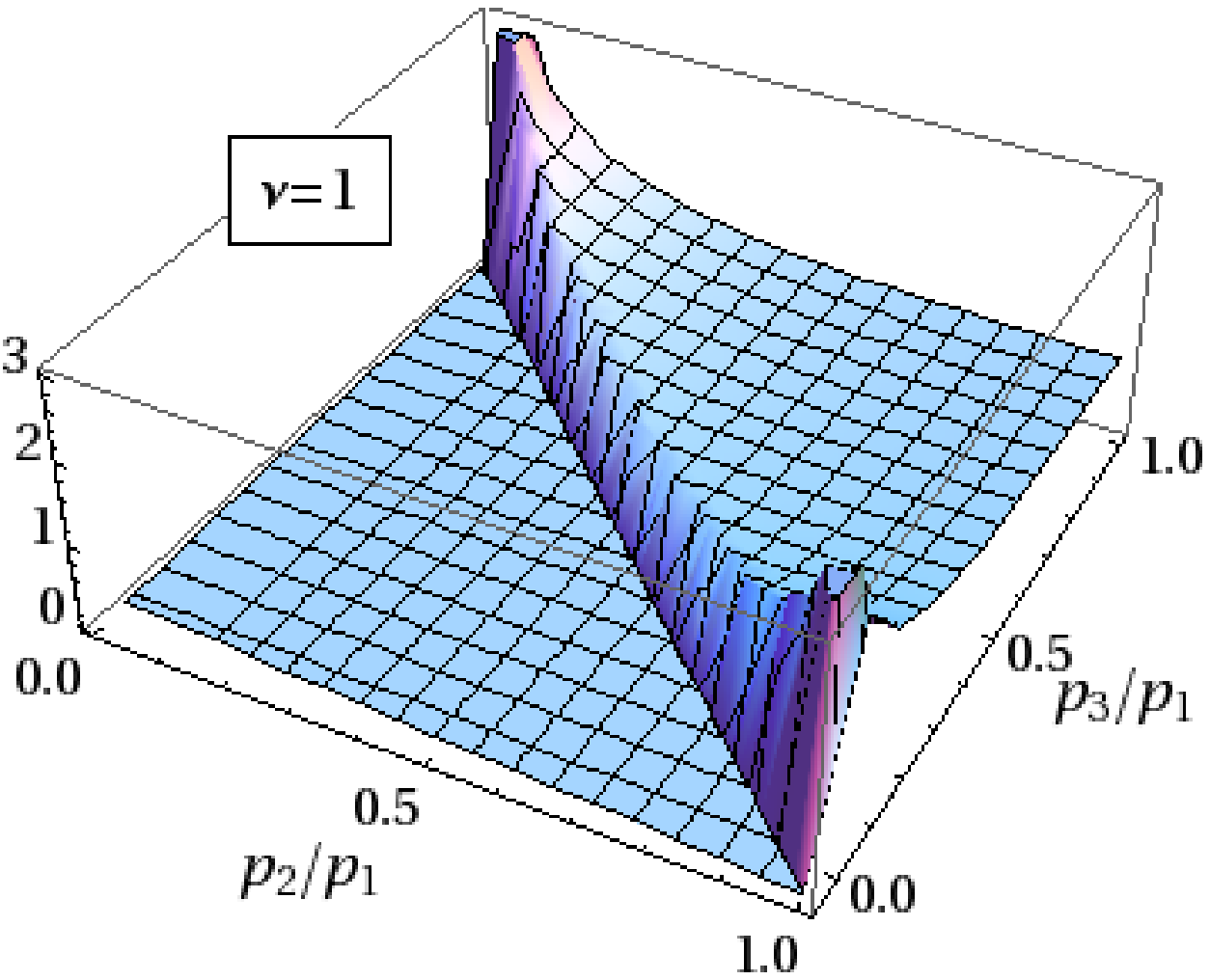, width=0.45\textwidth}
\end{center}
\caption{\label{Fig:nu0}
Shapes of bispectra with intermediate forms. We plot
$(p_1p_2p_3)^2 F$ with $\nu=0,~ 0.3,~ 0.5,~ 1$. The plot is
normalized such that $(p_1p_2p_3)^2 F=1$ for $p_1=p_2=p_3=1$.}
\end{figure}

Applying the technique of Wick rotation to the mixed form, we finally have a very efficient way to numerically compute the full shapes of the bispectra.
As an example, we present the plots for $\nu=0,~ 0.3,~ 0.5,~ 1$ in Fig.~\ref{Fig:nu0}.

To plot, we define the function $F$ as
\bea
\langle \zeta^3 \rangle \equiv F(p_1,p_2,p_3) P_\zeta^2 (2\pi)^7 \delta^3(\sum_i \bp_i) ~.
\label{F_def}
\eea
To illustrate the shape of a scale-invariant bispectrum, we conventionally normalize the amplitude $F$ by multiplying a factor of $(p_1p_2p_3)^2$. This makes it dimensionless and scale-independent.

From Fig.~\ref{Fig:nu0}, we can see that when $\nu$ is small, the shape looks more like an equilateral shape.
When $\nu$ gets larger, the shape looks more like a local shape.
In Sec.~\ref{Sec:squeezed}, we will study the analytical properties and explain the underlying physics of these shapes.

\begin{figure}[t]
\begin{center}
\epsfig{file=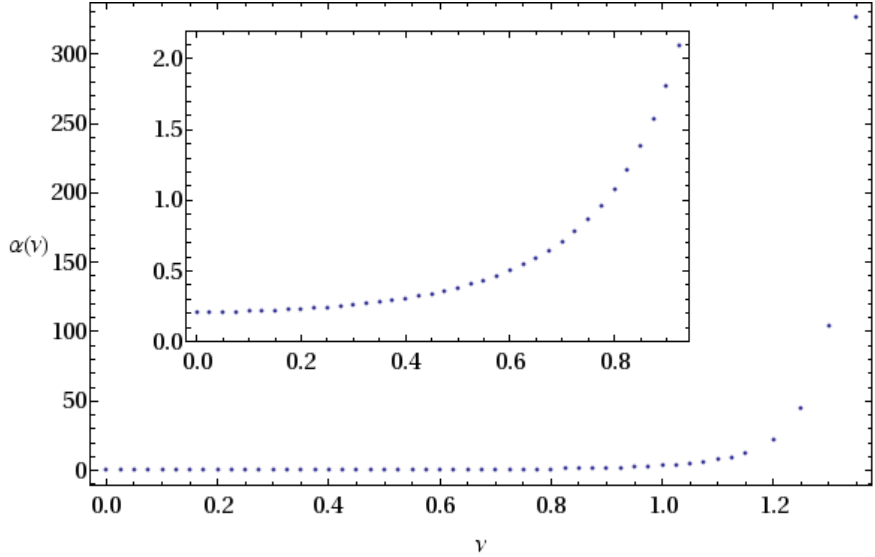, width=0.7\textwidth}
\end{center}
\caption{\label{Fig:eqlimit}
The numerical coefficient $\alpha(\nu)$ in $f_{NL}^{\rm int}$.}
\end{figure}

Finally, we would like to parameterize the magnitude of the non-Gaussianities in terms of an estimator $f_{NL}^{\rm int}$. According to the convention in the bispectrum literature, we define the number $f_{NL}^{\rm int}$ by matching with the
$f_{NL}^{\rm local}$ in the local shape ansatz in the equilateral limit. In other
words, we define $f_{NL}^{\rm int}$ so that, at $p_1=p_2=p_3$, the three point correlation function is
\begin{equation}
\langle \zeta({\bf p}_1) \zeta({\bf p}_2) \zeta({\bf p}_3) \rangle \to
(2\pi)^7\delta^3({\bf p}_1+{\bf p}_2+{\bf p}_3)P_\zeta^2
\left(\frac{9}{10}f_{NL}^{int}\right) \frac{1}{p_1^6}~.
\label{fnl_def}
\end{equation}
From this definition, we get
\begin{equation}\label{fnlcal}
  f_{NL}^{\rm int} = \alpha(\nu) P_\zeta^{-1/2} \left(\dot\theta_0/H\right)^3
 \left(-V'''/H\right)~,
\end{equation}
verifying the qualitative estimate given in the Introduction.
The numerical coefficient $\alpha(\nu)$ is plotted in
Fig. \ref{Fig:eqlimit}, and $P_\zeta\approx 6.1\times 10^{-9}$. $\dot\theta_0$ is positive in our
convention, so the sign of the $f_{NL}^{\rm int}$ is the opposite of
the $V'''(\sigma_0)$.

From Fig.~\ref{Fig:eqlimit}, we can see that
$\alpha(\nu)$ can get very large. For example, as $\nu$ varies from $0$ to $1.35$ (i.e.~mass varies from $1.5H$ to $0.65H$), $\alpha(\nu)$ grows from $0.2$ to $300$.
As $\nu\rightarrow 3/2$, $\alpha(\nu)$ blows up.
This divergence happens because we are using the constant turn assumption. When the effective mass of $\sigma$ becomes zero,
a fluctuation $\delta\sigma$ never decays at super-horizon. Then the transfer
from isocurvaton to curvaton lasts forever.
Practically, the upper bound of the conformal time
integration will not be zero. If the horizon crossing time of a perturbation mode is $N_f$
e-folds before the end of inflation (or the time when the inflaton trajectory becomes straight),
one needs to impose a cutoff
\begin{equation}
  \label{eq:cutoff}
  \tau_f\equiv -\frac{1}{He^{N_f}}~.
\end{equation}
As $\nu\to 3/2$, one can show that (\ref{ComForm3}) is dominated by the integrals that behavior as $\prod_{i=1}^4 \int d\tau_i/\tau_i \sim N_f^4$.
In principle, $N_f$ can be
as large as 60. But we caution that, although $\alpha(\nu)$ grows a lot as $\nu\to 3/2$, this does not mean that the non-Gaussianities can be enhanced by such a large factor, at least in the perturbative regime. The reason is the following. In this limit, $\CC(\nu)$ in (\ref{cdefinition}) scales as $N_f^2$ for the same reason. For large $N_f$, the perturbation theory requires $N_f^2 (\dot\theta_0/H)^2 \ll 1$ instead. Therefore, in the perturbative regime, the effective enhancement factor is only $N_f$.

\section{Squeezed limit of bispectra}\label{Sec:squeezed}
\setcounter{equation}{0}

In this section, we investigate the behavior of the three-point correlation function in the squeezed limit, $p_3 \ll p_1=p_2$. This is important for several reasons. First, in this special limit, analytical results for the shape functions are possible. These provide both useful checks on our numerical results and complimentary information. Second, numerical results are not useful in the construction of estimators that are often used in data analyses. We need to guess simple analytical expressions if they are not immediately available. Knowing the analytical results in the squeezed limit greatly helps in achieving this goal. Third, the scaling behavior of the squeezed limit is closely tied to the underlying physical mechanisms and we will use it to classify the shapes of bispectra.

We will investigate this limit in both the commutator and factorized form. We will see that they give equivalent results.
The following behavior of the Hankel function in the small argument limit, $x\ll 1$, will be useful in the analyses,
\bea
H_\nu^{(1)}(x) \to -i \frac{2^\nu \Gamma(\nu)}{\pi} x^{-\nu}
-i \frac{2^{-2+\nu}\Gamma(\nu)}{\pi (-1+\nu)} x^{-\nu+2}
+ \left( -i \frac{\cos(\pi \nu) \Gamma(-\nu)}{2^{\nu} \pi} + \frac{1}{2^\nu\Gamma(1+\nu)} \right) x^\nu + \cdots ~.
\label{Hankelsmalllimit}
\eea
Note that the real part starts from $\CO(x^\nu)$.

We start with the commutator form and first look at the contribution from the $A$ term (\ref{termA}),
\bea
&&\frac{3\pi^3}{2^6} \frac{c_2^3c_3}{HR^6} \frac{1}{p_1 p_2 p_3}
\cr
&\times&
{\rm Re} \left[ i \int_{-\infty}^0 d\tau_1 \int_{-\infty}^{\tau_1}
  d\tau_2 \int_{-\infty}^{\tau_2} d\tau_3 \int_{-\infty}^{\tau_3}
  d\tau_4~
(-\tau_1)^{-1/2} (-\tau_2)^{1/2} (-\tau_3)^{-1/2} (-\tau_4)^{-1/2}
\right.
\cr
&\times& \sin (-p_1\tau_1)
\left( H^{(1)}_\nu(-p_1\tau_1) H^{(2)}_\nu(-p_1\tau_2) - c.c. \right)
\left( H^{(2)}_\nu(-p_3\tau_2) H^{(1)}_\nu(-p_3\tau_4) e^{-i p_3\tau_4}
- c.c.\right)
\cr
&\times& \left.
H^{(1)}_\nu(-p_2\tau_2) H^{(2)}_\nu(-p_2\tau_3) e^{i p_2\tau_3}
\right] + {\rm 5~perm.} ~.
\label{termA_2}
\eea
We neglect the common factor $(2\pi)^3 \delta^3(\sum \bp_i)$ in this section.
Terms with momentum permutation behave differently in the squeezed
limit, and we exam each of them in the following.

For the term explicitly written in (\ref{termA_2}), we define
$x_i \equiv p_1 \tau_i$ ($i=1,2,3,4$)
and get
\bea
&&\frac{3\pi^3}{2^6} \frac{c_2^3c_3}{HR^6} \frac{1}{p_1^4 p_2 p_3}
\cr
&\times&
{\rm Re} \left[ i \int_{-\infty}^0 dx_1 \int_{-\infty}^{x_1}
  dx_2 \int_{-\infty}^{x_2} dx_3 \int_{-\infty}^{x_3}
  dx_4~
(-x_1)^{-1/2} (-x_2)^{1/2} (-x_3)^{-1/2} (-x_4)^{-1/2}
\right.
\cr
&\times& \sin (-x_1)
\left( H^{(1)}_\nu(-x_1) H^{(2)}_\nu(-x_2) - c.c. \right)
\left( H^{(2)}_\nu(-\frac{p_3}{p_1}x_2)
H^{(1)}_\nu(-\frac{p_3}{p_1}x_4) e^{-i\frac{p_3}{p_1}x_4} - c.c.\right)
\cr
&\times& \left.
H^{(1)}_\nu(-\frac{p_2}{p_1}x_2) H^{(2)}_\nu(-\frac{p_2}{p_1}x_3)
e^{i\frac{p_2}{p_1}x_3} \right] ~.
\label{termA_3}
\eea
The terms $H^{(2)}_\nu(-x_2p_3/p_1)$ in the
3rd line can be approximated in
the small $-x_2 p_3/p_1$ limit. The reason is as follows. If
$-x_2 p_3/p_1 \sim 1$, $|x_2| \gg 1$. Other terms in the
integrand have factors such
as $H^{(1)}(x_2)$. These factors become fast-oscillating
and hence suppress the integration.
However the terms
$H^{(1)}_\nu(-x_4 p_3/p_1)$ and $e^{-ix_4p_3/p_1}$ in the 3rd line cannot be approximated in
the
small $-x_4p_3/p_1$ limit, since there is no oscillatory term as $x_4$
gets large. We redefine $y_4 \equiv x_4 p_3/p_1$.
With this prescription we get
\bea
&& - \frac{3\pi^2}{2^{6-\nu}} \Gamma(\nu) \frac{c_2^3c_3}{H R^6}
\frac{1}{p_1^{\frac{7}{2}-\nu} p_2 ~p_3^{\frac{3}{2}+\nu}}
\cr
&\times& {\rm Re} \left[\int_{-\infty}^0 dx_1 \int_{-\infty}^{x_1}
  dx_2 \int_{-\infty}^{x_2} dx_3 \int_{-\infty}^{\frac{p_3}{p_1}x_3}
  dy_4 ~
(-x_1)^{-1/2} (-x_2)^{1/2} (-x_3)^{-1/2} (-y_4)^{-1/2} \right.
\cr
&\times&
\sin(-x_1) \left( H^{(1)}_\nu(-x_1) H^{(2)}_\nu(-x_2) - c.c.\right)
(-x_2)^{-\nu} H^{(1)}_\nu(-x_2) H^{(2)}_\nu(-x_3) e^{ix_3}
\cr
&\times& \left.
\left( H^{(1)}_\nu(-y_4) e^{-i y_4} + c.c.\right) \right] ~.
\label{termA_limit1}
\eea
In order not to be suppressed, $x_3 \lesssim 1$. So the upper limit of
the $y_4$ integral is effectively 0. Since it is also
convergent at $y_4\to 0$, this integral can be factored out.

We next look at the term with the permutation $p_1 \leftrightarrow
p_3$. With the same definition of $x_i$ ($i=1,2,3,4$), we get
\bea
&&\frac{3\pi^3}{2^6} \frac{c_2^3c_3}{HR^6} \frac{1}{p_1^4 p_2 p_3}
\cr
&\times&
{\rm Re} \left[ i \int_{-\infty}^0 dx_1 \int_{-\infty}^{x_1}
  dx_2 \int_{-\infty}^{x_2} dx_3 \int_{-\infty}^{x_3}
  dx_4~
(-x_1)^{-1/2} (-x_2)^{1/2} (-x_3)^{-1/2} (-x_4)^{-1/2}
\right.
\cr
&\times& \sin (-\frac{p_3}{p_1}x_1)
\left( H^{(1)}_\nu(-\frac{p_3}{p_1}x_1)
H^{(2)}_\nu(-\frac{p_3}{p_1}x_2) - c.c. \right)
\left( H^{(2)}_\nu(-x_2)
H^{(1)}_\nu(-x_4) e^{-ix_4} - c.c.\right)
\cr
&\times& \left.
H^{(1)}_\nu(-\frac{p_2}{p_1}x_2) H^{(2)}_\nu(-\frac{p_2}{p_1}x_3)
e^{i\frac{p_2}{p_1}x_3} \right] ~.
\label{termA_p1}
\eea
In the third line, the first three functions can be approximated
in the small $-x_i p_3/p_1$ ($i=1,2$) limit, and one of the Hankel functions
should be expanded to $\CO(x_i^\nu)$ in order to get
a non-zero result. Focusing on the scaling behavior of $p_i$, we see that it is proportional to
\bea
\sim \frac{1}{p_1^5 p_2} ~.
\label{Slimit_Aterm2}
\eea
Comparing to (\ref{termA_limit1}), this is negligible.

Finally, we look at the term with the permutation $p_2 \leftrightarrow
p_3$,
\bea
&&\frac{3\pi^3}{2^6} \frac{c_2^3c_3}{HR^6} \frac{1}{p_1^4 p_2 p_3}
\cr
&\times&
{\rm Re} \left[ i \int_{-\infty}^0 dx_1 \int_{-\infty}^{x_1}
  dx_2 \int_{-\infty}^{x_2} dx_3 \int_{-\infty}^{x_3}
  dx_4~
(-x_1)^{-1/2} (-x_2)^{1/2} (-x_3)^{-1/2} (-x_4)^{-1/2}
\right.
\cr
&\times& \sin (-x_1)
\left( H^{(1)}_\nu(-x_1)
H^{(2)}_\nu(-x_2) - c.c. \right)
\left( H^{(2)}_\nu(-\frac{p_2}{p_1}x_2)
H^{(1)}_\nu(-\frac{p_2}{p_1}x_4) e^{-i\frac{p_2}{p_1}x_4} - c.c.\right)
\cr
&\times& \left.
H^{(1)}_\nu(-\frac{p_3}{p_1}x_2) H^{(2)}_\nu(-\frac{p_3}{p_1}x_3)
e^{i\frac{p_3}{p_1}x_3} \right] ~.
\label{termA_p2}
\eea
In this case, we can also approximated the three functions in the
4th line in the small $-x_i p_3/p_1$ ($i=2,3$) limit.
For $\nu > 1/2$, we use the leading term for the two Hankel functions
and the subleading term for the exponential function and get
\bea
\sim \frac{1}{p_1^{5-2\nu} p_2
  p_3^{2\nu}} ~;
\label{termA_p2a}
\eea
and for $\nu < 1/2$, one of the Hankel functions should be expanded to $\CO(x_i^\nu)$ and we use the leading term for the exponential function,
\bea
\sim \frac{1}{p_1^4 p_2 p_3} ~.
\label{termA_p2b}
\eea
Both (\ref{termA_p2a}) and (\ref{termA_p2b}) are negligible comparing to (\ref{termA_limit1}), for $\nu<3/2$.

The other permutation $p_1\leftrightarrow p_2$ gives each term a factor
of 2.

We perform the similar analyses to the $B$ and $C$ terms.
Overall we find that the dominant contribution come from the $A$
and $B$ terms and their momentum permutation $p_1\leftrightarrow p_2$ only.
The final result is
\bea
\langle \zeta(\bp_1) \zeta(\bp_2)\zeta(\bp_3) \rangle
\xrightarrow{\rm p_3\ll p_1=p_2}
s(\nu) \frac{c_2^3c_3}{HR^6} \frac{1}{p_1^{\frac{7}{2}-\nu} p_2 ~p_3^{\frac{3}{2}+\nu}}
(2\pi)^3 \delta^3(\sum_i \bp_i)
~,
\label{Slimit_shape}
\eea
where
\bea
s(\nu) &\equiv& \frac{3\pi^2 \Gamma(\nu)}{2^{3-\nu}}
\cr
&\times& \int_{-\infty}^0 dx_1 \int_{-\infty}^{x_1}
  dx_2 \int_{-\infty}^{x_2} dx_3
\cr
&&
\left[
(-x_1)^{-1/2} (-x_2)^{1/2-\nu} (-x_3)^{-1/2} \sin(-x_1)
\right.
\cr
&&
{\rm Im} \left( H^{(1)}_\nu(-x_1) H^{(2)}_\nu(-x_2)
\right)
{\rm Im} \left( H^{(1)}_\nu(-x_2) H^{(2)}_\nu(-x_3) e^{ix_3} \right)
\cr
&&
+~ (-x_1)^{-1/2} (-x_2)^{-1/2} (-x_3)^{1/2-\nu} \sin(-x_1) \sin(-x_2)
\cr
&&
\left.
{\rm Im} \left( H^{(2)}_\nu(-x_1) H^{(2)}_\nu(-x_2)
\left( H^{(1)}_\nu(-x_3) \right)^2 \right)
\right]
\cr
&\times&
\int_{-\infty}^{0} dy_4
 (-y_4)^{-1/2}
{\rm Re} \left( H^{(1)}_\nu(-y_4) e^{-i y_4} \right) ~.
\label{Slimit_quasiloc}
\eea

Similar analyses can be applied to the ten terms of the factorized form listed in Appendix \ref{10terms_F}. We find
\bea
s(\nu)&=&
\frac{3\pi^2 \Gamma(\nu)}{2^{5-\nu}}
\cr
&\times&
{\rm Re} \left[ - i
\int_{-\infty}^0 dx_1 \int_{-\infty}^{x_1} dx_2
\left(
(-x_1)^{-1/2} (-x_2)^{1/2-\nu}
H^{(2)}_\nu(-x_1) e^{-i x_1}
\left( H^{(1)}_\nu(-x_2) \right)^2
\right. \right.
\cr
&& \quad ~~~~~~~~~~~~~~~~~~~~~~~~~~~~~
\left.
+ (-x_1)^{1/2-\nu} (-x_2)^{-1/2}
H^{(1)}_\nu(-x_1) H^{(2)}_\nu(-x_1)
H^{(1)}_\nu(-x_2) e^{-i x_2}
\right)
\cr
&& \quad ~~~~ \times
\int_{-\infty}^0 d\tx_1 ~ (-\tx_1)^{-1/2} H^{(2)}_\nu(-\tx_1)
e^{i\tx_1}
\cr
&& \quad ~~
+ i
\int_{-\infty}^0 dx_1 \int_{-\infty}^{x_1} dx_2 ~
(-x_1)^{-1/2}(-x_2)^{-1/2}
H^{(2)}_\nu(-x_1) e^{ix_1} H^{(2)}_\nu(-x_2) e^{ix_2}
\cr
&& \quad ~~~~ \times
\int_{-\infty}^0 d\tx_1 (-\tx_1)^{1/2-\nu}
\left( H^{(1)}_\nu(-\tx_1) \right)^2
\cr
&& \quad ~~
- i
\int_{-\infty}^0 dx_1 \int_{-\infty}^{x_1} dx_2 \int_{-\infty}^{x_2}
dx_3
\cr
&& \quad ~~~~~~
\left(
(-x_1)^{1/2-\nu} (-x_2)^{-1/2} (-x_3)^{-1/2}
\left( H^{(1)}_\nu(-x_1) \right)^2
H^{(2)}_\nu(-x_2)e^{ix_2} H^{(2)}_\nu(-x_3)e^{ix_3}
\right.
\cr
&& \quad ~~~~~~~
\left.\left.
+
(-x_1)^{-1/2} (-x_2)^{1/2-\nu} (-x_3)^{-1/2}
H^{(1)}_\nu(-x_1)e^{ix_1} H^{(1)}_\nu(-x_2) H^{(2)}_\nu(-x_2)
H^{(2)}_\nu(-x_3)e^{ix_3}
\right.\right.
\cr
&& \quad ~~~~~~~
\left. \left.
+
(-x_1)^{-1/2} (-x_2)^{-1/2} (-x_3)^{1/2-\nu}
H^{(1)}_\nu(-x_1)e^{ix_1} H^{(1)}_\nu(-x_2) e^{ix_2}
\left( H^{(2)}_\nu(-x_3) \right)^2
\right) \right]
\cr
&\times&
\int_{-\infty}^0 d\tx_2 (-\tx_2)^{-1/2}
{\rm Re} \left( H^{(1)}_\nu(-\tx_2)e^{-i \tx_2}\right)
~.
\label{Slimit_quasieq}
\eea

One can rewrite the expression ${\rm Re}[\cdots]$ in
(\ref{Slimit_quasieq}) in terms of
one time-ordered cubic integral, and this exactly reproduces
(\ref{Slimit_quasiloc}). Therefore the two expressions are equivalent.

\begin{figure}[t]
\center
\epsfig{file=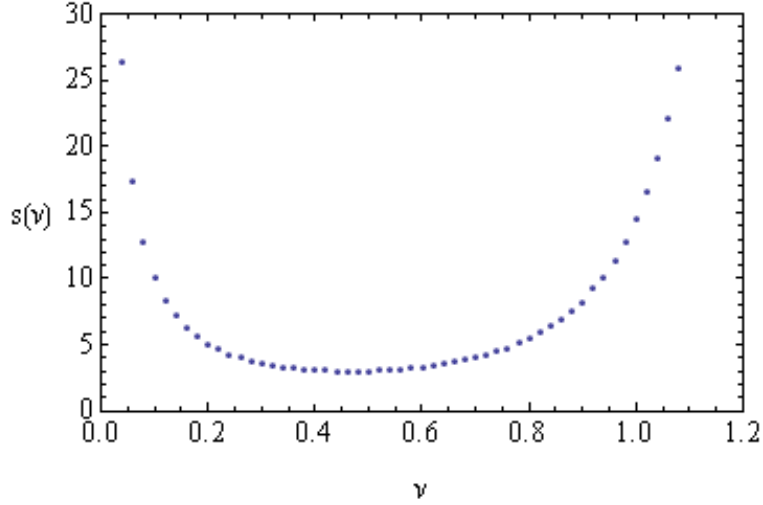, width=10cm}
\caption{The coefficient $s(\nu)$, defined in
  Eq. \eqref{Slimit_quasiloc}.}
\label{Fig:snu}
\end{figure}

The computational advantage of each form is the same as we have discussed in Sec.~\ref{Sec:trispectra}. The fastest way to evaluate $s(\nu)$ is to combine them into a mixed form, and do a Wick rotation in the UV part of this form. In this special case of the squeezed limit, it turns out that the expression (\ref{Slimit_quasiloc}) can also be evaluated by brute-force without these treatments. This is because the cubic integral is computationally less expensive than the quartic integral, and the UV behavior of this cubic integral is roughly $\int dx~ x^{-\nu-1/2} e^{\pm ix}$. For $1/2<\nu<3/2$, it converges without any regulation; for $0<\nu<1/2$ the oscillatory factors help it to converge, although the speed is increasingly slow towards $\nu=0$.
We plot $s(\nu)$ in Fig. \ref{Fig:snu}.

The squeezed limit as $\nu\to 0$ needs some extra care.
When we use (\ref{Hankelsmalllimit}) in the above analyses, the subleading term is suppressed by a factor of $(p_3/p_1)^{2\nu}$. This is the case only if
\bea
\frac{p_3}{p_1} \ll e^{-1/\nu} ~.
\eea
If $\nu$ is sufficiently close to zero so that the above condition is no longer satisfied,
the expansion of the Hankel function should be changed from
\bea
H^{(1)}_\nu(-\frac{p_3}{p_1} x_i) \to - i\frac{2^\nu \Gamma(\nu)}{\pi}
\left(\frac{p_3}{p_1}\right)^{-\nu} (-x_i)^{-\nu}
\eea
to
\bea
H^{(1)}_\nu(-\frac{p_3}{p_1} x_i) \to i\frac{2}{\pi} \ln
\frac{p_3}{p_1} ~.
\eea
If we fix $p_3/p_1$ while reducing $\nu$, as $\nu < -(\ln(p_3/p_1))^{-1}$, the factor
\bea
\frac{3\pi^2 \Gamma(\nu)}{2^{5-\nu}}
\frac{c_2^3c_3}{H R^6}
\frac{1}{p_1^{\frac{7}{2}-\nu} p_2 ~p_3^{\frac{3}{2}+\nu}}
\eea
in (\ref{Slimit_shape}) and (\ref{Slimit_quasiloc}) should be changed to
\bea
\frac{3\pi^2}{2^4}
\frac{c_2^3c_3}{H R^6}
\frac{\ln (p_3/p_1)}
{p_1^{\frac{7}{2}} p_2 ~p_3^{\frac{3}{2}}} ~;
\label{shape_nu=0}
\eea
and the two factors of $(-x_i)^{-\nu}$ inside the integrals
in (\ref{Slimit_quasiloc}) should be changed to $-1$.
In Fig.~\ref{Fig:snu}, we have assumed (\ref{Slimit_shape}). Therefore, the divergence as
$\nu$ approaches $0$ does not mean that
the non-Gaussianity is blowing up, rather signals the change of shape.

\medskip

We end this section by discussing two interesting physical aspects.

$\bullet$ {\em Quasi-equilateral and quasi-local shapes.}
Notice that the squeezed limit of our bispectra ($\sim p_3^{-3/2-\nu}$) lies between that of the equilateral ($\sim p_3^{-1}$) and local ($\sim p_3^{-3}$) bispectrum \cite{Babich:2004gb,Chen:2006nt}. We call these shapes the ``intermediate shapes''.
They are not super-position of any previous known shapes. For example the superposition of the local and equilateral shape gives a different scaling behavior at the squeezed limit.
For the constant turn case, these bispectra are scale-invariant.

Equilateral and local bispectra are two well-known types of scale-invariant non-Gaussianities that can become observably large. The underlying physics associated with these two shapes are as follows.

The large equilateral bispectrum is typically generated when the interacting modes are crossing the horizon around the same time. This happens for example in the single field inflation models with higher derivative interactions or certain multifield generalizations \cite{Chen:2006nt,Cheung:2007st,Li:2008qc,Langlois:2008mn}, such as DBI inflation \cite{Silverstein:2003hf,Chen:2004gc} or k-inflation \cite{ArmendarizPicon:1999rj,Garriga:1999vw}. Long wavelength modes that already crossed the horizon is frozen in single field inflation, and cannot have large correlations with the modes that are much shorter. As a consequence, the bispectrum, properly normalized, peaks at the equilateral limit.

On the contrary, the large local bispectrum is typically generated when the modes have already exited the horizon. This happens for example in special types of multi-field slow-roll models \cite{Sasaki:1995aw,Salopek:1990jq} or curvaton models \cite{curvaton}. Superhorizon curvature perturbations are not conserved in multi-field inflation models, and can receive contributions from isocurvature modes. Different patches of universe that are separated by inflationary horizons evolve independently, and so the non-Gaussianities come in locally in position space. Thus in momentum space the correlation becomes non-local and peaks in the squeezed triangle limit.

In the quasi-single field inflation, the isocurvaton has a mass that can vary around $\CO(H)$. For the heavier field $m>\sqrt{2}H$, i.e.~$\nu<1/2$, its amplitude decays faster after horizon-exit. Therefore large interactions happen during the horizon exit, and we get shapes that are closer to the equilateral type. Namely, the properly normalized bispectra peak roughly at the equilateral limit.\footnote{The fact that the shapes are in general flatter than the equilateral shape has to do with the fact that the interaction in this model originates from $V(\sigma)$ and is local in the position space to start with.} We call it ``quasi-equilateral". The numerical example of such a shape can be found in Fig.~\ref{Fig:nu0} ($\nu=0.3$).

For the lighter field $m<\sqrt{2}H$, i.e.~$3/2>\nu>1/2$, its amplitude decays slower. The conversion from the isocurvature to curvature mode is still continuing after the mode exits the horizon. Here the interactions among the isocurvature modes are local, and in addition the conversion will become increasingly local in the position space for the reason that we have explained. So we get shapes that are closer to the local type, and they peak at the squeezed limit. We call it ``quasi-local". See Fig.~\ref{Fig:nu0} ($\nu=1$).

In the limiting case $m^2=0$, i.e.~$\nu=3/2$, the squeezed limit of our bispectra coincide with that of the local type. In this limit, the isocurvaton fluctuations do not decay.
This is the reason that, in Fig.~\ref{Fig:Cnu}, \ref{Fig:eqlimit} and \ref{Fig:snu}, the amplitudes of ${\cal C}$, $f_{NL}^{\rm int}$ and $s$ approach infinity as $\nu\to 3/2$ for the constant turn case. As discussed in Sec.~\ref{Sec:bispectra}, infrared e-folds cutoff should be considered in this limiting case. Most of our analyses still apply in this limit, but we would like to distinguish the following two cases. First, if $V'''$ is still large, we can use (\ref{fnlcal}) but with infrared e-folds cutoff. The cutoff will introduce a running in $f_{NL}^{\rm int}$ because different modes correspond to different $N_f$ and $\alpha(\nu)$ is $N_f$-dependent. Second, if the potential in the isocurvature direction also becomes a slow-roll potential, $V'''$ is very small, $\sim \CO(\epsilon^{3/2}) H^2/M_p$. In this case, other terms in the cubic Lagrangian will become more important, but in terms of contributing to $f_{NL}^{\rm int}$ they are all small. As we discussed in Sec.~\ref{Sec:bispectra}, the enhancement factor from $\alpha(\nu)$ is only $N_f$ in the perturbative regime. So the bispectrum in this case is small. This is in accordance with the general findings of previous studies that it is very difficult to generate large non-Gaussianities in terms of turning trajectories in multifield slow-roll models \cite{Salopek:1990jq}, essentially because imposing slow-roll conditions in all directions are very restrictive.

As we have seen, the shapes of the non-Gaussianities depend very sensitively on the mass of the isocurvaton. The shape of bispectrum changes from quasi-equilateral to quasi-local as $m$ changes just from $1.47H$ ($\nu=0.3$) to about $1.1H$ ($\nu=1$). However on the other hand, as $\nu$ becomes close to $1.5$, the shapes of bispectra are very close to the local form, but $m$ is still of order $H$. For example, for $\nu=1.4$, $m$ is still $\approx 0.54H$; but the shape scales as $p_3^{-2.9}$, comparing to the local one $p_3^{-3}$. It is clear that $m$ is still much too heavy for the quasi-single field inflation to become two-field slow-roll inflation, hence the underlying models are very different. Therefore, it is an interesting question how good we can distinguish the quasi-local form from the local form experimentally.

$\bullet$ {\em Change of shapes in isocurvature-curvature conversion.}
The non-Gaussianities in this model originate from non-Gaussian fluctuations in the isocurvature direction. It is interesting to look at the shapes of the three-point correlation function of the isocurvature modes before it is transformed into that of the curvature modes,
\begin{align}
\langle \delta\sigma^3 \rangle
=&~ i \int_{t_0}^{t} dt_1 \langle [ H_I(t_1), \delta\sigma^3(t) ] \rangle
\cr
=&~ ic_3 \frac{\pi^6}{64}H^2 (-\tau)^{9/2} H^{(2)}_\nu(-p_1\tau) H^{(2)}_\nu(-p_2\tau) H^{(2)}_\nu(-p_3\tau)
\cr
& \int_{-\infty}^\tau d\tau_1 (-\tau_1)^{1/2}
H^{(1)}_\nu(-p_1\tau_1) H^{(1)}_\nu(-p_2\tau_1) H^{(1)}_\nu(-p_3\tau_1)
+ c.c. ~.
\end{align}
In the squeezed limit and for modes that exit the horizon, this is proportional to
\bea
\sim (-p_1 \tau)^{9/2-3\nu} p_3^{-2\nu} p_1^{-6+2\nu} ~.
\eea
We see that its amplitude is decaying and its
shape goes as $p_3^{-2\nu}$.
Therefore it is evident that the effect of the transfer vertex is not a simple projection. During the transfer, the shape of the correlation function has been changed, slightly towards the local type.
It is important to investigate such changes in other cases including the multi-field inflationary models.

\section{Shape ansatz}\label{Sec:shapeAns}
\setcounter{equation}{0}

As we have seen, the precise shapes of the bispectra are complicated for quasi-single field inflation and we have to perform numerical integration to see the full shapes. However, for the purpose of data analyses, simple analytical expressions which resemble closely to the precise shapes are desirable. So we would like to start with the analytical results in the squeezed limit, and construct such shape ansatz. For example, we find the following ansatz reproduces the squeezed limit behavior and has good overall match with our numerical results,
\bea
F= \frac{3^{7/2}}{10 N_\nu(\alpha/27)}
\frac{f_{NL}^{\rm int}}{(p_1p_2p_3)^{3/2}(p_1+p_2+p_3)^{3/2}}
N_\nu\left( \frac{\alpha ~  p_1p_2p_3}{(p_1+p_2+p_3)^3} \right) ~,
\label{ansatz1}
\eea
where $N_\nu$ is the Neumann Function and $F$ is defined in (\ref{F_def}). The parameter $\alpha$ can be adjusted to fit the ansatz with the numerical results, we found $\alpha \simeq 8$. For data analyses, $\alpha$ is no longer a free parameter.

\begin{figure}[htpb]
\begin{center}
\epsfig{file=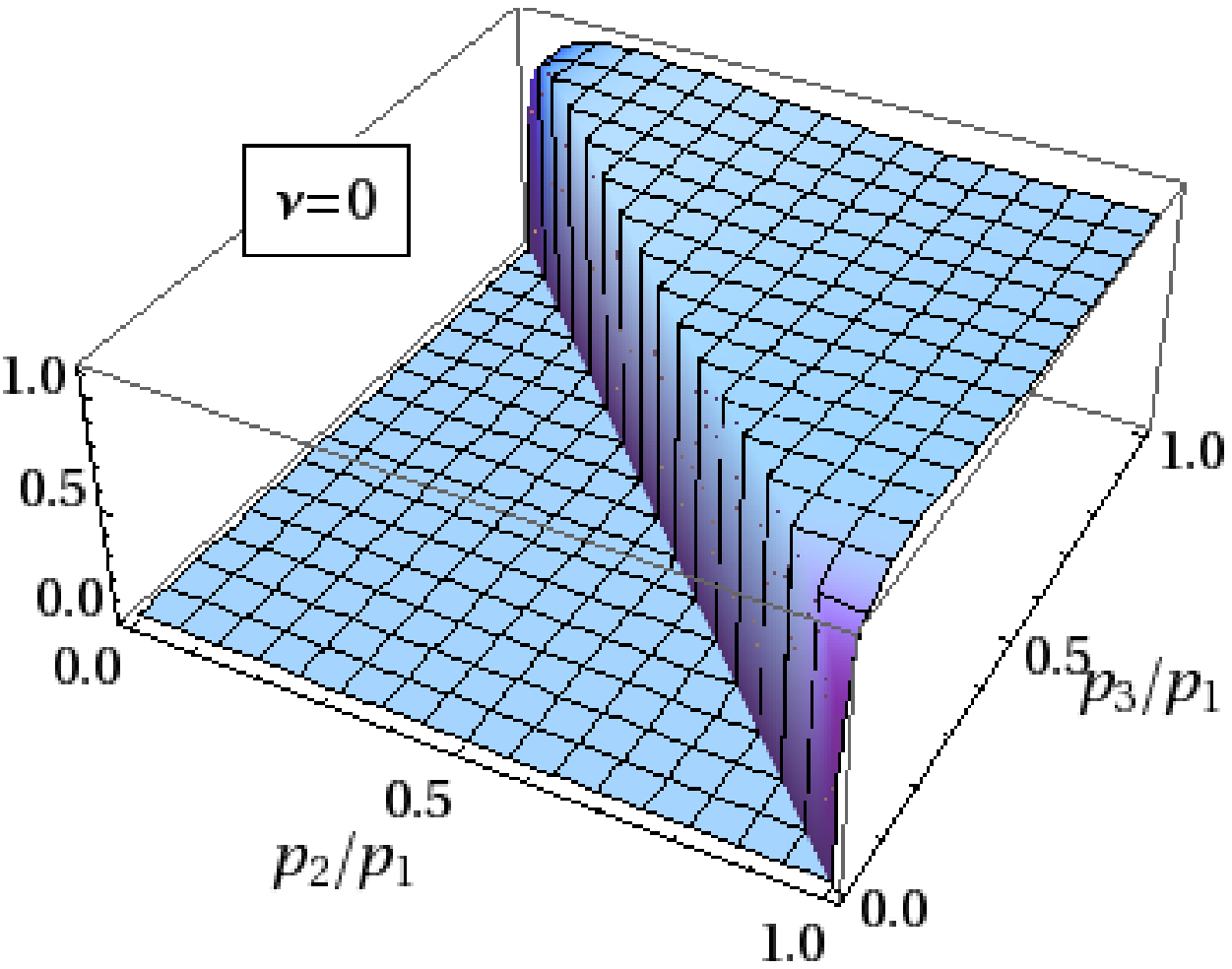, width=0.45\textwidth}
\epsfig{file=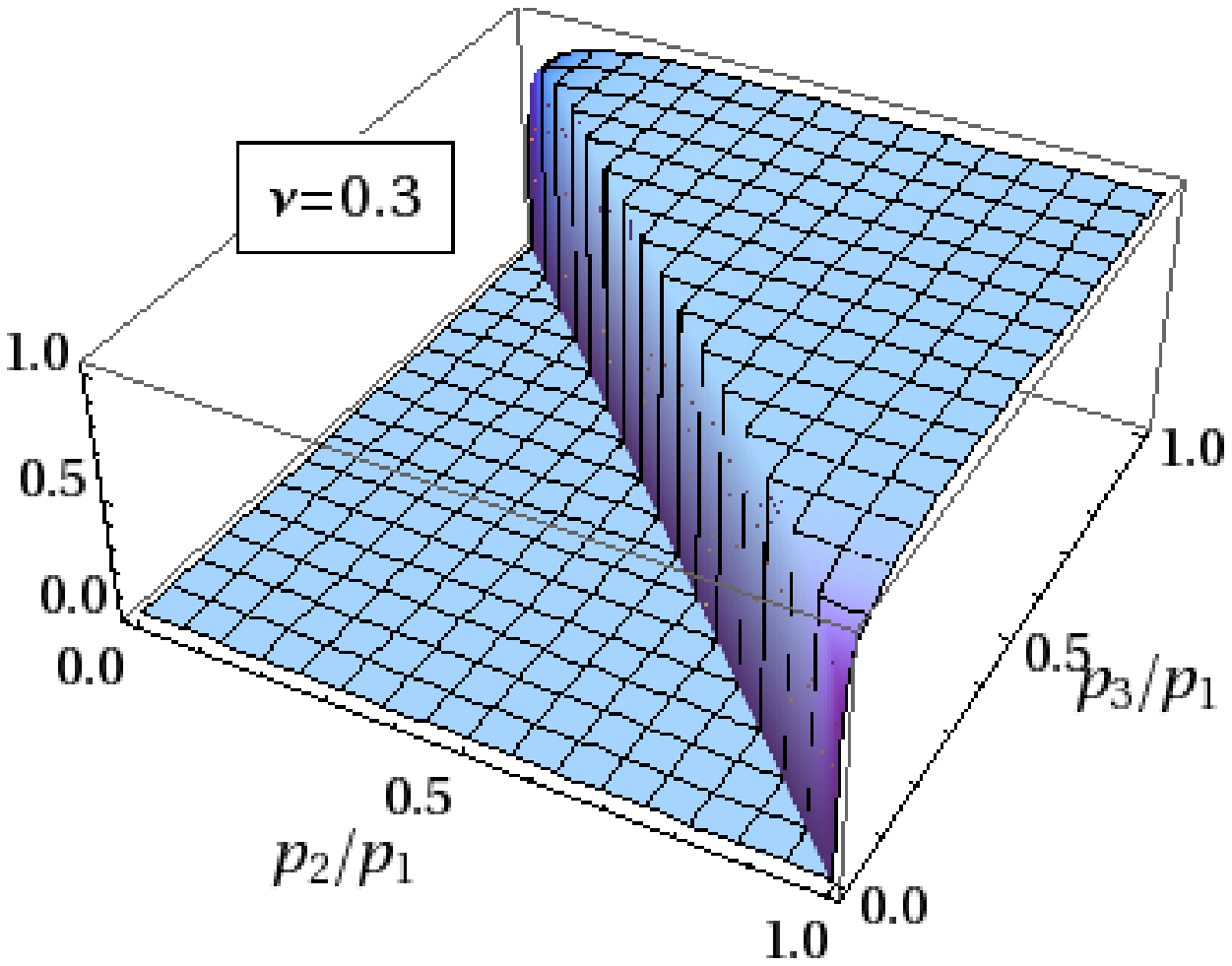, width=0.45\textwidth}
\epsfig{file=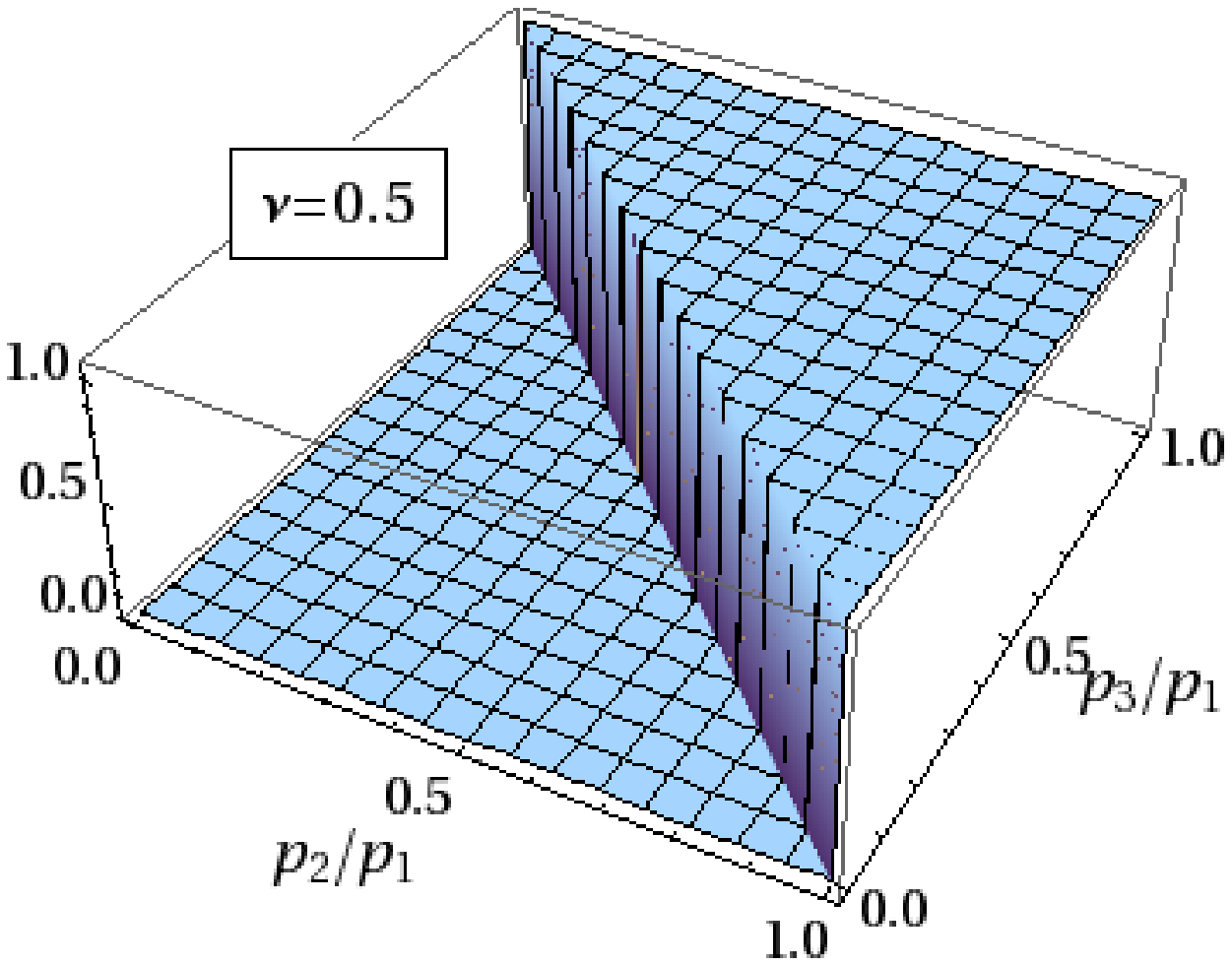, width=0.45\textwidth}
\epsfig{file=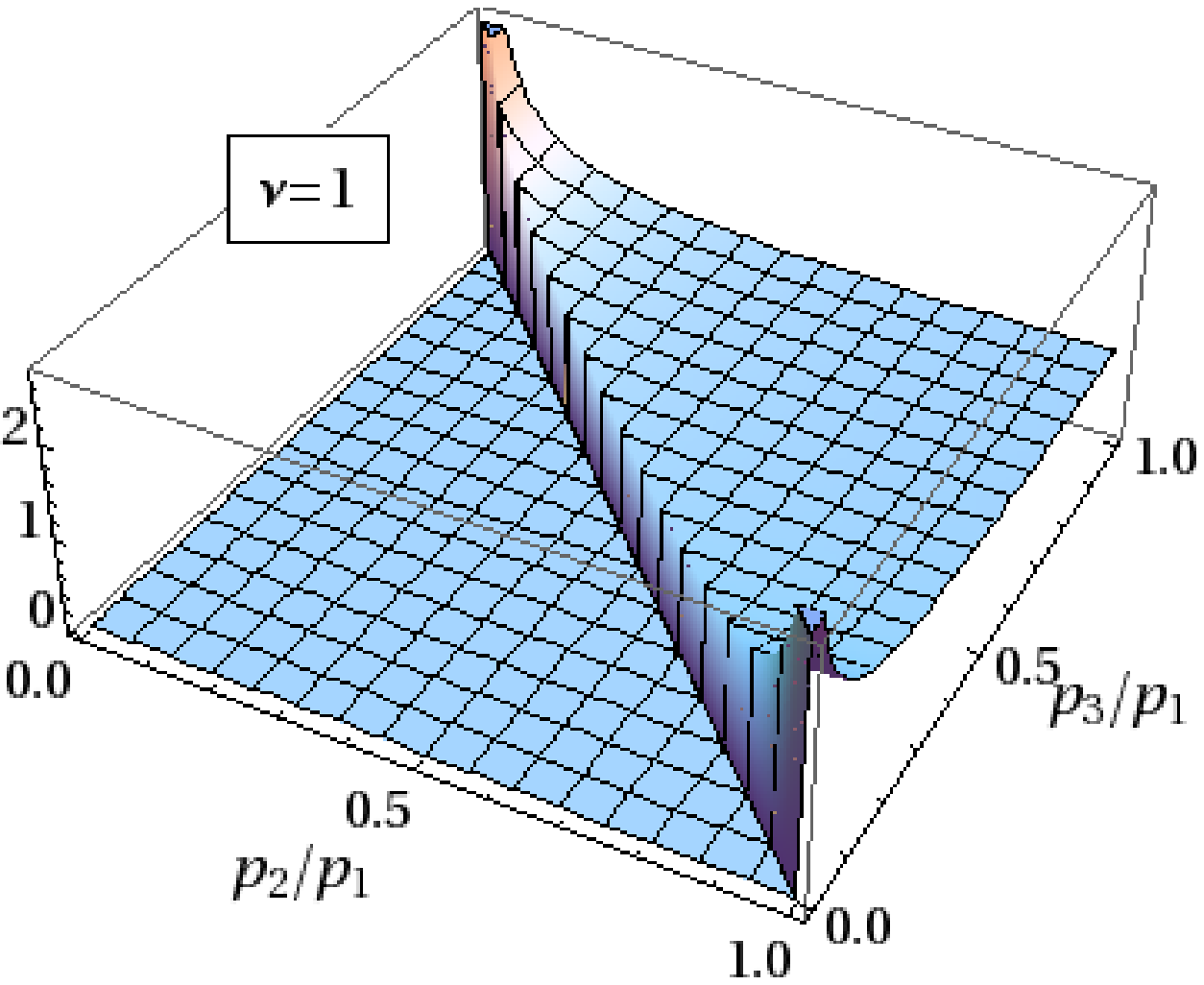, width=0.45\textwidth}
\end{center}
\caption{Shape ansatz \eqref{ansatz1} with $\alpha=8$. We plot $(p_1p_2p_3)^2 F$ for $\nu=0,~0.3,~0.5,~1$ respectively. The plot is
normalized such that $(p_1p_2p_3)^2 F=1$ for $p_1=p_2=p_3=1$.
\label{Fig:ansshapes}}
\end{figure}

We have compared this ansatz with the
numerical results of the full shapes for $\nu=0, 0.2, 0.3, 0.5, 1$,
and found good match.  Examples of the shape ansatz are shown in
Fig.~\ref{Fig:ansshapes}, and should be compared with the numerical
results in Fig.~\ref{Fig:nu0}. The overall shapes as a function of $\nu$
and the squeezed limits of both results match well. There are some
small differences in the bulk. For example, for the $\nu=1/2$ case,
although there is some slight growth towards the folded triangle limit
in the numerical results, we have numerically checked that the shape
remains very flat if we look at the more squeezed configuration,
consistent with the analytical results. So these small differences
should be due to the second order terms that we did not take into
account in the analytical computation of the squeezed limit.

\begin{figure}[htpb]
\begin{center}
\epsfig{file=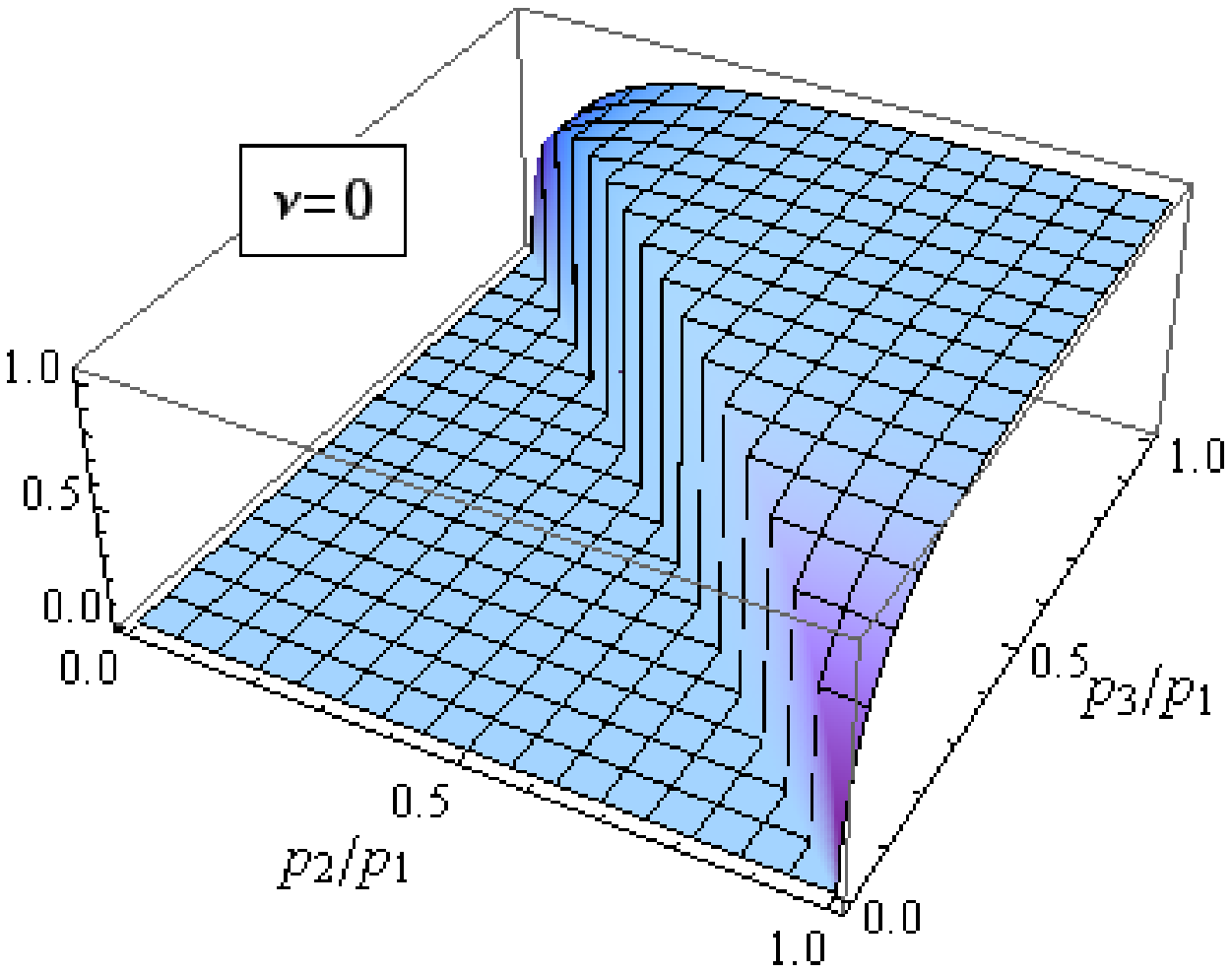, width=0.45\textwidth}
\epsfig{file=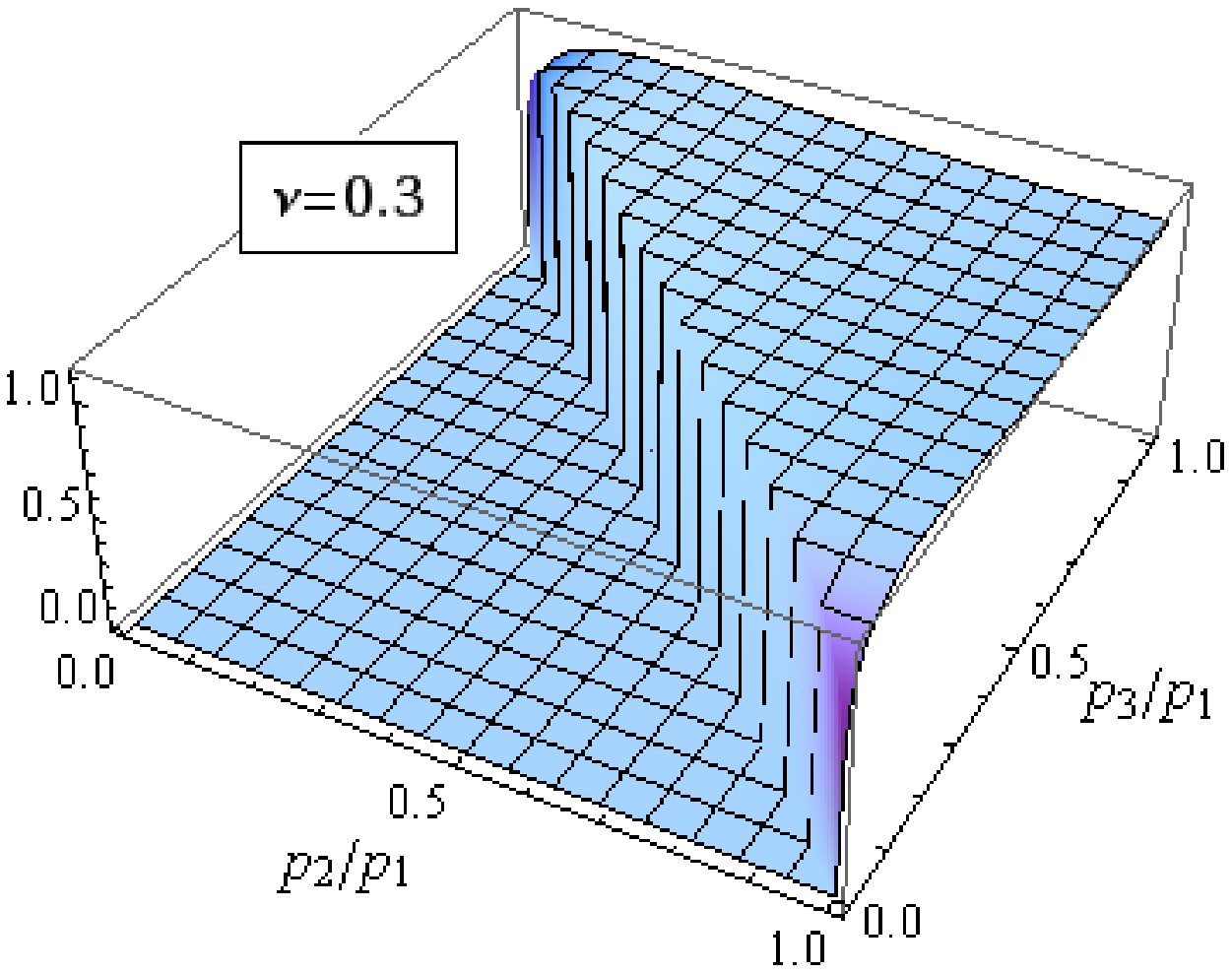, width=0.45\textwidth}
\epsfig{file=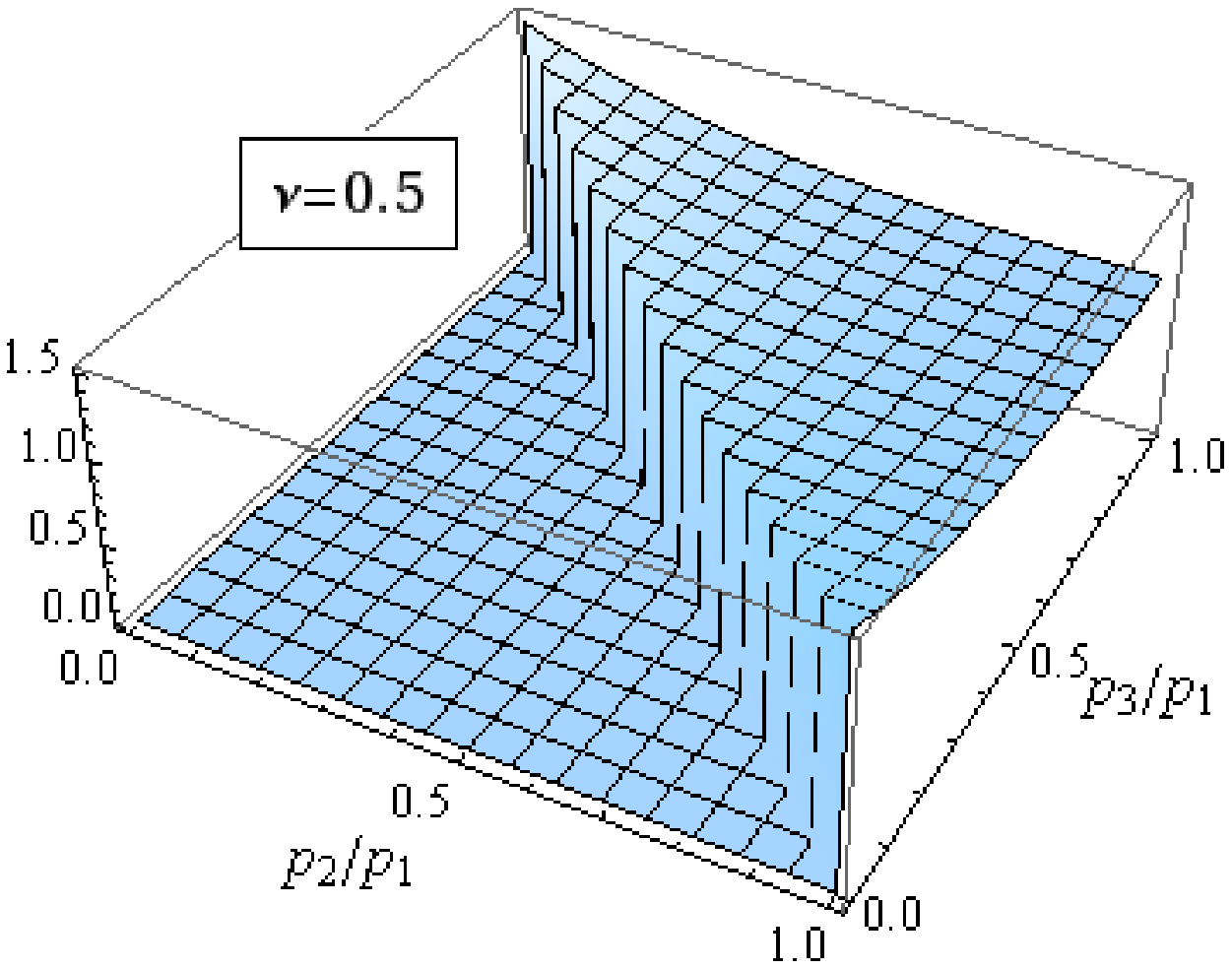, width=0.45\textwidth}
\epsfig{file=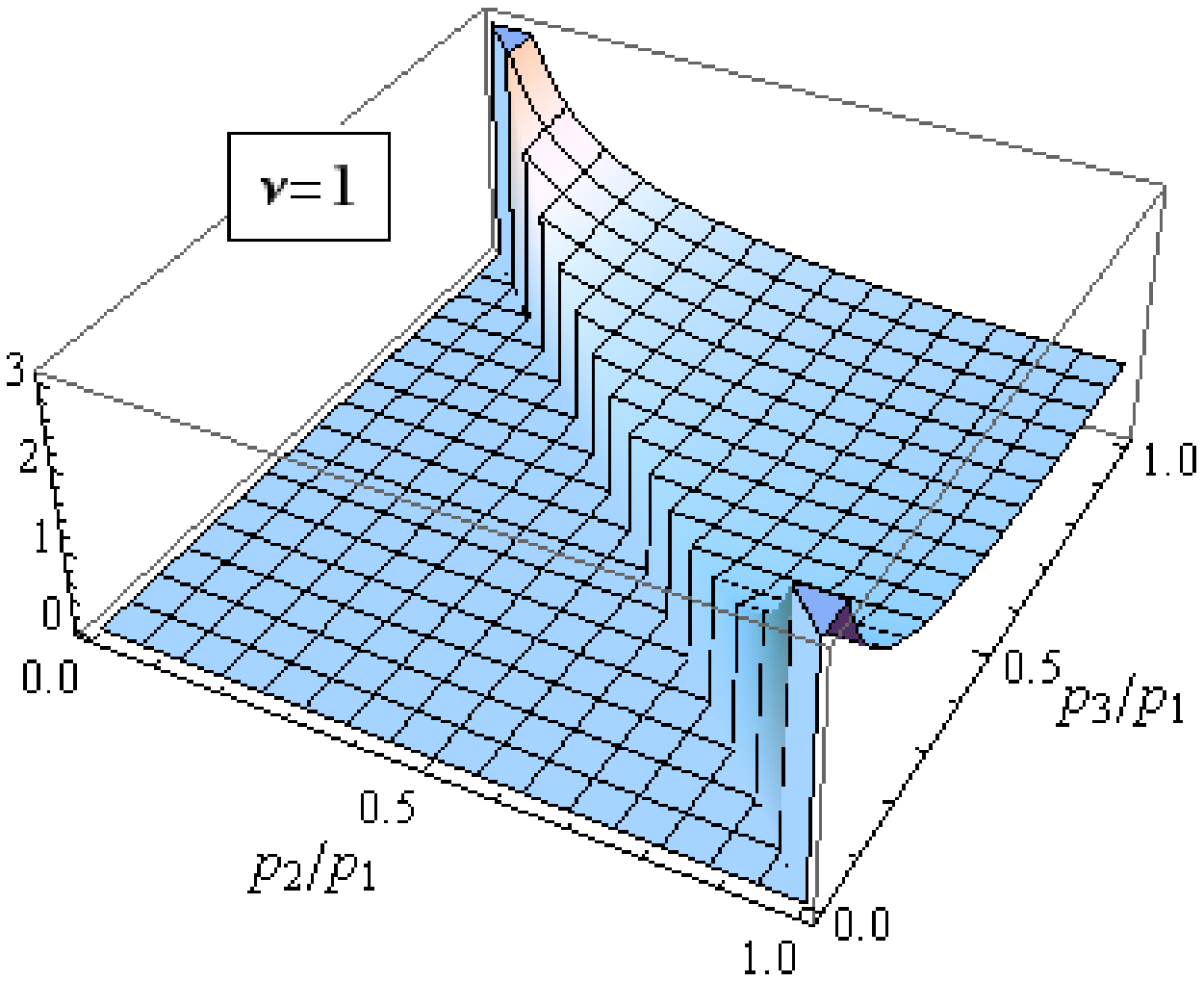, width=0.45\textwidth}
\end{center}
\caption{Shape templates \eqref{template1} with the same convention as in Fig.~\ref{Fig:ansshapes}.
\label{Fig:tplshapes}}
\end{figure}

In data analyses, the construction of the estimator involves triple integral of the shape function over the three momenta $p_i$. To have practical computational costs, it is necessary to factorize this integral into a multiplication of three integrals that involve individual $p_i$. So we would like to further approximate the shape ansatz by templates with simpler functions, for example,
\bea
F = \frac{3^{\frac{9}{2}-3\nu}}{10}
\frac{f_{NL}^{\rm int}(p_1^2+p_2^2+p_3^2)}{(p_1p_2p_3)^{\frac{3}{2}+\nu}(p_1+p_2+p_3)^{\frac{7}{2}-3\nu}} ~.
\label{template1}
\eea
These simpler templates are shown in Fig.~\ref{Fig:tplshapes}. They
reproduce the shape functions quite well
except near $\nu=0$. If necessary, one can come up with other templates that are more factorized, to get rid of $(p_1+p_2+p_3)^{-7/2+3\nu}$.

\section{Trispectra}\label{Sec:trispectra}
\setcounter{equation}{0}

Four-point correlation functions provide complementary information on inflationary dynamics, as well as redundant checks on the three-point functions. The study on the trispectra has recently attracted much attention in data analyses \cite{Kogo:2006kh} and model building \cite{Byrnes:2006vq,Seery:2008ax,Chen:2009bc,Gao:2009gd}.

In our model, there are two terms in the interacting
Hamiltonian that contribute to the leading trispectra, namely, the $V'''$ and
$V''''$ terms. There are two Feynman diagrams, the contact-interaction diagram (Fig.~\ref{Fig:fdiag4}(a)) and the scalar-exchange diagram (Fig.~\ref{Fig:fdiag4}(b)). Analogous to the $f_{NL}$ in the bispectra, we use the $t_{NL}$ to denote the magnitude of the trispectra for a given shape. For each shape component, we take the regular tetrahedron limit and define $t_{NL}$ as \cite{Chen:2009bc}
\bea
\langle \zeta^4 \rangle_{\rm component}
\xrightarrow[\rm limit]{\rm R.T.} (2\pi)^9 P_\zeta^3 \delta^3(\sum_i \bp_i)\frac{1}{p_1^9} t_{NL} ~.
\label{tNL_def}
\eea

For the contact-interaction diagram, the
contribution from the $V''''$ term can be calculated by expanding the
in-in formalism to the fifth order (Appendix \ref{App:inin}),
\begin{align}
  \langle \delta\theta^4 \rangle \supset - 2{\rm Re}\left[ i
\int_{t_0}^t d\tilde t_1 \int_{t_0}^{\tilde t_1}
d\tilde t_2
\int_{t_0}^t dt_1 \int_{t_0}^{t_1} dt_2 \int_{t_0}^{t_2}dt_3
~\langle H_I(\tilde t_2) H_I(\tilde t_1) ~\delta\theta_I^4~
H_I(t_1)H_I(t_2)H_I(t_3) \rangle \right]\nonumber\\
+2{\rm Re} \left[i
\int_{t_0}^{t}d\tilde t_1
\int_{t_0}^t dt_1 \int_{t_0}^{t_1} dt_2 \int_{t_0}^{t_2} dt_3
\int_{t_0}^{t_3} dt_4
~\langle H_I(\tilde t_1) ~\delta\theta_I^4~
H_I(t_1) H_I(t_2) H_I(t_3) H_I(t_4)\rangle
\right]\nonumber\\
-2{\rm Re} \left[i
\int_{t_0}^t dt_1 \int_{t_0}^{t_1} dt_2 \int_{t_0}^{t_2} dt_3
\int_{t_0}^{t_3} dt_4 \int_{t_0}^{t_4} dt_5
~\langle \delta\theta_I^4~
H_I(t_1) H_I(t_2) H_I(t_3) H_I(t_4) H_I(t_5)\rangle
\right] ~,
\end{align}
or alternatively using the commutator form
\begin{align}
  \langle \delta\theta^4 \rangle \supset
i \int_{t_0}^t dt_1 \cdots \int_{t_0}^{t_{4}}dt_5 \left\langle
\left[ H_I(t_5),\left[H_I(t_{4}),\cdots ,
\left[ H_I(t_1),Q_I(t) \right]
\cdots \right] \right]
\right\rangle ~,
\end{align}
where one of the $H_I$'s is to be replaced by $H_4^I\equiv \frac{1}{24}\int d^3 x
a^3 V'''' \delta\sigma^4$, and other $H_I$'s
are to be replaced by $H_2^I$.
The order of magnitude estimate for the size of this trispectrum is as follows.
The quartic interaction contributes a factor of $V''''$ to $t_{NL}$. The transfer efficiency is $\sim (\dot\theta/H)^4$, because we need four transfer-vertices. Keeping in mind that $\zeta \sim \CP_\zeta^{1/2}$ in $\langle \zeta^4 \rangle$, and comparing with the definition (\ref{tNL_def}), we get
\begin{equation}\label{tri4}
  t_{NL}^{\rm CI} \sim
  P_\zeta^{-1} \left(\dot\theta/H\right)^4
  V'''' ~,
\end{equation}
where the superscript ``CI" denotes contact-interaction.

For the scalar-exchange diagram, the contribution of the $V'''$ term
can be calculated by expanding the in-in formalism to the sixth order (Appendix \ref{App:inin}),
where two of the $H_I$'s are to be replaced by $H_3^I$, and other
$H_I$'s are to be replaced by $H_2^I$. Each cubic vertex contributes a factor of $(V'''/H)$ to $t_{NL}$, and the transfer efficiency is again $(\dot\theta/H)^4$. So overall,
\begin{equation}\label{tri3}
  t_{NL}^{\rm SE} \sim P_\zeta^{-1} \left(\dot\theta/H\right)^4
 \left(V'''/H\right)^2 ~,
\end{equation}
where the super-script ``SE" denote scalar-exchange.

Comparing \eqref{tri3} with the bispectra (\ref{fnlcal}), we have
\begin{align}
  t^{\rm SE}_{NL} \sim \left(H/ \dot \theta\right)^2 f_{NL}^2~.
\end{align}
In the slow turn case, $\left(\dot\theta/H\right)^2 \ll 1$, so
$t_{NL}^{} \gg f_{NL}^2$. Such a large trispectra may be a better probe
for quasi-single field inflation than bispectra.
Comparing \eqref{tri4} with the bispectra (\ref{fnlcal}), we have
\begin{align}
  t_{NL}^{\rm CI} \sim \left(H/ \dot \theta\right)^2
  \left(V''''H^2/(V''')^2\right) f_{NL}^2~.
\end{align}
$t_{NL}^{\rm CI}$ can be either larger or smaller than $f_{NL}^2$ and $t^{\rm SE}_{NL}$, depending
on the details of the potential $V'''$ and $V''''$.

It is important to study the integration numerically and analytically to see
the shapes of the trispectra, as well as the coefficients in
front of Eqs.~\eqref{tri4} and \eqref{tri3}.
This is beyond the
scope of the current work.

\section{Conclusion and discussion}\label{Sec:conclusion}
\setcounter{equation}{0}

To conclude, in this paper, we have investigated in detail a quasi-single
field inflation model. We find fields with mass of order $H$ can have important impacts on density perturbations through, for example, turning trajectories.
These effects can be computed perturbatively using transfer vertex and Feynman diagrams in the in-in formalism. A one-parameter family of potentially observable large bispectra arise. These new shapes are controlled sensitively by the mass of the isocurvaton and lie between the equilateral and local shape.
We also note that the sizes of the trispectra are even larger than those of the bispectra squared in this model.

There are a lot of issues remaining to be investigated in the quasi-single field inflation models. For example,

$\bullet$ {\em Data analysis}. Experimental constraints on non-Gaussianities depend on their detailed shapes and running. A variety of experimental methods have been applied to the local and equilateral bispectra \cite{CMB,LSS,otherCMB,lspace,running,Komatsu:2009kd}. It will be very
interesting to constrain the family of new shapes that we find here, or
to fit the parameter $\nu$, using the observational data.

$\bullet$ {\em Running of density perturbations}. In this paper we have only considered the constant turn case. More realistically, we expect parameters to vary along the trajectory. As we noted, such a non-constant
turn results in a running of the spectral index and non-Gaussianities. It is worth to
investigate this running effect in more details.

$\bullet$ {\em Trispectra}. In this paper, we obtained the order of magnitude estimate of the trispectra in quasi-single field inflation.
However, the $\nu$-dependent coefficients in the
trispectra, and more importantly the shape of the trispectra still
remain to be calculated.

$\bullet$ {\em Multiple isocurvature directions}. In this paper, we considered one massive isocurvaton. If there exist multiple massive isocurvature directions, instantaneously along the inflaton trajectory, one can find a two-dimensional hyper-surface where there is only one effective isocurvaton. If this hyper-surface is not changing with time, the model belongs to the two-field model such as the one we considered here. If the hyper-surface is changing with time, the situation becomes more complicated. It is worth to investigate the observational consequences of such a case.

$\bullet$ {\em Other quasi-single field inflaton models}. The turning trajectory model that we studied is one simple example of the quasi-single field inflation models. The coupling between the inflaton and isocurvatons can be introduced through the kinetic terms in a more general way which may or may not be described by turning trajectories.\footnote{We thank Jim Cline for helpful discussions on this point.} Such couplings can even be introduced through other types of couplings that do not involve the kinetic terms. Which correlation functions/non-Gaussianities are enhanced is determined by the structure of these couplings.

$\bullet$ {\em String cosmology}. As we discussed in the Introduction,
quasi-single field inflation is a natural picture for inflation in
string theory and supergravity. Fields with mass of order $H$ are ubiquitous and in fact are a common hazard to inflation model building. We have seen that, as isocurvatons, such fields can have important consequences on density perturbations. It thus becomes very interesting to build explicit quasi-single
field inflation models from string theory, in terms of either turning
trajectories or more general couplings. Because the shapes of the non-Gaussianities depend very sensitively on the mass, and the sizes depend on the strength of the couplings, we have the opportunities to probe such fields through experiments.

$\bullet$ {\em Generalizations}. As the number of fields increases, it becomes a logical possibility that we can have more than one inflaton fields. It is worth to generalize the formalism developed in the current work to these more general multiple field models.
It is also worth to apply the perturbative method used here to multifield inflation models.

\medskip
\section*{Acknowledgments}

We thank Robert Brandenberger, Bin Chen, Jim Cline, Andrew Frey, Alan Guth, Min-xin Huang,
Qing-Guo Huang, Kazuya Koyama, Miao Li, Michele Liguori, Subodh Patil, David Seery, Andrew Tolley, Paul Shellard, Meng Su, Bret Underwood, David Wands and Amit Yadav for helpful discussions. XC would like to thank Robert Brandenberger for the invitations to the workshops of ``Connecting fundamental physics to observations" and ``Holographic cosmology", and the hospitality of the KITPC at the Chinese Academy of Sciences, where part of this work was done. XC was supported by the Stephen Hawking advanced fellowship and US DOE under cooperative research
agreement DEFG02-05ER41360. YW was supported by NSERC and an IPP postdoctoral fellowship.

\appendix

\section{The full Lagrangian up to third order}
\label{App:thirdorderaction}
\setcounter{equation}{0}

In this section, we derive the full third order action in two different gauges mentioned in Sec.~\ref{Sec:twogauges}.
We also show that \eqref{CL3} is the leading order
interaction.
We set $M_p=1$ in this Appendix.

\subsection{Spatially flat gauge}
\label{Appgauge1}

The full action is
\bea
S=S_g+S_m ~,
\eea
where
\bea
S_g = \half \int d^4 x \sqrt{-g} {\cal R}
\eea
and
\bea
S_m=\int d^4x \CL_m
\eea
which is given in (\ref{ModelAction}).
Using the ADM metric,
\bea
ds^2=-N^2 dt^2 + h_{ij} (dx^i + N^i dt)(dx^j + N^j dt) ~,
\eea
the action becomes
\bea
S= \half \int dt dx^3 \sqrt{h} N (R^{(3)} + 2\CL_m)
+ \half \int dt dx^3 \sqrt{h} N^{-1} (E_{ij}E^{ij} -E^2) ~,
\eea
where the index of $N^i$ can be lowered by the 3d metric $h_{ij}$ and
$R^{(3)}$ is the 3d Ricci scalar constructed from $h_{ij}$. The
definition of $E_{ij}$ and $E$ are
\bea
E_{ij} &=& \half (\dot h_{ij} - \nabla_i N_j - \nabla_j N_i) ~,
\cr
E &=& E_{ij} h^{ij} ~.
\eea

We choose the following spatially flat gauge:
\bea
h_{ij}=a^2(t) \delta_{ij} ~, ~~~ \theta=\theta_0(t) + \delta \theta ~,~~~
\sigma= \sigma_0(t) + \delta \sigma ~.
\eea
For the constant turn case $\sigma_0(t)={\rm const.}$.

The constraint equations for the Lagrangian multiplers $N$ and $N_i$
are
\bea
R^{(3)} + 2 \CL_m + 2N \frac{\partial \CL_m}{\partial N} - \frac{1}{N^2}
(E_{ij} E^{ij} - E^2) &=& 0 ~,
\label{ADMconstr1}
\\
\nabla_i \left[ N^{-1} (E^{ij}-h^{ij}E) \right] + N \frac{\partial
  \CL_m}{\partial N_j} &=& 0 ~.
\label{ADMconstr2}
\eea

In the ADM formalism, to expand the action to the third order in
perturbations, it is
sufficient to solve the Lagrangian multipliers $N$ and $N_i$ to the
first order in perturbations. So we expand
\bea
N=1+\alpha_1~, ~~~
N_i = \partial_i \psi_1 + \tilde N_i^{(1)} ~,
\eea
where $\partial_i \tilde N_i^{(1)}=0$,
and plug them into (\ref{ADMconstr1}) and (\ref{ADMconstr2}). The
solutions with proper boundary conditions are
\bea
&& \alpha_1 = \frac{R^2 \dot\theta_0}{2H}\delta \theta ~, ~~~
\tilde N_i^{(1)} =0 ~,
\cr
&& \partial^2 \psi = -\frac{a^2}{2H}
\left(6H^2-R^2 \dot \theta_0^2\right)\alpha_1
- \frac{a^2}{2H} \left(R^2 \dot\theta_0 \dot{\delta\theta} +
R\dot\theta_0^2 \delta\sigma + V'_{sr} \delta\theta + V'\delta\sigma
\right) ~.
\eea
We then plug these solutions
into the action and expand up to the third order in
perturbations. The first order terms give the equation of motion for $\theta_0(t)$ and $\sigma_0(t)$. The quadratic order terms are
\begin{align}
\label{eq:L2_hc}
\CL_2=& \frac{a^3}{2}R^2 \dot{\delta\theta}^2 - \frac{a}{2} R^2(\partial_i \delta\theta)^2
-a^3 \left( \frac{V_{sr}''}{2R^2} - (3\epsilon-\epsilon^2+\epsilon\eta)H^2 \right)
R^2 \delta\theta^2
\cr
& +\frac{a^3}{2} \dot{\delta\sigma}^2 - \frac{a}{2} (\partial_i\delta\sigma)^2
-\frac{a^3}{2} (V''- \dot\theta_0^2) \delta\sigma^2
\cr
& +2 a^3 R\dot\theta_0 \dot{\delta\theta} \delta\sigma
- 2\epsilon a^3 R\dot\theta_0 H \delta\theta \delta\sigma ~.
\end{align}
The third order terms are
\begin{align}
&  \frac{{\cal L}_3}{a^3} =-\frac{V'''}{6}\delta\sigma^3 -
  \frac{\delta\theta^3}{48H^3M_p^6}\left(8H^3M_p^6
    V_{\rm sr}'''+12H^2M_p^4R^2V_{\rm sr}''\dot\theta_0-18H^2M_p^2R^6\dot\theta_0^3+3R^8\dot\theta_0^5
  \right)\nonumber\\&
-\frac{R^2}{a^2}\partial_i\psi\partial_i\delta\theta\dot{\delta\theta} +\dot\theta_0\delta\sigma^2\dot{\delta\theta} +\frac{R^4\dot\theta_0^2}{2a^2M_p^2H}\delta\theta^2\partial_i\partial_i\psi\nonumber\\&
+\frac{R^5\dot\theta_0^4\delta\theta^2\delta\sigma}{4H^2M_p^4}
+\frac{R^6\dot\theta_0^3}{4H^2M_p^4}\delta\theta^2\dot{\delta\theta}
+R\delta\sigma\dot{\delta\theta}^2\nonumber\\&
-\frac{R}{a^2}\delta\sigma\partial_i\delta\theta\partial_i\delta\theta
-\frac{2R\dot\theta_0}{a^2}\delta\sigma\partial_i\delta\theta\partial_i\psi -\frac{1}{a^2}\partial_i\psi\partial_i\delta\sigma \dot{\delta\sigma}\nonumber\\&
-\frac{R^2\dot\theta_0}{4a^4H}\delta\theta
\left[ \partial_i\partial_j\psi \partial_i\partial_j\psi-(\partial_i\partial_i\psi)^2\right] -\frac{R^4\dot\theta_0}{4a^2HM_p^2}\delta\theta\partial_i\delta\theta\partial_i\delta\theta\nonumber\\&
-\frac{R^2\dot\theta_0}{4a^2HM_p^2}\delta\theta\partial_i\delta\sigma\partial_i\delta\sigma +\frac{R^4\dot\theta^2_0}{2a^2HM_p^2}\delta\theta\partial_i\psi\partial_i\delta\theta
-\frac{\delta\sigma^2\delta\theta}{4HM_p^2}\left(R^2V''\dot\theta_0+R^2\dot\theta_0^3
\right)-\frac{R^3\dot\theta_0^2}{HM_p^2}\delta\theta\delta\sigma\dot{\delta\theta}\nonumber\\&
-\frac{R^4\dot\theta_0}{4HM_p^2}\delta\theta\dot{\delta\theta}^2-\frac{R^2\dot\theta_0}{4HM_p^2}\delta\theta\dot{\delta\sigma}^2~,
\end{align}
where
\begin{align}
  \partial^2\psi\equiv -a^2\frac{\dot\theta_0^2 R}{2H^2M_p^2}\left\{
     R\partial_t\left(\frac{H\delta\theta}{\dot\theta_0}\right)+2 H\delta\sigma \right\}~.
\end{align}

\subsection{Uniform inflaton gauge}
\label{Appgauge2}

In the uniform inflaton gauge,
\bea
\theta(\bx,t)=\theta_0(t)~, ~~~~~
\sigma(\bx,t)=\sigma_0 + \delta\sigma(\bx,t) ~,
\eea
\bea
h_{ij}(\bx,t) = a^2(t) e^{2\zeta(\bx,t)} \delta_{ij} ~.
\eea

Again solving the constraint equations to the linear order in perturbations, we get
\bea
\alpha_1 = \frac{\dot\zeta}{H} ~, ~~~~~
\tilde N^{(1)}_i =0 ~, ~~~~~
\psi = -\frac{\zeta}{H} + \chi ~,
\eea
where
\bea
\partial^2\chi = a^2 \epsilon (\dot\zeta - \frac{2H}{R}\delta\sigma ) ~.
\label{chidef}
\eea
Plug these into the Lagrangian and expand. The linear order terms are consistent with the Hubble equation and give the equation of motion for $\sigma_0(t)$. For the quadratic and cubic order perturbations,
after integrations by parts, we get
\begin{align}
\CL_2 = & \epsilon a^3 \dot\zeta^2 - \epsilon a (\partial\zeta)^2
\cr
& + \frac{a^3}{2} \dot{\delta\sigma}^2 - \frac{a}{2} \left(\partial_i \delta\sigma \right)^2 - \frac{a^3}{2}\left(V''-\dot\theta_0^2\right) \delta\sigma^2
\cr
& - 2a^3 \frac{R\dot\theta_0^2}{H} \delta\sigma \dot\zeta ~,
\end{align}
\begin{align}
\CL_3 = & -\frac{a^3}{6} V''' \delta\sigma^3
-\frac{a^3}{2} \frac{V''+\dot\theta_0^2}{H} \delta\sigma^2 \dot\zeta
-\frac{3}{2} a^3(V''-\dot\theta_0^2)\delta\sigma^2\zeta
\cr
& -\frac{a}{2} (\partial_i\delta\sigma)^2 (\frac{\dot\zeta}{H}-\zeta)
+a\dot{\delta\sigma} \partial_i \delta\sigma (\frac{\partial_i\zeta}{H}-\partial_i\chi)
+\frac{a^3}{2} \dot{\delta\sigma}^2 (-\frac{\dot\zeta}{H}+3\zeta)
\cr
& -2\epsilon a \frac{1}{R} \delta\sigma(\partial_i\zeta)^2
+ 4\epsilon a^3 \frac{1}{R} \dot{\delta\sigma} \dot\zeta \zeta
+ 2\epsilon a^3 \frac{1}{R} \delta\sigma \dot\zeta^2
\cr
& +(4\epsilon -\epsilon^2) a \frac{H}{R} \delta\sigma \partial_i\zeta
\partial_i \chi
+(-4\epsilon^2+2\epsilon\eta) a^3 \frac{H}{R} \delta\sigma \dot\zeta \zeta
-\epsilon \dot\eta a^3 \frac{H}{R} \delta\sigma \zeta^2
\cr
& +\epsilon^2 a^3 \dot\zeta^2\zeta + \epsilon^2 a \zeta(\partial_i\zeta)^2
-(2\epsilon-\frac{\epsilon^2}{2}) a \dot\zeta \partial_i\zeta \partial_i\chi
\cr
& + \frac{1}{2} \epsilon \dot\eta a^3 \zeta^2 \dot\zeta
+ \frac{\epsilon}{4a} \partial^2\zeta (\partial_i\chi)^2
+ f(\zeta,\delta\sigma) \frac{\delta\CL}{\delta \zeta}\Big|_1 ~,
\end{align}
where
\begin{align}
f(\zeta,\delta\sigma) = & -\frac{\eta}{4} \zeta^2 - \frac{\dot\zeta \zeta}{H}
-\frac{1}{4a^2H^2} [\partial^{-2}\partial_i\partial_j
(\partial_i\zeta\partial_j\zeta)-(\partial_i\zeta)^2]
\cr
& -\frac{1}{2a^2H^2} \partial^{-2}
(\partial_i\partial_j\zeta \partial_i\partial_j\chi-\partial^2\zeta\partial^2\chi) ~,
\\
\frac{\delta\CL}{\delta\zeta}\Big|_1 =
& -2a \left( \frac{d}{dt}\partial^2\chi + H\partial^2\chi -
\epsilon \partial^2\zeta \right) ~,
\end{align}
and $\chi$ is defined in (\ref{chidef}).

\subsection{Estimate all cubic terms}

As we have mentioned in the Introduction, the contribution of the
$\delta\sigma^3$ term to $f_{NL}$ can be estimated as follows.
$V'''/H$ and $\dot\theta_0/H$ are the
dimensionless coupling constants for the cubic isocurvature interaction and the transfer vertex, respectively. Note that $\zeta\sim \sqrt{P_\zeta}$. Then the three-point function is
\begin{equation}
  \langle \zeta^3\rangle \sim \frac{V'''}{H}
  \left(\frac{\dot\theta_0}{H}\right)^3 P_\zeta^{3/2} \sim
  \frac{1}{\sqrt{P_\zeta}}\frac{V'''}{H}\left(
  \frac{\dot\theta_0}{H}\right)^3 P_\zeta^{2} ~.
\end{equation}
According to the definition (\ref{fnl_def}),
\begin{equation}
  f_{NL}\sim \frac{1}{\sqrt{P_\zeta}}\frac{V'''}{H}\left(
  \frac{\dot\theta_0}{H}\right)^3 ~.
\end{equation}

Other terms can be estimated similarly. For example, the $\delta\theta^3$ term in the spatially flat gauge contributes ${\cal O}(\epsilon^2)
P_\zeta^{1/2}$ to $\langle \zeta^3 \rangle$, where ${\cal O}(\epsilon^2)$ collectively denote
second order in slow-roll parameters. So the $\delta\theta^3$ terms
leads to $f_{NL}\sim {\cal O}(\epsilon^2)$, which is negligible. In Table
\ref{tab_fnl_1} and \ref{tab_fnl_2}, we
summarize the orders of magnitude of $f_{NL}$ contributed from each term in the two gauges.

\begin{table}[htp]
\begin{center}
\begin{tabular}{ | c | c | c | c | c | c | c | c | c | c |}
    \hline
     term &$\delta\sigma^3$ & $\delta\theta^3$ &
     $\partial_i\psi\partial^i\delta\theta\dot{\delta\theta}$
     & $\delta\sigma^2\dot{\delta\theta}$ &
     $\delta\theta^2\partial_i\partial^i\psi$ & $\delta\theta^2\delta\sigma$ &
     $\delta\theta^2\dot{\delta\theta}$ & $\delta\sigma\dot{\delta\theta}^2$
     & $\delta\sigma\partial_i\delta\theta\partial^i\delta\theta$\\
    \hline
    $f_{NL}^{\rm term}$ & $\frac{1}{\sqrt{P_\zeta}}\frac{V'''}{H}(\frac{\dot\theta_0}{H})^3 $ & $\epsilon^2$
     & $\epsilon$ & $(\frac{\dot\theta_0}{H})^4$ & $\epsilon^2$ & $\epsilon^2 (\frac{\dot\theta_0}{H})^2$
     & $\epsilon^2$ & $(\frac{\dot\theta_0}{H})^2$ & $(\frac{\dot\theta_0}{H})^2$\\
     \hline
\end{tabular}

\vspace{3mm}

\begin{tabular}{ | c | c | c | c | c | c | c |}
    \hline
     term &$\delta\sigma\partial_i\delta\theta\partial^i\psi$ &
     $\partial_i\psi\partial^i\delta\sigma \dot{\delta\sigma}$ &
     $\delta\theta\left[(\partial_i\psi\partial_j\psi)^2-(\partial_i\partial^i\psi)^2\right]$
     & $\delta\theta\partial_i\delta\theta\partial^i\delta\theta$ &
     $\delta\theta\partial_i\delta\sigma\partial^i\delta\sigma$ &
     $\delta\theta\partial_i\psi\partial^i\delta\theta$ \\
    \hline
    $f_{NL}^{\rm term}$ & $\epsilon (\frac{\dot\theta_0}{H})^2$ & $\epsilon (\frac{\dot\theta_0}{H})^2$
     & $\epsilon^2$ & $\epsilon$ & $\epsilon (\frac{\dot\theta_0}{H})^2$ & $\epsilon^2$
     \\
     \hline
\end{tabular}

\vspace{3mm}

\begin{tabular}{ | c | c | c | c | c |  }
    \hline
     term &$\delta\sigma^2\delta\theta$ &
     $\delta\theta\delta\sigma\dot{\delta\theta}$ &
     $\delta\theta\dot{\delta\theta}^2$ &
     $\delta\theta\dot{\delta\sigma}^2$\\
    \hline $f_{NL}^{\rm term}$&
    $\epsilon (\frac{\dot\theta_0}{H})^2 \left( \frac{m^2+\dot\theta_0^2}{H^2} \right) $ &
    $\epsilon (\frac{\dot\theta_0}{H})^2$ &
    $\epsilon$ &
    $\epsilon (\frac{\dot\theta_0}{H})^2$ \\
     \hline
\end{tabular}

\end{center}
\caption{The order of magnitude estimation for the contributions to $f_{NL}$ from each term in the Lagrangian. The first row lists terms in the spatially flat gauge. The second
  row lists the order of magnitude for $f_{NL}$ contributed from the
  corresponding term. In this table, $\epsilon$ collectively denotes all slow
  roll parameters.}
\label{tab_fnl_1}
\end{table}

\begin{table}[htp]
\begin{center}
\begin{tabular}{ | c | c | c | c | c | c |}
    \hline
     term &
     $\delta\sigma^3$ &
     $\delta\sigma^2\dot\zeta$ &
     $\delta\sigma^2 \zeta$ &
     $(\partial_i\delta\sigma)^2(\frac{\dot\zeta}{H}-\zeta)$ & $\dot{\delta\sigma} \partial_i\delta\sigma (\frac{\partial_i\zeta}{H}-\partial_i\chi)$ \\
    \hline
     $f_{NL}^{\rm term}$ & $\frac{1}{\sqrt{P_\zeta}}\frac{V'''}{H}(\frac{\dot\theta_0}{H})^3 $ & $\left(\frac{\dot\theta}{H}\right)^2 \frac{V''+\dot\theta_0^2}{H^2}$ &
     $\left(\frac{\dot\theta}{H}\right)^2 \frac{V''-\dot\theta_0^2}{H^2}$ & $\left(\frac{\dot\theta_0}{H}\right)^2$ & $\left(\frac{\dot\theta_0}{H}\right)^2 + \epsilon\left(\frac{\dot\theta_0}{H}\right)^4$\\
     \hline
\end{tabular}

\vspace{3mm}

\begin{tabular}{ | c | c | c | c | c | c | c |}
    \hline
     term &
     $\dot{\delta\sigma}^2 \left(-\frac{\dot\zeta}{H}+3\zeta \right)$ &
     $\delta\sigma(\partial_i\zeta)^2$ &
     $\dot{\delta\sigma}\dot\zeta\zeta$ &
     $\delta\sigma\dot\zeta^2$ &
     $\delta\sigma\partial_i\zeta\partial_i\chi$ &
     $\delta\sigma\dot\zeta\zeta$
     \\
    \hline
    $f_{NL}^{\rm term}$ &
    $\left(\frac{\dot\theta_0}{H}\right)^2$ &
    $\left(\frac{\dot\theta_0}{H}\right)^2$ &
    $\left(\frac{\dot\theta_0}{H}\right)^2$ &
    $\left(\frac{\dot\theta_0}{H}\right)^2$ &
    $\epsilon\left(\frac{\dot\theta_0}{H}\right)^2 +\epsilon\left(\frac{\dot\theta_0}{H}\right)^4$ &
    $\epsilon \left(\frac{\dot\theta_0}{H}\right)^2$
     \\
     \hline
\end{tabular}

\vspace{3mm}

\begin{tabular}{ | c | c | c | c | c | c | c | c | }
    \hline
     term &
     $\delta\sigma\zeta^2$ &
     $\dot\zeta^2 \zeta$ &
     $\zeta(\partial_i\zeta)^2$ &
     $\dot\zeta \partial_i\zeta \partial_i\chi$ &
     $\zeta^2\dot\zeta$ &
     $\partial^2\zeta(\partial_i\chi)^2$ &
     $f\frac{\delta\CL}{\delta\zeta}|_1$
     \\
    \hline $f_{NL}^{\rm term}$&
     $\frac{\dot\eta}{H} \left(\frac{\dot\theta_0}{H} \right)^2$ &
     $\epsilon$ &
     $\epsilon$ &
     $\epsilon (1+\frac{\dot\theta_0^2}{H^2})$ &
     $\frac{\dot\eta}{H}$ &
     $\epsilon^2 (1+\frac{\dot\theta_0^2}{H^2})^2$ &
     $\eta + \eta \frac{\dot\theta_0^2}{H^2}$
     \\
     \hline
\end{tabular}

\end{center}
\caption{The same table for the uniform inflaton gauge. In the last entry, the 1st term is due to the field redefinition in 3pt, the 2nd is due to an extra cubic term from the field redefinition in the transfer vertex.}
\label{tab_fnl_2}
\end{table}

As we can see the only term that can make $f_{NL}\gg 1$ is the $\delta\sigma^3$ term.
The rest of terms are corrections suppressed
by either the slow-roll parameters or $P_\zeta^{1/2}\sim
H/\sqrt{\epsilon}\sim
H^2/(R\dot\theta_0)$.

New cubic terms will arise after we convert the Lagrangian density to the interaction Hamiltonian density. Here we show that they do not change the order of magnitude estimate performed above.
The Lagrangian density in our case is of the form
\bea
\CL_2 &=& f_0 \dot \alpha^2 + \tilde f_0 \dot \beta^2 + f_2 ~,
\cr
\delta \CL_2 &=& g_1 \dot \alpha ~,
\cr
\CL_3 &=& h_1 \dot\alpha^2 + h_2 \dot\alpha + j_1 \dot\beta^2 + j_2 \dot\beta + h_3 ~,
\label{Lterms}
\eea
where $\alpha$ and $\beta$ represent $\zeta$ and $\delta\sigma$ respectively, and the subscripts in various functions denote the orders of $\alpha$ and $\beta$ in these functions. Following the same description in Sec.~\ref{Sec:model}, we get the following interaction Hamiltonian density,
\bea
\CH_2 &=& f_0 \dot\alpha_I^2 + \tilde f_0 \dot\beta_I^2 - f_2 + \frac{g_1^2}{2f_0} ~,
\cr
\delta\CH_2 &=& -g_1 \dot\alpha_I ~,
\cr
\CH_3 &=& -h_1 \dot\alpha_I^2 - (h_2-\frac{g_1h_1}{f_0}) \dot\alpha_I - j_1 \dot\beta_I^2 - j_2 \dot\beta_I - h_3 + \frac{g_1h_2}{2f_0} ~.
\label{Hterms}
\eea
Comparing to (\ref{Lterms}), the new terms appeared in $\CH_2$ is
\bea
\frac{g_1^2}{2f_0} \sim \left(\frac{\delta\CL_2}{\CL_2}\right)^2 f_0\dot\alpha_I^2 ~,
\eea
and in $\CH_3$ are
\bea
\frac{g_1h_1}{f_0}\dot\alpha_I &\sim& \left( \frac{\delta\CL_2}{\CL_2} \right) h_1 \dot\alpha_I^2 ~,
\\
\frac{g_1 h_2}{2f_0} &\sim& \left( \frac{\delta\CL_2}{\CL_2} \right) h_2 \dot\alpha_I ~.
\eea
As long as $\delta\CL_2/\CL_2 \lesssim 1$, these new terms will not be more important than the original ones.

\section{Details of terms in various forms}
\label{App:allterms}
In this appendix we give the details of the terms in the factorized, commutator and mixed forms.

\subsection{All 10 terms in the factorized form}
\label{10terms_F}
\setcounter{equation}{0}

The three-point function is the sum of the following ten terms,
\bea
\langle \zeta^3 \rangle = -\left(\frac{H}{\dot\theta_0}\right)^3 \sum_{i=1}^{10} (i) ~.
\eea
For simplicity, we only indicate the conformal time variables, such as
$\tau_1$ or
$\tilde\tau_1$, once in each integrand. It applies to each factor
before it.

\bea
(1) &=& -12 u_{p_1}^* u_{p_2} u_{p_3}(0)
\cr &\times&
{\rm Re} \left[ \int_{-\infty}^0 d\ttau_1~ a^3 c_2 v_{p_1}^*
u'_{p_1}(\ttau_1)
\int_{-\infty}^{\ttau_1} d\ttau_2~ a^4 c_3 v_{p_1}
v_{p_2} v_{p_3}(\ttau_2) \right.
\cr &\times& \left.
\int_{-\infty}^0 d\tau_1~ a^3 c_2 v_{p_2}^*
u_{p_2}^{\prime *} (\tau_1)
\int_{-\infty}^{\tau_1} d\tau_2~ a^3 c_2 v_{p_3}^*
u_{p_3}^{\prime *} (\tau_2) \right]
\cr &\times&
(2\pi)^3 \delta^3(\sum_i \bp_i) + {\rm 5~ perm.}
\\ \nonumber \\
(2) &=& -12 u_{p_1}^* u_{p_2} u_{p_3}(0)
\cr &\times&
{\rm Re} \left[ \int_{-\infty}^0 d\ttau_1~ a^4 c_3 v_{p_1}^*
v_{p_2} v_{p_3}(\ttau_1)
\int_{-\infty}^{\ttau_1} d\ttau_2~ a^3 c_2 v_{p_1}
u'_{p_1}(\ttau_2) \right.
\cr &\times&
\left. \int_{-\infty}^0 d\tau_1~ a^3 c_2 v_{p_2}^*
u_{p_2}^{\prime *} (\tau_1)
\int_{-\infty}^{\tau_1} d\tau_2~ a^3 c_2 v_{p_3}^*
u_{p_3}^{\prime *} (\tau_2) \right]
\cr &\times&
(2\pi)^3 \delta^3(\sum_i \bp_i) + {\rm 5~ perm.}
\\ \nonumber \\
(3) &=& 12 u_{p_1} u_{p_2} u_{p_3}(0)
\cr &\times& {\rm Re} \left[
\int_{-\infty}^0 d\ttau_1~ a^4 c_3 v_{p_1}
v_{p_2} v_{p_3}(\ttau_1) \right.
\cr &\times& \left.
\int_{-\infty}^0 d\tau_1~ a^3 c_2 v_{p_1}^*
u_{p_1}^{\prime *} (\tau_1)
\int_{-\infty}^{\tau_1} d\tau_2~ a^3 c_2 v_{p_2}^*
u_{p_2}^{\prime *} (\tau_2)
\int_{-\infty}^{\tau_2} d\tau_3~ a^3 c_2 v_{p_3}^*
u_{p_3}^{\prime *} (\tau_3)
\right]
\cr &\times&
(2\pi)^3 \delta^3(\sum_i \bp_i) + {\rm 5~ perm.}
\\ \nonumber \\
(4) &=& 12 u_{p_1}^* u_{p_2} u_{p_3}(0)
\cr &\times& {\rm Re} \left[
\int_{-\infty}^0 d\ttau_1~ a^3 c_2 v_{p_1}
u_{p_1}^{\prime} (\ttau_1)
\right.
\cr &\times& \left.
\int_{-\infty}^0 d\tau_1~ a^4 c_3 v_{p_1}^*
v_{p_2} v_{p_3}(\tau_1)
\int_{-\infty}^{\tau_1} d\tau_2~ a^3 c_2 v_{p_2}^*
u_{p_2}^{\prime *} (\tau_2)
\int_{-\infty}^{\tau_2} d\tau_3~ a^3 c_2 v_{p_3}^*
u_{p_3}^{\prime *} (\tau_3)
\right]
\cr &\times&
(2\pi)^3 \delta^3(\sum_i \bp_i) + {\rm 5~ perm.}
\\ \nonumber \\
(5) &=& 12 u_{p_1}^* u_{p_2} u_{p_3}(0)
\cr &\times& {\rm Re} \left[
\int_{-\infty}^0 d\ttau_1~ a^3 c_2 v_{p_1}
u_{p_1}^{\prime} (\ttau_1)
\right.
\cr &\times& \left.
\int_{-\infty}^0 d\tau_1~ a^3 c_2 v_{p_2}
u_{p_2}^{\prime *} (\tau_1)
\int_{-\infty}^{\tau_1} d\tau_2~ a^4 c_3 v_{p_1}^*
v_{p_2}^* v_{p_3}(\tau_2)
\int_{-\infty}^{\tau_2} d\tau_3~ a^3 c_2 v_{p_3}^*
u_{p_3}^{\prime *} (\tau_3)
\right]
\cr &\times&
(2\pi)^3 \delta^3(\sum_i \bp_i) + {\rm 5~ perm.}
\\ \nonumber \\
(6) &=& 12 u_{p_1}^* u_{p_2} u_{p_3}(0)
\cr &\times& {\rm Re} \left[
\int_{-\infty}^0 d\ttau_1~ a^3 c_2 v_{p_1}
u_{p_1}^{\prime} (\ttau_1)
\right.
\cr &\times& \left.
\int_{-\infty}^0 d\tau_1~ a^3 c_2 v_{p_2}
u_{p_2}^{\prime *} (\tau_1)
\int_{-\infty}^{\tau_1} d\tau_2~ a^3 c_2 v_{p_3}
u_{p_3}^{\prime *} (\tau_2)
\int_{-\infty}^{\tau_2} d\tau_3~ a^4 c_3 v_{p_1}^*
v_{p_2}^* v_{p_3}^*(\tau_3)
\right]
\cr &\times&
(2\pi)^3 \delta^3(\sum_i \bp_i) + {\rm 5~ perm.}
\\ \nonumber \\
(7) &=& -12 u_{p_1} u_{p_2} u_{p_3}(0)
\cr &\times& {\rm Re} \left[
\int_{-\infty}^0 d\tau_1~ a^4 c_3 v_{p_1}
v_{p_2} v_{p_3}(\tau_1)
\int_{-\infty}^{\tau_1} d\tau_2~ a^3 c_2 v_{p_1}^*
u_{p_1}^{\prime *} (\tau_2)
\right.
\cr && \left.
\int_{-\infty}^{\tau_2} d\tau_3~ a^3 c_2 v_{p_2}^*
u_{p_2}^{\prime *} (\tau_3)
\int_{-\infty}^{\tau_3} d\tau_4~ a^3 c_2 v_{p_3}^*
u_{p_3}^{\prime *} (\tau_4)
\right]
\cr &\times&
(2\pi)^3 \delta^3(\sum_i \bp_i) + {\rm 5~ perm.}
\\ \nonumber \\
(8) &=& -12 u_{p_1} u_{p_2} u_{p_3}(0)
\cr &\times& {\rm Re} \left[
\int_{-\infty}^0 d\tau_1~ a^3 c_2 v_{p_1}
u_{p_1}^{\prime *} (\tau_1)
\int_{-\infty}^{\tau_1} d\tau_2~ a^4 c_3 v_{p_1}^*
v_{p_2} v_{p_3}(\tau_2)
\right.
\cr && \left.
\int_{-\infty}^{\tau_2} d\tau_3~ a^3 c_2 v_{p_2}^*
u_{p_2}^{\prime *} (\tau_3)
\int_{-\infty}^{\tau_3} d\tau_4~ a^3 c_2 v_{p_3}^*
u_{p_3}^{\prime *} (\tau_4)
\right]
\cr &\times&
(2\pi)^3 \delta^3(\sum_i \bp_i) + {\rm 5~ perm.}
\\ \nonumber \\
(9) &=& -12 u_{p_1} u_{p_2} u_{p_3}(0)
\cr &\times& {\rm Re} \left[
\int_{-\infty}^0 d\tau_1~ a^3 c_2 v_{p_1}
u_{p_1}^{\prime *} (\tau_1)
\int_{-\infty}^{\tau_1} d\tau_2~ a^3 c_2 v_{p_2}
u_{p_2}^{\prime *} (\tau_2)
\right.
\cr && \left.
\int_{-\infty}^{\tau_2} d\tau_3~ a^4 c_3 v_{p_1}^*
v_{p_2}^* v_{p_3}(\tau_3)
\int_{-\infty}^{\tau_3} d\tau_4~ a^3 c_2 v_{p_3}^*
u_{p_3}^{\prime *} (\tau_4)
\right]
\cr &\times&
(2\pi)^3 \delta^3(\sum_i \bp_i) + {\rm 5~ perm.}
\\ \nonumber \\
(10) &=& -12 u_{p_1} u_{p_2} u_{p_3}(0)
\cr &\times& {\rm Re} \left[
\int_{-\infty}^0 d\tau_1~ a^3 c_2 v_{p_1}
u_{p_1}^{\prime *} (\tau_1)
\int_{-\infty}^{\tau_1} d\tau_2~ a^3 c_2 v_{p_2}
u_{p_2}^{\prime *} (\tau_2)
\right.
\cr && \left.
\int_{-\infty}^{\tau_2} d\tau_3~ a^3 c_2 v_{p_3}
u_{p_3}^{\prime *} (\tau_3)
\int_{-\infty}^{\tau_3} d\tau_4~ a^4 c_3 v_{p_1}^*
v_{p_2}^* v_{p_3}^*(\tau_4)
\right]
\cr &\times&
(2\pi)^3 \delta^3(\sum_i \bp_i) + {\rm 5~ perm.}
\eea

\subsection{All 3 terms in the commutator form}
\label{3terms_C}

For convenience, we also list all three terms in the commutator form here.
\bea
\langle \zeta^3 \rangle &=& -12 \left( \frac{H}{\dot\theta_0} \right)^3 u_{p_1} u_{p_2} u_{p_3}(0)
\cr
&& {\rm Re} \left[ \int_{-\infty}^0 d\tau_1 \int_{-\infty}^{\tau_1}
  d\tau_2 \int_{-\infty}^{\tau_2} d\tau_3 \int_{-\infty}^{\tau_3}
  d\tau_4~
\prod_{i=1}^4 \left( a^3(\tau_i) c_2(\tau_i) \right) \right.
\cr
&&\left. \left( a(\tau_2) \frac{c_3(\tau_2)}{c_2(\tau_2)} A
+ a(\tau_3) \frac{c_3(\tau_3)}{c_2(\tau_3)} B
+ a(\tau_4) \frac{c_3(\tau_4)}{c_2(\tau_4)} C \right) \right]
\nonumber \\
&\times& (2\pi)^3 \delta^3(\sum_i \bp_i)
+ {\rm 5~perm.} ~,
\eea
where
\bea
A &=& \left(u_{p_1}^{\prime}(\tau_1) - c.c. \right)
\left( v_{p_1}(\tau_1) v_{p_1}^*(\tau_2) - c.c. \right)
\left( v_{p_3}(\tau_2) v_{p_3}^*(\tau_4) u_{p_3}^{\prime *}(\tau_4) -
c.c. \right)
\cr
&& v_{p_2}(\tau_2) v_{p_2}^*(\tau_3) u_{p_2}^{\prime *}(\tau_3) ~,
\label{Aterm} \\
B &=& \left(u_{p_1}^{\prime}(\tau_1) - c.c. \right)
\left( u_{p_2}^\prime (\tau_2) - c.c. \right)
\left( v_{p_1}^*(\tau_1) v_{p_2}^*(\tau_2) v_{p_1}(\tau_3)
v_{p_2}(\tau_3) - c.c. \right)
\cr
&& v_{p_3}(\tau_3) v_{p_3}^*(\tau_4) u_{p_3}^{\prime *}(\tau_4) ~,
\label{Bterm} \\
C &=& -\left( u_{p_1}^\prime(\tau_1) - c.c. \right)
\left( u_{p_2}^\prime(\tau_2) - c.c. \right)
\left( u_{p_3}^\prime(\tau_3) - c.c. \right)
\cr
&& v_{p_1}^*(\tau_1) v_{p_2}^*(\tau_2) v_{p_3}^*(\tau_3)
v_{p_1}(\tau_4) v_{p_2}(\tau_4) v_{p_3}(\tau_4) ~.
\label{Cterm}
\eea

\subsection{All terms in the mixed form}
\label{App:mixedform}

We introduce a cutoff $\tau_c$ to separate the IR and UV region. A
quartic integral becomes
\bea
&&
\int_{-\infty}^0 d\tau_1 \int_{-\infty}^{\tau_1} d\tau_2
\int_{-\infty}^{\tau_2} d\tau_3 \int_{-\infty}^{\tau_3} d\tau_4
~f(\tau_1,\tau_2,\tau_3,\tau_4)
\cr
&=& \int_{\tau_c}^0 d\tau_1 \int_{\tau_c}^{\tau_1} d\tau_2
\int_{\tau_c}^{\tau_2} d\tau_3 \int_{\tau_c}^{\tau_3} d\tau_4
~f(\tau_1,\tau_2,\tau_3,\tau_4)
\cr
&+& \int_{\tau_c}^0 d\tau_1 \int_{\tau_c}^{\tau_1} d\tau_2
\int_{\tau_c}^{\tau_2} d\tau_3
\left[ \int_{-\infty}^{\tau_c} d\tau_4 \right]
~f(\tau_1,\tau_2,\tau_3,\tau_4)
\cr
&+& \int_{\tau_c}^0 d\tau_1 \int_{\tau_c}^{\tau_1} d\tau_2
\left[
\int_{-\infty}^{\tau_c} d\tau_3 \int_{-\infty}^{\tau_3} d\tau_4
\right]
~f(\tau_1,\tau_2,\tau_3,\tau_4)
\cr
&+& \int_{\tau_c}^0 d\tau_1
\left[
\int_{-\infty}^{\tau_c} d\tau_2
\int_{-\infty}^{\tau_2} d\tau_3 \int_{-\infty}^{\tau_3} d\tau_4
\right]
~f(\tau_1,\tau_2,\tau_3,\tau_4)
\cr
&+& \int_{-\infty}^{\tau_c} d\tau_1 \int_{-\infty}^{\tau_1} d\tau_2
\int_{-\infty}^{\tau_2} d\tau_3 \int_{-\infty}^{\tau_3} d\tau_4
~f(\tau_1,\tau_2,\tau_3,\tau_4) ~.
\label{mixed}
\eea
We label the above five terms as $(a)$, $(b)$, $(c)$, $(d)$ and $(e)$,
respectively.

We now write the three-point function in a mixed form. In the IR
region, we write them in terms of the commutator form, so that the
cancellation of the leading terms, especially the cancellation of the
IR spurious
divergences, is manifest; in the UV region, we write them in terms of
the factorized form, so that in each factor
the fast UV convergence can be achieved by a Wick rotation and there is no spurious divergence
for any momentum configuration.

We start with the three terms in the commutator form
(\ref{Aterm})-(\ref{Cterm}). We expand the expressions at a particular
layer when the factorized form is need. To
factorize the integral, combinations of different terms from all three
terms, $A$, $B$ and $C$, including the complex conjugates and the
momentum permutations, are needed. It is always possible to make the
required part of the integral
factorize to a level indicated by the ten terms in Appendix
\ref{10terms_F}, which cannot be further factorized.
Hence the UV behavior becomes as good as those
10 terms. At the same time, the IR convergence is as good as the three terms in Appendix \ref{3terms_C}.

For the $(a)$-term in (\ref{mixed}), we use the integrand in the
commutator form. So we have 3 terms, which we label as
$(a1)$, $(a2)$ and $(a3)$.

For the $(b)$-term, the separation is simple, we get two terms from
(\ref{Aterm}) and one each from (\ref{Bterm}) and (\ref{Cterm}). We
label as $(b1)$, $(b2)$, $(b3)$ and $(b4)$.

For the $(c)$ term, we have
\bea
(c1)&=& 12 u_{p_1} u_{p_2} u_{p_3}(0)
\cr
&\times& {\rm Re} \left[
\int_{\tau_c}^0 d\tau_1 \int_{\tau_c}^{\tau_1}
  d\tau_2~
  a^3 c_2(\tau_1) ~ a^4 c_3(\tau_2)
\left(u_{p_1}^\prime(\tau_1) - c.c.\right)
\left(v_{p_1}(\tau_1) v_{p_1}^*(\tau_2) - c.c.\right)
v_{p_2} v_{p_3}(\tau_2) \right.
\cr
&& \left.
\quad \times
\int_{-\infty}^{\tau_c} d\tau_3 \int_{-\infty}^{\tau_3} d\tau_4~
a^3 c_2 v_{p_2}^* u_{p_2}^{\prime *}(\tau_3) ~
a^3 c_2 v_{p_3}^* u_{p_3}^{\prime *}(\tau_4) \right]
\cr
&\times& (2\pi)^3 \delta^3(\sum_i \bp_i) + {\rm 5~perm.}
~,
\\
(c2)
&=& - 6 u_{p_1} u_{p_2} u_{p_3}(0)
\cr
&\times&
\int_{\tau_c}^0 d\tau_1 \int_{\tau_c}^{\tau_1}
  d\tau_2~
  a^3 c_2(\tau_1) ~ a^4 c_3(\tau_2)
\left( u_{p_1}^\prime(\tau_1) - c.c. \right)
\left( v_{p_1}(\tau_1) v_{p_1}^*(\tau_2) - c.c. \right)
 v_{p_2} v_{p_3}^*(\tau_2)
\cr
&\times& \int_{-\infty}^{\tau_c} d\tau_3~ a^3 c_2
v_{p_2}^* u_{p_2}^{\prime *}(\tau_3)
\int_{-\infty}^{\tau_c} d\tau_4~ a^3 c_2
v_{p_3} u_{p_3}^{\prime}(\tau_4)
\cr
&\times& (2\pi)^3 \delta^3(\sum_i \bp_i) + {\rm 5~perm.}
~,
\eea
which comes from (\ref{Aterm}) and we have used the
complex conjugation to factorize $(c2)$;
\bea
(c3,4,5)&=& 12 u_{p_1} u_{p_2} u_{p_3}(0)
\cr
&\times&
{\rm Re}
\left[ \int_{\tau_c}^0 d\tau_1 \int_{\tau_c}^{\tau_1}
  d\tau_2~
a^3 c_2
\left(u_{p_1}^\prime - c.c.\right)
v_{p_1}^*(\tau_1) ~
a^3 c_2
\left(u_{p_2}^\prime - c.c.\right)
v_{p_2}^*(\tau_2) \right.
\cr
&& \quad\times
\left(
\int_{-\infty}^{\tau_c} d\tau_3~ a^4 c_3
v_{p_1} v_{p_2} v_{p_3}(\tau_3)
\int_{-\infty}^{\tau_c} d\tau_4~ a^3 c_2
v_{p_3}^* u_{p_3}^{\prime *}(\tau_4)
\right.
\cr
&& \quad ~
- \int_{-\infty}^{\tau_c} d\tau_3 \int_{-\infty}^{\tau_3} d\tau_4 ~
a^4 c_3 v_{p_1} v_{p_2} v_{p_3}^*(\tau_3) ~
a^3 c_2 v_{p_3} u_{p_3}^{\prime}(\tau_4)
\cr
&& \quad ~
\left. \left.
- \int_{-\infty}^{\tau_c} d\tau_3 \int_{-\infty}^{\tau_3} d\tau_4 ~
a^3 c_2 v_{p_3}^* u_{p_3}^{\prime}(\tau_3) ~
a^4 c_3 v_{p_1} v_{p_2} v_{p_3}(\tau_4)
\right) \right]
\cr
&\times& (2\pi)^3 \delta^3(\sum_i \bp_i) + {\rm 5~perm.} ~,
\eea
where the contributions from (\ref{Bterm}) and (\ref{Cterm}) are
combined to factorize the 1st term above, and the 2nd and 3rd terms
(which cannot be further factorized)
come from (\ref{Bterm}) and (\ref{Cterm}), respectively.

For the $(d)$-term, we have
\bea
(d1-6)&=& 12 u_{p_1} u_{p_2} u_{p_3}(0)
\cr
&\times&
{\rm Re} \left[ \int_{\tau_c}^0 d\tau_1 ~ a^3 c_2
\left( u_{p_1}^\prime - c.c.\right) v_{p_1}(\tau_1)
\right.
\cr
&& \quad \times
\left(
- \int_{-\infty}^{\tau_c} d\tau_2~ a^3 c_2 v_{p_2}
u_{p_2}^{\prime}(\tau_2)
\int_{-\infty}^{\tau_c} d\tau_3 \int_{-\infty}^{\tau_3} d\tau_4~
a^4 c_3 v_{p_1}^* v_{p_2}^* v_{p_3}(\tau_3) ~
a^3 c_2 v_{p_3}^* u_{p_3}^{\prime *}(\tau_4)
\right.
\cr
&& \quad ~
- \int_{-\infty}^{\tau_c} d\tau_2 ~ a^3 c_2
v_{p_2} u_{p_2}^{\prime}(\tau_2)
\int_{-\infty}^{\tau_c} d\tau_3 \int_{-\infty}^{\tau_3} d\tau_4 ~
a^3 c_2 v_{p_3} u_{p_3}^{\prime *}(\tau_3) ~
a^4 c_3 v_{p_1}^* v_{p_2}^* v_{p_3}^*(\tau_4)
\cr
&& \quad ~
+\int_{-\infty}^{\tau_c} d\tau_2 ~ a^4 c_3
v_{p_1}^* v_{p_2}^* v_{p_3}^*(\tau_2)
\int_{-\infty}^{\tau_c} d\tau_3 \int_{-\infty}^{\tau_3} d\tau_4 ~
a^3 c_2 v_{p_2} u_{p_2}^{\prime}(\tau_3) ~
a^3 c_2 v_{p_3} u_{p_3}^{\prime}(\tau_4)
\cr
&& \quad ~
+\int_{-\infty}^{\tau_c} d\tau_2 \int_{-\infty}^{\tau_2} d\tau_3
\int_{-\infty}^{\tau_3} d\tau_4 ~
a^4 c_3 v_{p_1}^* v_{p_2} v_{p_3}(\tau_2) ~
a^3 c_2 v_{p_2}^* u_{p_2}^{\prime *}(\tau_3) ~
a^3 c_2 v_{p_3}^* u_{p_3}^{\prime *}(\tau_4)
\cr
&& \quad ~
- \int_{-\infty}^{\tau_c} d\tau_2 \int_{-\infty}^{\tau_2} d\tau_3
\int_{-\infty}^{\tau_3} d\tau_4 ~
a^3 c_3 v_{p_2}^* u_{p_2}^{\prime}(\tau_2) ~
a^4 c_2 v_{p_1} v_{p_2} v_{p_3}^*(\tau_3) ~
a^3 c_2 v_{p_3} u_{p_3}^{\prime}(\tau_4)
\cr
&& \quad ~
\left. \left.
+ \int_{-\infty}^{\tau_c} d\tau_2 \int_{-\infty}^{\tau_2} d\tau_3
\int_{-\infty}^{\tau_3} d\tau_4 ~
a^3 c_2 v_{p_2} u_{p_2}^{\prime *}(\tau_2) ~
a^3 c_2 v_{p_3} u_{p_3}^{\prime *}(\tau_3) ~
a^4 c_3 v_{p_1}^* v_{p_2}^* v_{p_3}^*(\tau_4)
\right) \right]
\cr
&\times& (2\pi)^3 \delta^3(\sum_i \bp_i) + {\rm 5~perm.} ~,
\eea
where the contributions from (\ref{Aterm}), (\ref{Bterm}) and
(\ref{Cterm}) are
combined to factorize the first three terms above, and the last three
terms (which cannot be further factorized)
come from (\ref{Aterm}), (\ref{Bterm}) and (\ref{Cterm}),
respectively.

For the $(e)$-term, we use the integrand in the factorized from in
Appendix \ref{10terms_F}. So we
have 10 terms, which we label as $(e1)$, $\dots$, $(e10)$.

\section{Numerical integration and Wick rotation}
\label{App:wick}
\setcounter{equation}{0}

In the $\tau \to -\infty$ end of various
integrations that we encounter in
the in-in formalism, we slightly rotate
$\tau$ to the imaginary plane, $\tau \to -\infty(1\pm i\epsilon)$, to
achieve the convergence. If we have to compute the integration
numerically,
this is not easy to implement \cite{Chen:2006xjb}. There are several
ways to improve this situation.

One way is to use the integration by part to increase the convergence
speed of the integrand at the $\tau\to -\infty$ end
\cite{Chen:2008wn}.
Additional terms appeared in the integration
by part are the integrand and its derivatives evaluated at a boundary.
One may also choose a cutoff at some point $\tau_0$ and replace the
integrand by an analytical approximation for $\tau<\tau_0$.

In this paper, we use another way that is more efficient in
terms of numerical integration.
For an integral
\bea
\int_{-\infty}^0 d\tau_1 \int_{-\infty}^{\tau_1} d\tau_2 \cdots
\int_{-\infty}^{\tau_{n-1}} d\tau_n f(\tau_1,\tau_2, \cdots, \tau_n)
~,
\label{int_orig}
\eea
we analytically
continue $\tau_i \to i x_i$, so that the integral is equal to\footnote{There are no singularities in the complex plane in our case. But it is an interesting question how this procedure will be modified if there are.}
\bea
i^n
\int_{\pm\infty}^0 dx_1 \int_{\pm\infty}^{x_1} dx_2
\cdots
\int_{\pm\infty}^{x_{n-1}} dx_n f(ix_1,ix_2,\cdots,ix_n) ~.
\label{int_Wick}
\eea
After this Wick rotation, the original oscillating behavior of
the integrands at the
lower end becomes the exponential decay. Note that
the lower limit for $x_i$ is chosen to be either $+\infty$ or
$-\infty$ to make sure that the corresponding
integrand is decaying as $x_i\to
-\infty$, and gives zero when evaluating the indefinite integration at
$x_i= -\infty$.

The reason that the above two expressions are equivalent is the
following. We first look at the inner-most integral in
(\ref{int_orig}). Denoting the indefinite integration as
$F_{n-1}(\tau_1,\cdots,\tau_{n-1},\tau_n)$, we get
$F_{n-1}(\tau_1,\cdots,\tau_{n-1},\tau_{n-1}) -
F_{n-1}(\tau_1,\cdots,\tau_{n-1},-\infty)$.
For the type of integration that we have in the
in-in formalism, as we take
$\tau_n\to -\infty$, the lower end of $F_{n-1}$ is quickly oscillating
around a constant which we shift to zero in the definition of $F_{n-1}$.
The slight rotation $\tau_n \to -\infty(1\pm
i\epsilon)$ is chosen to make this oscillating piece go away, and the
integration becomes
$F_{n-1}(\tau_1,\cdots,\tau_{n-1},\tau_{n-1})$. After
the Wick rotation, such an oscillating behavior becomes an exponential
behavior and is zero when we evaluate it at either $\infty$ or
$-\infty$.
For the method to work, it is important that none of the other terms
that originally goes to
zero at infinity becomes non-zero at infinity after Wick rotation.
The upper limit
$F_{n-1}(ix_1,\cdots,ix_{n-1},ix_{n-1})$ is the analytical
continuation of
$F_{n-1}(\tau_1,\cdots,\tau_{n-1},\tau_{n-1})$. Performing the rest of
the integration in the same fashion, at the outer-most integral,
because the upper limit is $0$, we obtain $F_1(0,\cdots,0)$ which is
the same before and after the analytical continuation.

For example, for $a_1>a_2>0$,
\bea
\int_{-\infty}^0 d\tau_1 e^{-i a_1 \tau_1} \int_{-\infty}^{\tau_1}
d\tau_2 e^{i a_2\tau_2} = \frac{1}{a_2(a_1-a_2)} ~.
\label{NoWick}
\eea
After the analytical continuation, we have
\bea
i^2 \int_{-\infty}^0 dx_1 e^{a_1 x_1} \int_{+\infty}^{x_1} dx_2
e^{-a_2 x_2} = \frac{1}{a_2(a_1-a_2)} ~.
\label{WithWick}
\eea
The integration (\ref{WithWick}) is much easier to do numerically than (\ref{NoWick}).

To apply the above method to quasi-single field inflation and more general cases, some comments and discussions are in order here.

$\bullet$ The Wick rotation is not only useful to deal with the UV behavior of the integration, but also IR in some cases. If in the IR regime ($\tau\rightarrow 0$), there are both oscillation and exponential suppression in the integrand, caused by the mode function (\ref{mode_v2}), we can also perform a Wick rotation to find the total suppression factor. Instead of rotating $\pi/2$ in the complex plane, we can rotate another angle so that the integration after the Wick rotation only contains exponential suppression factor. This is useful when dealing with the $m/H>3/2$ case. In this case, in the IR regime of the correlation functions, one needs to integrate
\begin{equation}
  \int a d\tau~ (-\tau)^{-i a \tilde\nu+b} \cdots \propto \int dt~ e^{(i a \tilde\nu-b)Ht} \cdots~,
\end{equation}
where $a$ and $b$ are some real numbers and we have approximated $\tau\propto e^{-Ht}$.
We rotate
\begin{equation}\label{2ptsuppression}
  t\rightarrow \frac{(ia\tilde\nu+b)}{\left|ia\tilde\nu-b\right|} t ~.
\end{equation}
Then the exponential factor in the integration becomes a suppression factor $\sim e^{-a mt}$ as $m\gg H$. We also note that, when the modes are sub-horizon, they start to oscillate and their contributions are cancelled out. This corresponds to the scale $t\sim 1/H$. Combining both effects, we get factors of Boltzmann suppression $e^{-m/H}$ for the correlation functions involving very massive modes.

$\bullet$ If the upper limit of the outmost integral is non-zero, say $\tau_c$, we make a shift $\tau_i \to \tau_i+\tau_c$ before the Wick rotation. This is used in the calculation of the mixed form.

$\bullet$ This method is particularly easy if we know the analytical form of the
integrand. This is the case in our
paper. However, even if we do not,
one can do the same Wick rotation for the equations of motion, such as
(\ref{modefun_u}) and (\ref{modefun_v}),
\bea
\frac{d^2 u_k}{dx^2} - \frac{2}{x} \frac{d u_k}{dx} -
k^2 u_k &=& 0 ~,
\\
\frac{d^2 v_k}{dx^2} - \frac{2}{x} \frac{d v_k}{dx} -
k^2 v_k + \frac{m^2}{H^2 x^2} v_k &=& 0 ~,
\eea
and solve
the mode functions with the boundary condition
\bea
Ru_k~,~v_k \to
- \frac{H}{\sqrt{2k}} x e^{k x} ~, ~~~{\rm as}~ x\to -\infty ~.
\eea

\section{Perturbation series in the in-in formalism}
\label{App:inin}
\setcounter{equation}{0}

In this appendix, we list various forms of in-in formalism. We start from
\cite{inin}

\begin{equation}\label{iningeneral}
  \langle Q(t) \rangle \equiv
\langle 0| \left[ \bar T \exp\left( i\int_{t_0}^t dt' H_I(t')\right)
  \right] Q_I(t)
  \left[T \exp\left( -i\int_{t_0}^t dt' H_I(t')\right)
  \right] |0\rangle~.
\end{equation}
One can expand Eq.~\eqref{iningeneral} into series in two forms, namely, the
factorized form and the commutator form. The $n$th order contribution
of the factorized form is
\begin{align}\label{ininfactorizedeven}
&i^n (-1)^{n/2}\int_{t_0}^{t}d\bar t_1 \int_{t_0}^{\bar t_1}d\bar{t}_2 \cdots
\int_{t_0}^{\bar t_{n/2-1}}d\bar{t}_{n/2}
\int_{t_0}^{t}dt_1 \int_{t_0}^{t_1}dt_2 \cdots
\int_{t_0}^{t_{n/2-1}}dt_{n/2} \nonumber\\
\times&
\langle H_I(\bar t_{n/2})\cdots
 H_I(\bar t_{1}) Q_I(t) H_I(t_1)\cdots H_I(t_{n/2}) \rangle
 \nonumber\\
+&2{\rm Re}\sum_{m=1}^{n/2} i^n (-1)^{m+n/2} \int_{t_0}^{t}d\bar t_1
 \int_{t_0}^{\bar t_1}d\bar{t}_2\cdots
\int_{t_0}^{\bar t_{n/2-1-m}}d\bar{t}_{n/2-m}
\int_{t_0}^{t}dt_1 \int_{t_0}^{t_1}dt_2 \cdots
\int_{t_0}^{t_{n/2-1+m}}dt_{n/2+m} \nonumber\\ \times&
\langle H_I(\bar t_{n/2-m})\cdots
 H_I(\bar t_{1}) Q_I(t) H_I(t_1)\cdots H_I(t_{n/2+m}) \rangle
\end{align}
for even $n$, and
\begin{align}\label{ininfactorizedodd}
&2{\rm Re}\sum_{m=1}^{(n+1)/2} i^{n}(-1)^{m+(n-1)/2} \int_{t_0}^{t}d\bar t_1
 \cdots
\int_{t_0}^{\bar t_{(n-1)/2-m}}d\bar{t}_{(n+1)/2-m}
\int_{t_0}^{t}dt_1 \cdots
\int_{t_0}^{t_{(n-3)/2+m}}dt_{(n-1)/2+m} \nonumber\\ \times&
\langle H_I(\bar t_{(n+1)/2-m})\cdots
 H_I(\bar t_{1}) Q_I(t) H_I(t_1)\cdots H_I(t_{(n-1)/2+m}) \rangle
\end{align}
for odd $n$.

The $n$th order contribution of the commutator form takes the form
\cite{Weinberg:2005vy}
\begin{align}\label{inincommutator}
  i^n \int_{t_0}^t dt_1 \int_{t_0}^{t_1} dt_2 \cdots \int_{t_0}^{t_{n-1}}dt_n \left\langle
\left[ H_I(t_n),\left[H_I(t_{n-1}),\cdots ,
\left[ H_I(t_1),Q_I(t) \right]
\cdots \right] \right]
\right\rangle~.
\end{align}
In the factorized form, all
the integrands are under a unique integration.

Each form has its computational advantages and disadvantages, as we
discussed in detail in Sec.~\ref{Sec:bispectra}.

We can also rewrite all the integrals in terms of anti-time-ordered
integral instead of time-ordered. Namely, the factorized form can be
written as
\begin{align}\label{tmp1}
&i^n(-1)^{n/2}\int_{t_0}^{t}d\bar t_1 \int^{t}_{\bar t_1}d\bar{t}_2 \cdots
\int^{t}_{\bar t_{n/2-1}}d\bar{t}_{n/2}
\int_{t_0}^{t}dt_1 \int^{t}_{t_1}dt_2 \cdots
\int^{t}_{t_{n/2-1}}dt_{n/2} \nonumber\\
\times&
\langle H_I(\bar t_{1})\cdots
 H_I(\bar t_{n/2}) Q_I(t) H_I(t_{n/2})\cdots H_I(t_{1}) \rangle
 \nonumber\\
+&2{\rm Re}\sum_{m=1}^{n/2} i^{n}(-1)^{n/2+m} \int_{t_0}^{t}d\bar t_1
 \int^{t}_{\bar t_1}d\bar{t}_2\cdots
\int^{t}_{\bar t_{n/2-1-m}}d\bar{t}_{n/2-m}
\int_{t_0}^{t}dt_1 \int^{t}_{t_1}dt_2 \cdots
\int^{t}_{t_{n/2-1+m}}dt_{n/2+m} \nonumber\\ \times&
\langle H_I(\bar t_{1})\cdots
 H_I(\bar t_{n/2-m}) Q_I(t) H_I(t_{n/2+m})\cdots H_I(t_{1}) \rangle
\end{align}
for even $n$, and
\begin{align}\label{tmp2}
&-2{\rm Re}\sum_{m=1}^{(n+1)/2} i^n (-1)^{m+(n-1)/2} \int_{t_0}^{t}d\bar t_1
 \cdots
\int^{t}_{\bar t_{(n-1)/2-m}}d\bar{t}_{(n+1)/2-m}
\int^{t}_{t_0}dt_1 \cdots
\int^{t}_{t_{(n-3)/2+m}}dt_{(n-1)/2+m} \nonumber\\ \times&
\langle H_I(\bar t_1)\cdots
 H_I(\bar t_{(n+1)/2-m}) Q_I(t) H_I(t_{(n-1)/2+m})\cdots H_I(t_1) \rangle
\end{align}
for odd $n$. And the commutator form can be written as
\begin{align}\label{tmp3}
  i^n \int_{t_0}^t dt_1 \int_{t_1}^{t} dt_2 \cdots
  \int_{t_{n-1}}^{t}dt_n
\left\langle
\left[ H_I(t_1),\left[H_I(t_2),\cdots ,
\left[ H_I(t_n),Q_I(t) \right]
\cdots \right] \right]
\right\rangle~.
\end{align}

\end{document}